\documentclass[doublecolumn]{aa}

\usepackage{graphicx}
\usepackage[intlimits,sumlimits]{amsmath}
\usepackage{deluxetable}
\usepackage{times,epsfig} 
\usepackage{amssymb}
\usepackage{mathrsfs}  
\usepackage{natbib}
\bibpunct{(}{)}{;}{a}{}{,} 
\usepackage{caption}
\usepackage{amsmath}
\usepackage[english]{babel}
\usepackage{longtable}
\usepackage{url}
\usepackage{hyperref}

\begin{document}

\title{{\em Carnegie Supernova Project}: Observations of Type~IIn supernovae
\thanks{Based on observations
 collected at the European Organisation for Astronomical Research in the Southern Hemisphere, Chile (ESO Programs 076.A-0156, 078.D-0048 and 082.A-0526). This paper includes data gathered using the 6.5 m Magellan Telescopes, which are located at Las Campanas Observatory, Chile.}}

\author{F. Taddia\inst{1}
\and M.~D. Stritzinger\inst{2}
\and J. Sollerman\inst{1}
\and M.~M. Phillips\inst{3}
\and J.~P. Anderson\inst{4}
\and L. Boldt\inst{5}
\and A. Campillay\inst{3}
\and S. Castell\'{o}n\inst{3}
\and C. Contreras\inst{3}
\and G. Folatelli\inst{6}
\and M. Hamuy\inst{4}
\and E. Heinrich-Josties\inst{7}
\and W. Krzeminski\inst{8}
\and N. Morrell\inst{3}
\and C.~R. Burns\inst{7}
\and W.~L. Freedman\inst{7}
\and B.~F. Madore\inst{7}
\and S.~E. Persson\inst{7}
\and N.~B. Suntzeff\inst{9,10}}

\institute{The Oskar Klein Centre, Department of Astronomy, Stockholm University, AlbaNova, 10691 Stockholm, Sweden\\ (\email{francesco.taddia@astro.su.se})
\and
 Department of Physics and Astronomy, Aarhus University, Ny Munkegade 120, DK-8000 Aarhus C, Denmark
 \and
Carnegie Observatories, Las Campanas Observatory, Casilla 601, La Serena, Chile
\and
Departamento de Astronomia, Universidad de Chile, Casilla 36D, Santiago, Chile
\and
Argelander Institut f\"ur Astronomie, Universit\"at Bonn, Auf dem H\"ugel 71, D-53111 Bonn, Germany
\and
Kavli Institute for the Physics and Mathematics of the Universe (IPMU), University of Tokyo, 5-1-5 
Kashiwanoha, Kashiwa, Chiba 277-8583, Japan
\and
  Observatories of the Carnegie Institution for Science, 813 Santa Barbara Street, Pasadena, CA, USA
\and
N. Copernicus Astronomical Center, ul. Bartycka 18, 00-716 Warszawa, Poland
\and
Department of Physics and Astronomy, Texas A\&M University, College Station, TX 77845, USA
\and
The Mitchell Institute for Fundamental Physics and Astronomy, Texas A\&M University, College Station, TX 77845, USA}

\date{Received 29/01/2013 / Accepted 10/04/2013}

\abstract
{}
{The observational diversity 
displayed by various Type~IIn supernovae (SNe~IIn) is explored and quantified. 
In doing so, a more coherent picture ascribing the variety of observed SNe~IIn types to particular progenitor scenarios is sought.}
{{{\em Carnegie Supernova Project}} (CSP) optical and near-infrared light curves and visual-wavelength spectroscopy  of the Type IIn SNe~2005kj, 2006aa, 2006bo, 2006qq, and 2008fq are presented. Combined with previously published observations of the Type~IIn SNe 2005ip and 2006jd, the full CSP sample is used to derive physical parameters that describe the nature of the interaction between the expanding SN ejecta and the circumstellar material (CSM).}
{For each SN of our sample, we find counterparts, identifying objects similar to SNe~1994W (SN~2006bo), 1998S (SN~2008fq), and 1988Z (SN~2006qq). We present the unprecedented initial $u$-band plateau of SN~2006aa, and its peculiar late-time luminosity and temperature evolution. For each SN, mass-loss rates of 10$^{-4}-$10$^{-2}$~$M_{\odot}$~yr$^{-1}$ are derived, assuming the CSM was formed by steady winds. Typically 
wind velocities of a few hundred km~s$^{-1}$ are also computed.}
{The CSP SN~IIn sample seems to be divided into subcategories rather than to have exhibited a continuum of observational properties. The wind and mass-loss parameters would favor a luminous blue variable progenitor scenario. However
the assumptions made to derive those parameters strongly influence the results, and therefore, other progenitor channels behind SNe~IIn cannot be excluded at this time.}


\maketitle

\section{Introduction}

\citet{schlegel90} was the first to introduce the narrow-line Type~IIn supernova 
(SN~IIn) subclass to accommodate objects like SN~1987F \citep{wegner96} and SN~1988Z \citep{stathakis91,turatto93}, 
which exhibit prevalent narrow-emission features. 
 Today, members of this SN subclass cover a broad range of observational properties; however, they share the 
 commonality of early phase spectra 
 containing narrow Balmer emission features placed over a blue continuum. 
 The most prominent features (e.g. H$\alpha$) typically exhibit multiple components, which
 consist of narrow cores (full-width-at-half-maximum FWHM$_n$ of tenths to hundreds of km~s$^{-1}$) that are often situated on 
 top of intermediate-velocity components (FWHM$_i \sim 1000$~km~s$^{-1}$) and broad bases (FWHM$_b$ of a few thousand km~s$^{-1}$).

The multicomponent emission features 
are powered through a combination of SN emission and 
circumstellar interaction (CSI) processes.
CSI occurs when the SN blast-wave shocks circumstellar material (CSM) \citep[e.g.][]{chugai94,fransson02}, which originates in prevalent winds and/or violent eruptions
 of the progenitors during their pre-SN evolutionary phase \citep{kudritzki00}. 

In general, one finds that a forward shock is generated and propagates through the pre-SN wind as the outward moving SN blast-wave shocks CSM, while a reverse shock forms and propagates back through the expanding SN ejecta. 
 At the interface between the forward and reverse shocks, a cold dense shell
 (CDS) of gas forms and rapidly cools through collisional de-excitation processes. 
 Within the shock region, high-energy radiation is produced, which subsequently ionizes the un-shocked pre-SN wind. As this gas recombines, narrow emission lines are produced that can span a wide range in ionization potential 
 \citep[e.g.][]{chevalier94,chugai94,smith09_05ip}.
 
 Within this general picture, the broad-line component is typically associated with 
SN emission, although one cannot rule out thermal broadening of the narrow core 
for an optically thick wind \citep{chugai01}. The intermediate-component originates in gas emitting in the post-shock region. Strong CSI also generates
 X-ray and radio emission \citep[e.g.][]{nymark06,dwarkadas12,chandra12}
and may also have an important influence on producing 
the necessary conditions to form dust, which has been detected in 
a handful of SNe~IIn \citep[see e.g.][]{fox11}. 

Over the last two decades, observations have led to the realization 
of a remarkable heterogeneity between the photometric 
 and spectroscopic properties, which have been exhibited by various SNe~IIn \citep[e.g.][]{pastorello02,kiewe12}. 
Today, the SNe~IIn class consists of objects that show a large range in peak luminosities. At the very bright end of the distribution are 
SN~2006gy-like objects {\citep{smith07}}, which can reach peak absolute 
magntiudes of 
$M_R\lesssim$ $-$21, while at the faint end are objects like SN~1994aj, which 
reach peak values of $M_R\gtrsim$ $-$17 mag \citep{benetti98}.

Some SNe~IIn are observable in the optical for years past discovery (e.g. SN~1995N, \citealp{fransson02}),
while others fade to obscurity within months (e.g SN~1994W, \citealp{sollerman98}). 
Moreover, objects discovered soon after explosion indicate 
rise times ranging from several days to tens of days 
\citep[see][]{kiewe12}.
Around peak brightness, a handful of SNe~IIn show Gaussian-shaped light curves
(e.g. SN~2006gy), whereas others like SN~1994W exhibit a 
 plateau light-curve phase that is reminiscent of Type IIP SNe \citep{kankare12,94Wlike}. 
 The evolution of SN~IIn light curves post-maximum also shows a 
 large degree of heterogeneity. 
 
Close examination of SN~IIn spectra not only reveals the presence of multicomponent emission features but also a variety of emission line profiles, ranging from symmetric to highly asymmetric.  These line profile asymmetries are often observed to grow over time.
For example, prevalent attenuation of the red wing of H$\alpha$ in the cases of SNe~2006jd \citep[][, hereafter S12]{stritzinger12} and 2010jl \citep{smith12}
 was observed to increase over the first six months from discovery. 
 Depending on the nature of the SN-CSM interaction, this behavior 
could be linked to increasing absorption due to dust 
condensation or to the occultation of radiation by a thick photosphere and/or surrounding CSM (e.g. a thick disk). 

CSI has also been observed to occur in hydrogen poor core collapse (CC) SNe, which interact with helium-rich CSM. The prototype of this phenomenon 
is the Type~Ibn SN~2006jc \citep{pastorello08}.
Furthermore, some thermonuclear explosions may occur in a hydrogen-rich environment, producing Type~IIn-like SNe~Ia. 
Examples of these events include SNe~2002ic
\citep{hamuy03}, 2005gj \citep{aldering06,prieto07}, 2008J \citep{taddia12}
and PTF~11kx \citep{dilday12}.

Progenitor stars of SNe~IIn must eject enough mass to enable efficient CSI.
A luminous blue variable (LBV) star, which was identified in pre-SN explosion images, has been linked to the Type~IIn SN~2005gl \citep{galyam09}.
More recently, SN~2009ip has also been linked to an LBV \citep{mauerhan12}, which has exhibited episodic rebrightening events three years prior to core collapse.
However, \citet{pastorello12} and \citet{soker12} have suggested that the last bright event of SN~2009ip is due to other mechanisms rather than to core-collapse.
LBVs are known to suffer significant mass loss ($\dot{M}\sim 10^{-2}-10^{-3}~M_\odot~yr^{-1}$) through 
episodic outbursts that occur on time scales ranging from months to years.
This behavior is well-documented through detailed observations of Eta Carinae 
and other ``SN impostors" \citep[e.g. see][]{smith11}.

Nevertheless, the variety of SN~IIn properties suggests that their origins may lie 
within a diversity of progenitor stars.
For example, \citet{fransson02} and \citet{smith09} have suggested that red supergiant (RSG) stars, which experience a pre-SN mass-loss phase driven by a superwind, could be viable progenitors for a large fraction of observed SNe~IIn. This behavior is observed in VY Canis Majoris.

 A proper determination of the progenitor mass-loss history is crucial to discern the nature of the progenitor itself. Estimates of mass-loss rates found in the literature often assume that the progenitor wind is steady, even though X-ray observations demonstrate that this is likely not the case for the majority of objects \citep{dwarkadas11}. 
 Ad hoc assumptions on the density profiles of progenitor pre-SN winds
 can therefore lead to incorrect mass-loss rate estimates. 
 From the theoretical point of view, steady, line-driven winds 
  are the only mass-loss mechanism that is reasonably understood \citep[but see][and reference therein]{smith11rates}, whereas the physics behind episodic and eruptive mass loss behavior is still in its infancy \citep[see e.g.][]{smith06,yoon10}.

In this paper, we present and analyze optical and near-infrared (NIR) light curves and visual-wavelength spectroscopy of five SNe~IIn observed by 
the {\em Carnegie Supernova Project} (CSP, \citealp{hamuy06}).
These new objects are SNe 2005kj, 2006aa, 2006bo, 2006qq, and 2008fq. 
Basic information of these SNe~IIn is provided in 
Tables~\ref{sample} and \ref{cbet}. 
In addition to these five objects, the full CSP SN~IIn sample includes detailed observations of SNe~2005ip and 2006jd, which are presented by S12.

The organization of this paper is as follows:
Sect.~\ref{sec:reduction} contains details
concerning data acquisition and reduction;
Sect.~\ref{sec:sample} describes the CSP SN~IIn sample;
Sect.~\ref{sec:phot} reports the 
broad-band optical and near-infrared photometry;
Sect.~\ref{sec:spectra} describes the 
spectroscopic observations; Sect.~\ref{sec:discussion} presents
the discussion and the interpretation of the data; and finally, Sect.~\ref{sec:conclusions} presents the conclusions.
 
\section{Observations}
\label{sec:reduction}

Optical and NIR photometry of the five SNe~IIn, which are presented in this paper,
 were computed from images obtained 
with the Henrietta Swope 1 m and the Ir\'en\'ee du Pont 2.5 m telescopes
located at the Las Campanas Observatory (LCO).
The vast majority of optical imaging were taken with the Swope, which was equipped with the
 direct imaging camera known as ``Site3" and a
set of Sloan $ugri$ and Johnson $BV$ filters.
A small amount of optical imaging was also obtained for two objects, SNe~2006aa and 2006bo, using the du Pont, which was equipped with a direct imaging camera named after its chip Tek-5, also known as ``Tek-cinco". 
NIR imaging was performed with both the Swope ($+$ RetroCam) and du Pont 
($+$ Wide Field IR Camera, WIRC), which were equipped with a set of $YJH$ filters. 
A small amount of $K_s$-band imaging was also obtained with the du Pont for SNe~2005kj and 2008fq.
 The Magellan (Baade) telescope at LCO was also used to get three epochs of $K_s$-band imaging for SN~2005kj.
Complete details concerning the instruments and passbands used by the CSP can be found
in \citet{hamuy06}, \citet{contreras10} and \citet{stritzinger11}.

A detailed description of CSP observing techniques and 
data reduction pipelines can be found 
in \citet{hamuy06} and \citet{contreras10}. 
In short, each optical image undergoes a series of reduction steps consisting of
(i) bias subtraction, (ii) flat-field correction, and (iii) the application of
a shutter time and CCD linearity corrections. 
Given the nature of observing in the NIR, the basic reduction process 
for these images consists of (i) dark subtraction, (ii) flat-field correction, (iii) 
sky subtraction, (iv) non-linearity correction, and finally, (v) the geometrical alignment and combination of each 
dithered frame. 

Deep template images of each SN host galaxy were obtained using the
du Pont telescope under excellent seeing conditions.
The templates are used to subtract away the host contamination 
 at the position of the SN from each science image \citep[see][for details]{contreras10}.
{In Table~\ref{tabtemplate}, the UT date of each template observation is reported.
In some cases (e.g. SNe~2005kj and 2006qq), we had to wait several years after discovery to obtain useful templates without SN contamination because of the long-lasting CSI and late-time dust emission. SNe~2005ip and 2006jd are still bright, and templates are not available (S12).}

PSF photometry of each SN was computed differentially with respect to a local photometric sequence of stars.
 The calibration of the photometric sequences was done with respect to 
 \citet{smith02} ($ugri$), \citet{landolt92} ($BV$), 
 and \citet{persson98} ($YJHK_s$) standard fields observed over a minimum of 
 three photometric nights. 
The local sequences for each CSP SN~IIn in the 
 standard photometric systems are provided in Table~\ref{tabstars}, 
 while final SN optical and NIR photometry {\em in the natural system} are listed in 
 Table~\ref{tabphot} and Table~\ref{tabphotIR}, respectively.

Visual-wavelength spectroscopy of the CSP SN~IIn sample was obtained using the Du Pont and Magellan (Baade ad Clay) telescopes at LCO and the ESO New 
Technology Telescope (NTT) at the La Silla Observatory.
A spectrum of SN~2005kj was obtained with the 60-inch telescope
at the Cerro Tololo Inter-American Observatory (CTIO).
These data were reduced in a standard manner as described by 
\citet{hamuy06}, which included: (i) overscan correction, (ii) bias subtraction, (iii) flat-fielding, 
(iv) 1-D spectrum extraction, (v) wavelength calibration
(vi) telluric corrections, (vii) flux calibration, and (viii) combination of
multiple exposures. 
Finally, the flux of each spectrum
was adjusted to match broad-band
photometry.
The journal of the spectroscopic observations is provided in 
Table~\ref{tabspectra}.

\section{The CSP SN~IIn sample}
\label{sec:sample}

 Table~\ref{sample} lists the properties of the full CSP SN~IIn sample, which include the 
 SN coordinates, its host galaxy name, type, distance, the estimates of 
 total absorption [$E(B-V)_{MW}$ $+$ $E(B-V)_{host}$] 
 and the equivalent width (EW) of interstellar \ion{Na}{i}~D absorption. 
 Finding charts for the five new CSP SNe~IIn are given in Fig.~\ref{fc}.
 As most of the SNe followed by the CSP were discovered by targeted surveys,
 it is not surprising that the hosts of the CSP SN~IIn sample are 
bright, nearby spiral galaxies in the redshift range of 0.007$\leq$z$\leq$0.029. 
 Quoted distances to each event correspond to the luminosity distances that are provided by 
 NED, which include corrections for peculiar motion (GA+Shapley+Virgo) and are based on cosmological parameters obtained by WMAP5 \citep{komatsu09}. 
 Galactic absorption values were taken from NED \citep{schlafly11},
while estimates of $E(B-V)_{\rm host}$ were determined through the combination of
EW measurements of interstellar \ion{Na}{i}~D absorption lines
 at the host-galaxy redshift and the relation $E(B-V)_{\rm host}=0.16 \cdot\rm EW({\rm \ion{Na}{i}~D})_{\rm h}$ \citep{turatto03}. 
 Only two objects display significant \ion{Na}{i}~D absorption.
SN~2006qq exhibits clear \ion{Na}{i}~D absorption at the host rest frame in the spectrum taken on January 1, 2007. We note that the host galaxy of this object is remarkably tilted, exhibiting an inclination between the line of sight and its polar axis of 85.6~deg.
SN~2008fq, which is located close to the center of its host, shows prominent narrow absorption lines of \ion{Na}{i}~D in each spectrum, although the individual D1 and D2 components are not resolved. 
In the case of SN~2008fq, we take the mean of the EW measurements as obtained from 
our series of spectra. We determined $E(B-V)_{\rm host}$ $=$ 0.12 and 0.46~mag for SN~2006qq and 2008fq, respectively. We note, however, that the EW of \ion{Na}{i}~D from low resolution spectra is a debatable proxy for the extinction of extragalactic sources, and it does not work for SNe~Ia \citep{poznanski11}.

Table~\ref{cbet} indicates that several of the objects were discovered within a few days to weeks past the last non-detection. Specifically for three SNe~IIn, we have early observations and good constraints on their explosion epoch 
(see column 5 in Table~\ref{cbet}). Only SNe~2005kj and 
2006bo have poorly constrained explosion dates. As noted in S12, SNe~2005ip and 2006jd are believed to have been discovered soon after explosion. 
In what follows, the discovery date is used to define the temporal phase of each object.

Plotted in Fig.~\ref{lc} are the optical and NIR light curves of the five new CSP SNe~IIn. 
Typically, each SN was observed for approximately 100 days with a cadence of several days, however, the SN~2005kj observations extend to over 180 days.
Due to scheduling constraints, the NIR coverage typically is with a lower cadence. 

As indicated in Table~\ref{tabspectra} the number of optical spectra for this sample of five SNe~IIn vary from two (SN~2006bo) to 15 (SN~2005kj) epochs.
Spectroscopic follow-up started from a maximum of 19 days after discovery (i.e. SN~2006aa) down to a minimum of 2 days after discovery (i.e. SN~2008fq). 
Depending on the exact instrument used, the wavelength coverage typically ranges 
from 3800~\AA\ to 9300~\AA\ (see Table~\ref{tabspectra}).
The spectroscopic sequences of these five SNe~IIn are presented in 
Fig.~\ref{spec1} and Fig.~\ref{spec2}. Included in each panel as colored diamonds are 
photometric broad-band measurements derived from interpolated light curves.

Our dataset is also complemented with data published in the literature.
Several of the SNe~IIn in our sample were observed in the mid-infrared (MIR) with 
the {\em Spitzer Space Telescope} \citep{fox11}. 
In particular, late-phase Spitzer observations of SN~2006qq  
(approximately 1050 days past discovery) found this object to be 
quite bright at $\lambda>$3.6~$\mu$m, while similar imaging of 
SNe~2006aa and 2006bo showed no MIR emission.

\section{Photometry}
\label{sec:phot}
\subsection{Light curves}
\label{subsec:lcs}

SN~2005kj was observed only after maximum light, and its explosion date is 
not well constrained by pre-explosion images. Its optical light curves (Fig.~\ref{lc}) show a decline of approximately 3.5~mag
in the first 100 days in the $u$ band, whereas the decline is slower for the redder bands in the same time interval ($i$ band becomes only 0.8~mag fainter).
After 100 days the light curve slope steepens, the $i$ band
decays by 1.4~mag in the following 70 days, and the bluer bands show a higher rate of decline. The NIR light curves exhibit a somewhat slower evolution at all phases. The SN was detected even at 382 days in $Y$ and $J$ bands (see Table~\ref{tabphotIR}). 

SN~2006aa was discovered relatively young and has an uncertainty of $\pm$8 days on the explosion date. 
Its light curves are characterized by a ``Gaussian'' shape with an initial rise that culminates at maximum light approximately 50 days after discovery for all the filters. Interestingly, the super-luminous SNe~II \citep{galyam12} and the peculiar SN~Ib/c 2005bf \citep{folatelli06} peak at 50-60 days. However, no sign of interaction was seen in the case of SN~2005bf, and its long rise time can be attributed to the delay in the diffusion of photons through the ejecta. The rise in the $u$ band follows an initial 10 day ``plateau''. This feature is observed for the first time in the SN~IIn class for which a small number of early time observations in the near ultraviolet exists. We note that the SN~IIn 2011ht did not present this ``plateau", although it was discovered soon after explosion \citep{roming12}. After maximum light, the light-curve decline is steeper in the bluer bands, as seen in SN~2005kj, with $B$ and $i$ bands fading by approximately 2~mag 
 and 1~mag respectively in about 80 days.

SN~2006bo was discovered after maximum light (its explosion date is not well-constrained from pre-SN images), and its optical light curves 
are characterized by an initial linear decline (in mag), which is more pronounced in the bluer bands. In the first 40 days, the $u$ band faded 
by approximately 2~mag, whereas the $i$ band faded by only 0.5~mag.
The last optical observation that was obtained approximately 150 days after discovery and 3 months subsequent to the previous 
 optical observation reveals an increase in the decline rate some point after 60 days. The NIR light curves cover only the first tens of days after discovery and are 
nearly flat, showing only a marginal decline in brightness.

SN~2006qq was observed after its maximum light in the optical bands and before its maximum light in the NIR. Each optical band is characterized by a decline, which becomes progressively less steep with time and wavelength. The NIR light 
curves show a maximum emission at 20--30 days after discovery. The SN was detected in $H$ band later than 700 days after discovery (see Table~\ref{tabphotIR}). 

SN~2008fq was discovered prior to 9 days after explosion.
It has a light curve that exhibits a fast rise, which culminates in an early maximum (approximately 10 days in $B$ band, 20 days in $i$ band, and a few days later in the NIR). Then a 
steep decline lasts approximately 2 weeks, which is followed by a slow, linear decline until the end of the observations (approximately day 70). 

We determined the slopes of the optical light curves by dividing the difference in magnitude of contiguous photometric epochs by their temporal difference and then fitting the obtained slopes with low-order polynomials.
The slope evolutions confirm that bluer bands tend to decline faster than redder bands after maximum light (this is not true for the long-lasting SNe~2005ip, 2006jd, and for SN~2006qq). The decline after maximum light is not linear at least for those objects which were observed for less than 100 days, whereas SN~2006qq shows an almost linear decline in each band after 100 days. For SNe~2005kj and 2006aa, the decline rate increases with time; for SNe~2005ip, 2006jd and 2006bo, the decline rate decreases. When we compare the decay rate of $^{56}$Co (0.0098~mag~day$^{-1}$) with that of the 
$V$ band for the five SNe that are observed after 100 days since discovery, we find that only SN~2006qq (and marginally SN~2005ip) has a slope 
compatible with radioactive decay. SN~2006bo might also be characterized by a late-time decline rate, which is compatible with radioactive decay. Unfortunately, the data do not allow us to confirm it. Decay rates at 100 days are summarized in Table~\ref{tabslope}.

\subsection{Absolute magnitudes and colors}

Armed with the distances and $E(B-V)_{tot}$ color excess values listed in Table~\ref{sample}, absolute magnitude light curves were computed for the entire CSP SN~IIn sample and were plotted in Fig.~\ref{absmag} (top panels).
In doing so, a standard reddening law 
characterized by an $R_V$ $=$ 3.1 \citep{cardelli89} was used to convert 
$E(B-V)_{tot}$ to absorption in each photometric bandpass. We assume that the host reddening is zero for the SNe with no detectable \ion{Na}{i}~D.
 
Since it is approximately 1~mag more luminous than SNe~2005kj, 2006aa, and 2006qq, SN~2008fq turns out to be the intrinsically brightest object in the sample.
We note that SN 2008fq was corrected for a substantial amount of extinction ($A_V$ $=$ 1.6 mag), which is derived from \ion{Na}{i}~D absorption that is known to provide poor 
constraints on reddening \citep{blondin09,folatelli10,poznanski11}. 
Nevertheless, when adopting the \ion{Na}{i}~D-based color excess, 
the $r$-band light curve of SN~2008fq 
reaches a maximum value only 0.2 mag brighter than that of SN~1998S \citep{liu00}, which exhibits similar light curve shape and 
spectral properties (see bottom panel of Fig.~\ref{absmag} and Sect.~\ref{sec:spectracomp}).
 With this color excess, the intrinsic broadband colors of SN~2008fq also appear consistent to those of SN~1998S, as revealed in the top panel of Fig.~\ref{color}.
 Peaking at $V$ $=$ $-$19.3~mag, SN~2008fq belongs to the bright-end of the ``normal" SN~IIn category, whereas the super-luminous SNe~IIn exhibit absolute magnitudes brighter than $-21$ \citep{galyam12}. 

Since it is approximately 2~mag less luminous than SN~2008fq at peak and never brighter than $V$ $\approx-$17 mag, SN~2006bo is the faintest object within the full CSP SN~IIn sample.
The $r$-band light curve evolves along a plateau phase over the first two months of monitoring and, as revealed in Fig.~\ref{absmag}, is similar in overall shape to the 1994W-like SN~2011ht \citep{94Wlike} but with a somewhat fainter luminosity along the plateau.
Since the explosion epoch of SN~2006bo is poorly constrained, it may have been discovered weeks after explosion. This possibility would then imply a higher tail luminosity as compared to that of SN~2011ht.

SNe~2005kj, 2006aa, and 2006qq have peak $V$- and $r$-band 
absolute magntiudes between $-$17.7 and $-$18.5 mag.
Among them, SN~2006qq decays similarly to the slow-declining and long-lasting 
SNe~2005ip and 2006jd (S12), although it is 
brighter than those objects (see Fig.~\ref{absmag}). 
In addition, SN~2005kj exhibits a relatively high luminosity for more than 180 days, but its decline 
rate beyond 100 days is larger than that of SN~2006qq. 
The evolution and brightness of the $r$-band light curve of SN~2006aa closely resembles that of 
SN~2005db \citep{kiewe12}.
 However, the rise time for SN~2006aa is particularly long (approximately 50 days).
 SN~2006aa also displays faint emission in the NIR, since it is approximately 1~mag fainter than the bright SN~2008fq.

Estimates of the peak absolute magnitudes, derived from the individual light curves of each SN, are given in Table~\ref{max}. 
These values have been determined by fitting the light curves with low-order polynomials.
For completeness, the peak absolute magnitudes of SNe~2005ip and 2006jd are also 
included in Table~\ref{max} (S12).

Figure~\ref{color} contains the comparison of intrinsic $B-V$, $V-i$, and $J-H$ colors of the full CSP SN~IIn sample. For comparison, the panel containing the $B-V$ color curve evolution also includes several well-observed objects taken from the literature. 
Both the $B-V$ and $V-i$ color curves indicate that SNe~IIn 
evolve to the red over time, and in this sample,
SNe~2008fq and 2006bo show the largest $B-V$ color evolution ranging from 0.2~mag to 1.0~mag over a period of approximately 50 days.
 SN~2006bo appears to be much redder than the 1994W-like SN~2011ht, even though the light curve shapes are similar.
SNe~2005kj and 2006qq experience a slow cooling, which is comparable to that of the long-lasting SNe~IIn 2005ip and 2006jd. 
In particular, SN~2005kj takes approximately 120 days to reach a $B-V$ $=$ 1.0~mag.
Interestingly, SN~2006aa first evolves to the blue, then reaches a minimum value 
 in the optical colors at the time of maximum light, and then evolves
 to the red over time.

The NIR ($J-H$) color curves follow more flat trajectories compared to the optical color curves. 
However, the $J-H$ color curve clearly exhibits values that increase over time in the case of SNe~2005ip and 2006jd, and this evolution is likely driven by 
emission related to newly formed dust \citep[][S12]{fox11}.

\subsection{Bolometric properties}

The photometric dataset of our CSP SN~IIn sample is well-suited to 
construct quasi-bolometric light curves. 
To do so requires the interpolation (or when necessary extrapolation) of NIR magnitudes at epochs that coincide with the optical observations. 
Next, broad-band magnitudes were converted to fluxes at the effective wavelength of each bandpass. After correcting each flux point for reddening, a spectral energy distribution 
(SED) at each observed epoch was fit with a spline, which resulted in a nearly 
complete UV-optical-NIR (UVOIR) SED. 
The SEDs were then integrated over wavelength space 
spanning from $u$ to $H$ band, and then the final 
 UVOIR luminosity was computed by multiplying the UVOIR flux
 with 4$\pi$D$_L^2$ (where D$_L$ is the luminosity distance).
The error on $L$ was obtained through simulating hundreds of SEDs (according to the uncertainties on the specific fluxes) and taking the standard deviation of their integrals. 

The final UVOIR light curves for the full CSP SN~IIn sample are plotted in the 
top panel of Fig.~\ref{bolo}. As observed in the $r$-band absolute magnitude
light curves, SN~2008fq and SN~2006bo are the brightest and faintest
objects within the CSP SN~IIn sample, respectively, while SN~2006aa and
SN~2008fq appear to have been discovered before reaching maximum.
The peak bolometric luminosities range from those 
of normal CC SNe~IIP to values 10 times higher, which are
comparable to those of luminous SNe~II \citep{inserra12}.

Inspection of the late phase portion of the UVOIR light curves reveals
that the sample exhibits a diversity of decline rates, ranging from values 
both above and below the expeced decline rate of the full trapping of energy produced by
the radioactive decay chain $^{56}$Co $\rightarrow$ $^{56}$Fe. 
SNe~2005kj and 2006aa exhibit steeper slopes than those expected for full trapping of gamma-rays that are produced 
from the $^{56}$Co decay, 
while the slopes of SNe~2005ip, 2006jd, and 2006qq are more shallow.
UVOIR light curve decline rates at 100 days after discovery are 
reported in Table~\ref{tabslope}. 
The inconsistency between observed and $^{56}$Co decline rates suggests radioactivity is not the prevalent energy source for these objects.

The SEDs were also used to estimate temperature ($T$) and radius ($R$) 
of the emitting region, by using black-body (BB) fits. When performing the BB fits, 
 $u$- and $B$-band flux points were excluded, because they sample portions of the optical spectrum that suffer a significant reduction in continuum flux because of line blocking effects, particularly at late epochs. 
 In addition, the $r$ band was also excluded from the BB fits
given the presence of prevalent H$\alpha$ emission.

The time evolution of $T$ and $R$ for the full CSP SN~IIn sample 
is plotted in the bottom portion of Fig.~\ref{bolo}.
During early epochs, $T$ monotonically drops over time for each SN except for SN~2006aa, which shows a peak at approximately 9000~K at the time of maximum light.
At epochs past 50 days after discovery, all objects decrease in $T$ except for SNe~2005ip and 2006jd,
which both show enhanced thermal emission related to dust condensation (S12). 
The maximum $T$ values range from approximately 7000 to 11,500~K and decline over time to values no less than 5500~K by about day 60. These values are consistent with SNe~IIn presented in the literature \citep[see e.g.][]{smith09_05ip}.

For all the SNe with a SED that is well fit by a single BB component (i.e. all objects, except SNe~2005ip and 2006jd), Fig.~\ref{bolo} also reveals that $R$ slowly increases over time until it reaches an almost constant value.  Typical radii extend to approximately 10$^{15}$~cm, which is also typical for many SNe~IIn in the literature (e.g. SN~1998S, \citealp{fassia01}). 
In a SN~IIn, most of the continuum emission is likely to come from an expanding thin cold-dense shell. This shell continually expands to
larger radii. However, the black-body (BB) radius appears to stall or in some cases, even to shrink, because the shell-covering factor decreases as the optical depth drops. This was discussed by \citet{smith08}, where they present SN~2006tf.

\section{Spectroscopy}
\label{sec:spectra}

In this section, the visual-wavelength spectra of the five new CSP SNe~IIn are analysed and compared 
to other objects in the literature. 
A special emphasis is placed on constraining the evolution of both luminosity and line profile shape for the H$\alpha$ and H$\beta$ emission features.
We note that the resolution of our spectra (see Table~\ref{tabspectra}) allow us to resolve H$\alpha$ FWHM velocities down to only 320~km~s$^{-1}$ for SNe~2008fq, while H$\alpha$ FWHM velocities are resolved down to 180~km~s$^{-1}$ for the other objects.

\subsection{SN~2005kj}

The spectral sequence of SN~2005kj (Fig.~\ref{spec1}, top panel) covers approximately 150 days
of evolution. Each spectrum exhibits prevalent, symmetric Balmer emission lines.
The left panel of Fig.~\ref{ha_5kj} displays the H$\alpha$ and H$\beta$ emission features, which were fit by multiple Lorentizian components, while the right panel contains the flux and velocity evolution of both features. Flux and velocity measurements are also reported in Tables~\ref{tab_flux05kj} and \ref{tab_vel05kj}.

 During the first 70 days of evolution, H$\alpha$ and H$\beta$ are characterized by a narrow FWHM$_{\rm n}$ $\approx$ 1000~km~s$^{-1}$ and
a broad FWHM$_{\rm b}$ $\approx$ 2500~km~s$^{-1}$ component.
Around 70 days, the broad component becomes difficult to 
detect, while H$\beta$ begins to form a 
P-Cygni profile with an absorption minimum of $v_{min}$ $=-$600--800~km~s$^{-1}$.

H$\beta$ is initially very bright and almost comparable to H$\alpha$ in luminosity. However, the Balmer decrement 
increases over time, and H$\beta$ is found to be much fainter at late epochs. The total flux of both H$\alpha$ and 
H$\beta$ drops over time.
The H$\gamma$ and H$\delta$ lines are also detected:
 both exhibit a P-Cygni profile that shows a minimum 
absorption with similar velocity to H$\beta$. After approximately 35 days, they are barely discernible from the noise.

\ion{He}{i}~$\lambda$5876 is visible until $\approx$ 20 days, showing narrow and broad components in emission.
Concerning the narrow absorption lines, \ion{Na}{i}~D is not detected during early epochs; however, the presence of \ion{Ca}{ii} H\&K 
$\lambda\lambda$3933,3968 is clear. 
Numerous \ion{Fe}{ii} and \ion{Ti}{ii} P-Cygni profiles are visible in the blue portion of the spectra of SN~2005kj from day 
$+$32.8. At epochs $\gtrsim$70 days, the \ion{Ca}{ii} triplet and \ion{O}{i}~$\lambda$8447 emerge, and at later times ($\gtrsim$100 
days), a P-Cygni profile identified with \ion{Na}{i}~D is also visible.
The continuum flux, initially dominated by UV and blue photons, progressively becomes red, as expected from the decreasing 
temperature (see Fig.~\ref{bolo}, middle panel).

\subsection{SN~2006aa}
The seven spectra that we obtained for SN~2006aa (Fig.~\ref{spec1}, bottom panel) are typical SN~IIn spectra, which are mainly 
characterized by Balmer emission lines (in particular H$\alpha$ and H$\beta$, but also H$\gamma$ and H$\delta$). As 
shown in Fig.~\ref{ha_6aa}, their continuum-subtracted, extinction-corrected profiles are well fit by the sum of a broad 
(FWHM$_{\rm b}\approx$2000~km~s$^{-1}$) and a narrow 
(FWHM$_{\rm n}\lesssim$700~km~s$^{-1}$) Lorentzian in emission. 
 In addition, a narrow Lorentzian in absorption is fit at 
$\approx-$800~km~s$^{-1}$ from the central wavelength, in order to reproduce the narrow absorption feature. The H$\alpha$ and H$\beta$ broad-component fluxes reach maximum values 
when the SN is at maximum light, whereas the 
narrow component exhibits nearly constant emission over time.
{Fluxes and velocities of the most important hydrogen lines are reported in Tables~\ref{tab_flux06aa} and \ref{tab_vel06aa}.}

\ion{Fe}{ii} features are observed in all the spectra (between 20 and 80 days). As in the case of SN~2005kj, narrow 
lines of \ion{Na}{i}~D in absorption are not detected at the host galaxy rest frame, whereas narrow \ion{Ca}{ii} H\&K absorption 
lines are observed. The spectral continuum is well fit by a BB function with temperature T$_{BB}$ $=$ 8000-10000~K.

\subsection{SN~2006bo}

Only two early-time spectra were collected for SN~2006bo (Fig.~\ref{spec2}, top panel). However, these reveal an interesting 
H$\alpha$ profile (see Fig.~\ref{ha_6bo}), which is characterized by a prevalent narrow emission (FWHM$_{\rm n}\approx$1000~km~s$^{-1}$) feature and an absorption component 
at $\approx-$600~km~s$^{-1}$. 
Identical features are present in H$\beta$ and in the faint 
H$\gamma$ line, where the absorption is even more pronounced. 
This is a case of a SN~IIn that shows no clear signatures of a broad component.
{H$\alpha$ and H$\beta$ fluxes and velocities are given in Tables~\ref{tab_flux06bo} and \ref{tab_vel06bo}.}

No \ion{Na}{i}~D absorption features are found in the spectra at the redshift of 
SN~2006bo; however, features attributed to \ion{Ca}{ii} H\&K are discernible. 
\ion{Fe}{ii} and \ion{Ti}{ii} P-Cygni profiles are also observed, while
weak traces of the \ion{Ca}{ii} NIR triplet characterize the red end of the spectrum.
The spectral continuum is well reproduced by a BB function with temperatures T$_{BB}$ $=$ 7000-8000~K.

\subsection{SN~2006qq}
The spectral sequence of SN~2006qq covers the first 70 days (Fig.~\ref{spec2}, middle panel). H$\alpha$ is the most conspicuous feature in emission. Besides its narrow, unresolved component (FWHM$_{\rm n}\lesssim$ 200~km~s$^{-1}$), a broad, strongly blue-shifted component emerges at 30 days, reaching maximum flux at approximately 60 days. At this epoch, both H$\alpha$ and H$\beta$ have an almost parabolic shape (see Fig.~\ref{ha_6qq}).
The blue and red velocities at zero intensity (BVZI and RVZI) of the asymmetric H$\alpha$ profile are
approximately 10,000~km~s$^{-1}$ and 3000~km~s$^{-1}$.
We note that the narrow Balmer lines are likely contaminated by the emission of an underlying \ion{H}{ii} region.
In the last two spectra, the broad Balmer emission significantly drops, indicating that after 70 days the ejecta-CSM interaction is weaker. This is consistent with the $r$-band ``jump" at similar epochs (see Fig.~\ref{lc}). 
{All the Balmer line fluxes and velocities are given in Tables~\ref{tab_flux06qq} and \ref{tab_vel06qq}.}

Beside the Balmer lines, the spectra are characterized by broad features in the blue, especially in the first two spectra. H$\beta$ falls between two of these broad features, which we identify as \ion{Fe}{ii} (similar broad features were identified in SN~1996L by \citealp{benetti99}).
At late times, the spectrum is flatter, and only Balmer lines are clearly identifiable on a continuum that is well represented by a
6000~K BB function.
In the spectrum taken on 1st January 2007, narrow \ion{Na}{i}~D absorption (EW$=$0.76$\pm$0.21~\AA) and traces of \ion{Ca}{ii} H\&K are present.
Narrow emission of [\ion{N}{ii}]~$\lambda$5755 are clearly observed until $\approx$ 70 days. In the first spectrum, narrow emission lines of \ion{He}{i}~$\lambda$5876 and $\lambda$7065 are also present.

\subsection{SN~2008fq}

The spectral sequence of SN~2008fq (Fig.~\ref{spec2}, bottom panel) shows strong evolution within the first month of follow-up, which includes the appearance of a broad H$\alpha$ emission line that dominates the last spectra. The analysis of the Balmer lines is presented in Fig.~\ref{ha_8fq}, where H$\alpha$ and $H\beta$ are shown after low-order polynomial continuum-subtraction and extinction correction. H$\alpha$ lines have also been fit by the sum of several Lorentzian functions to measure fluxes and typical velocities {(see also Tables~\ref{tab_flux08fq} and \ref{tab_vel08fq})}.
The first three spectra exhibit a narrow H$\alpha$ P-Cygni profile, characterized by unresolved emission (FWHM$_{\rm n}<$ 200--400~km~s$^{-1}$) and blue-shifted absorption with velocity $v_{min}$ $=-$ 400--500~km~s$^{-1}$. 
H$\beta$ presents a similar shape. 
A narrow [\ion{N}{ii}] emission is also present, indicating that the narrow Balmer lines are contaminated by the flux of an underlying \ion{H}{ii} region.
A faint and broad H$\alpha$ P-Cygni profile is also observable in the first two spectra with FWHM$_{\rm b}\approx$4000~km~s$^{-1}$ in emission and an absorption minimum at $v_{min}$ $=-$7000-8000~km~s$^{-1}$.
H$\beta$ also shows broad features, but the presence of nearby, broad \ion{Fe}{ii} lines 
makes it difficult to fit these components.
In the first two spectra ,narrow emission lines of \ion{He}{i}~$\lambda$5876 are also observed. This line is located close to the
 narrow \ion{Na}{i}~D absorption line, which is present in all five spectra. 
 \ion{Ca}{ii} H\&K is also detected in all of the spectra. 
 In the first spectrum, the emission feature at 4640$-$4690~\AA\ is likely due to \ion{C}{iii}~$\lambda$4648, \ion{N}{iii}~$\lambda$4640, and \ion{He}{ii}~$\lambda$4686.
 Given the strong similarity between SN~2008fq and SN~1998S (see Sect.~\ref{sec:spectracomp}),
  we follow the line identification that was proposed for SN~1998S by \citet{fassia01}. We note that high-ionization lines of carbon and nitrogen
are commonly found in the spectra of Wolf-Rayet stars.
At $V_{\rm max}$ (14 days after discovery), the broad H$\alpha$ emission increases 
in brightness, while the faint, broad absorption continues to be present. 
At the same time, the continuum in the blue portion of the spectrum exhibits some P-Cygni features that we attribute to \ion{Fe}{ii}, \ion{Sc}{ii}, and \ion{Na}{i}~D lines,
while broad \ion{Ca}{ii} triplet features emerge in the red portion of the SED.
After $\approx$ 30 days, H$\alpha$ shows a prevalent, broad emission (FWHM$_{\rm b}\approx$7000~km~s$^{-1}$). The faint and broad blue absorption is characterized by $v_{min}$ $\approx-$7000~km~s$^{-1}$). The narrow 
Balmer P-Cygni absorption is difficult to identify, whereas the narrow Balmer emission (FWHM$_{\rm n}\approx$200~km~s$^{-1}$) is still 
present.

The spectral continuum is affected by a significant amount of extinction (through the EW of
 \ion{Na}{i}~D, we estimate $E(B-V)_{\rm host}$ $=$ 0.46$\pm$0.03~mag), which makes the blue 
 portion of the spectrum very faint.

\subsection{Spectral comparison}
\label{sec:spectracomp}

Each of our five objects discussed in the previous sections has prevalent H$\alpha$ and 
H$\beta$ features, which serve as the basis for their classification as bonafide 
SNe~IIn. 
In three of them (SNe~2005kj, 2006aa, and 2006bo), the narrow 
Balmer lines show P-Cygni profiles that sit on top of broad and symmetric bases.
In the three panels of Fig.~\ref{wingsH}, the Balmer lines of these three objects 
are compared after continuum-subtraction, extinction correction, and peak normalization. 
It turns out that for shorter wavelengths, the narrow absorption deepens.
The broad component is likely produced by fast moving material that is associated with the underlying SN ejecta, but it might also be due to the electron scattering of the narrow line photons.

Concerning the narrow Balmer lines, their narrow P-Cygni profiles suggest that they arise from the recombination of a slow CSM that surrounds the SN rather than 
from an underlying \ion{H}{ii} region. 
Only for SN~2006qq, the narrow Balmer absorption is not detected, and in these spectra, we also observe narrow emission from [\ion{N}{ii}], which is
typical of \ion{H}{ii} regions. However, the absence of strong oxygen and sulfur emission lines suggests that the narrow emission in SN~2006qq is not 
only due to contamination, but also to ionization of the CSM. 

Plotted in the left panel of Fig.~\ref{ew} are the H$\alpha$ EW values for the full 
CSP SN~IIn sample, along with those of several objects from the literature. 
For some objects, this parameter tends to increase over time, suggesting 
that the bulk of the radiation at early phases is related to continuum flux, whereas 
line emission dominates at later epochs. 
In particular, SNe~1998S and 2008fq standout, both showing a strong H$\alpha$ EW enhancement of 10~\AA\ to 250~\AA\ within the first 70 days of evolution. 
Other events, like SNe~2005kj and 2006aa, exhibit an almost constant H$\alpha$ EW. In Table~\ref{tab_EW}, we report the EW values of our five objects.

We also examine the total H$\alpha$ luminosity, which is shown in the 
right panel of Fig.~\ref{ew}. 
$L$(H$\alpha$) values were computed by multiplying the de-reddened fluxes (presented in 
Figs.~\ref{ha_5kj}--\ref{ha_8fq}) by 4$\pi$D$_L^2$, where D$_L$ is the luminosity distance listed in Table~\ref{sample}.
H$\alpha$ luminosities of our objects are comparable with the typical values found for SNe~IIn in the literature. 

Concerning other lines, it is known that SNe~IIn can exhibit \ion{Fe}{ii} and \ion{Ti}{ii} P-Cygni profiles. Following the
line identification presented by \citet{kankare12}, these
 features are identified in moderately young spectra of SNe~2005kj, 2006aa, and 2006bo and shown in Fig.~\ref{FeCaII} (left panel).
In SNe~2006qq and 2008fq, \ion{Fe}{ii} features are present as well, although they appear heavily blended.
As shown in the right panel of Fig.~\ref{FeCaII}, the \ion{Ca}{ii} NIR triplet is observed in neither SNe~2006aa nor 2006qq, is barely detected 
in SNe~2005kj and 2006bo, and is clearly observed in SN~2008fq, whose red portion of the spectrum is  characterized by broad P-Cygni profiles.

All of our objects are found to be spectroscopically similar to other events that were previously reported in literature. 
In Fig.~\ref{compnormalIIn}, we present a spectral comparison of SNe~2005kj and 2006aa to SNe~2005db \citep{kiewe12} and 
1995G \citep{pastorello02}. Each spectrum is corrected for reddening and presented in its rest frame. The spectra
are mainly characterized by a combination of narrow and broad Balmer features. \ion{Fe}{ii} lines are also present, modifying the bright blue continuum.
The red continuum is almost featureless. Basically, the spectra of SNe~2005kj and 2006aa follow the definition of SN~IIn 
provided by \citet{schlegel90} and are similar to those of many other SNe~IIn in the literature (e.g. SNe~1999eb and 1999el; see Fig.~8 of \citealp{pastorello02}).

SN~2006bo appears to fit well within the recently suggested 
Type~IIn-P subclass \citep[see][]{94Wlike}. 
The light curves presented in Sect.~\ref{sec:phot} show a luminosity drop after a plateau phase, which was also observed in SNe~1994W \citep{sollerman98}, 2009kn 
\citep{kankare12}, and 2011ht \citep{94Wlike}. 
The spectral comparison to SN~2009kn confirms that
SN~2006bo is a similar object, as shown in Fig.~\ref{comp06bo}. 
SN~2006bo spectra clearly resemble those of
SN~2009kn $\approx$ 2 months after explosion (during the plateau phase). 
The common features are the narrow Balmer emission 
lines, along with the narrow Balmer absorption lines, and the \ion{Fe}{ii} P-Cygni profiles.

SN~2006qq is a peculiar object with its very broad emission features and its strong asymmetry in the H$\alpha$ 
profile. In Fig.~\ref{asym}, its H$\alpha$ profile is compared to those of SNe~1988Z \citep{turatto93} and 2006jd 
(S12). Even though SN~2006qq clearly exhibits slower velocities in the broad component than those of SNe~1988Z 
and 2006jd, the profile is similarly characterized by a strong suppression of the red-wing. The fact that the broad 
H$\alpha$ component significantly strengthens weeks after explosion is similar to what was observed for 
SN~2006jd but on different timescales (for SN~2006jd, H$\alpha$ became remarkably brighter about 400 days after 
explosion).

SN~2008fq can be defined as a SN~1998S-like event. 
SN~1998S is among the best-studied interacting SNe. 
A comparison of the spectral evolution of SNe~1998S and 2008fq is presented in 
Fig.~\ref{comp98S08fq}. 
Here, the spectra were de-reddened, and the listed epochs are 
relative to the time of $V$-band maximum. 
The similarity between the two objects is remarkable at all epochs. 
We note that the light curves are also very similar, as demonstrated in Sect.~\ref{sec:phot}.
The first example of SNe similar to 1998S and 2008fq is actually SN~1983K \citep{niemela85,phillips90}, but a less extensive set of spectral and photometric observations exists for this object. SN~2005gl also
 shows a spectral evolution that resembles that of SN~1998S \citep{galyam07,galyam09}.
 Another object, which shares similar properties to SNe~1998S 
and 2008fq is SN~2007pk, which has recently been presented by \citet{inserra12}.
We note that SNe~2007pk and 2008fq show traces of \ion{C}{iii} and \ion{N}{iii} emission that were strong at early 
epochs for both SNe~1983K and 1998S.
Finally, the late-time spectra ($\geq$232 days) of SN~2007od \citep{andrews10} are similar to those of SN~1998S. 
Interestingly, the absolute magnitudes of the above-mentioned events
are also similar, ranging between $V_{\max}\sim$ $-$18.5~mag (SN~2007pk) and 
$V_{\max}\sim$ $-$19.3~mag (SN~2008fq). These values are affected by large uncertainties on 
the extinction.

\section{Discussion}
\label{sec:discussion}

In the following, we discuss the data for each of the five new CSP SNe~IIn to 
determine the likely scenario underlying each object.
Consistent with the interpretation of other SNe~IIn, we assume that the ejecta of the exploding progenitor stars are 
experiencing strong interaction with surrounding CSM. 
In the following, constraints are placed on the mass-loss rate, the density, and the extent of the CSM for each SN.
Estimates derived in the following section are summarized in Table~\ref{tabpar}, which
 also includes the corresponding values derived from observations of SNe~2005ip and 
 2006jd. We stress that the mass-loss rate estimates are based on the uncertain assumption that the progenitor 
 wind was steady. Therefore, the more robust density wind parameter $w$ = $\dot{M} v_{w}^{-1}$ is also reported in 
 Table~\ref{tabpar}.

\subsection{SN~2005kj}

The explosion date of SN~2005kj is not well constrained by pre-explosion images (see Table~\ref{cbet}).
However, its light curves were probably observed only a few days after maximum, which can be inferred from their flattening 
at early epochs. The high (approximately 10,000~K) early-time temperatures determined by BB fits and presented in Fig.~\ref{bolo} (middle panel) also suggest
 that the observations began soon after explosion.
Based on the temperature comparison between SN~2005kj to SN~2008fq, we assume here that 
the explosion of SN~2005kj occurred 8 days prior to its discovery. 

The high-energy radiation produced via shock interaction ionizes the slow moving CSM and gives rise to observed
 narrow (FWHM$_{\rm n}$ $\approx$ 1000~km~s$^{-1}$) Balmer emission lines through recombination. 
 The narrow Balmer absorption feature, appearing a couple of 
months after discovery (well visible in H$\beta$), is also likely produced within the 
slow-moving CSM ($v_{min}$ $\approx$ 600~km~s$^{-1}$).

SN~2005kj presents iron and hydrogen P-Cygni profiles that are rounded and symmetric. The
same properties were observed in the spectra of SN~1994W \citep{sollerman98} and SN~1995G \citep{pastorello02}. 
Because of that, line optical depths of $\tau$ $>$1 were inferred for both objects, 
 which allows the use of the expression n$_{\rm H}$ $>$ 3 $\times$ 10$^{8}$ v$_{3}$ 
r$_{15}^{-1}$~cm$^{-3}$ \citep{mihalas78} to estimate the electron density of the CSM. Here, r$_{15}$ is the CSM shell radius, which is close to the 
photospheric radius (expressed in units of 10$^{15}$~cm, from BB fits to the SED); v$_{3}$ is the CSM shell velocity 
measured from the narrow P-Cygni absorption minimum and given in units of 10$^{3}$~km~s$^{-1}$.
We use the same expression for SN~2005kj (v$_{3}$ $\approx$ 0.6 and r$_{15}$ $\approx$ 1), obtaining a lower limit for the CSM electron 
density: n$_{\rm H}$ $>$ 1.8$\times$10$^{8}$~cm$^{-3}$.

Clear signs of CSI are observed from the epoch of the first spectrum to the epoch of the last one (approximately 170 days since explosion), which means the ejecta did not overtake the outer radius of the CSM in 170 days. 
If the fastest ejecta were moving at $v^{\rm max}_{\rm ejecta}$ $\approx$ 3000~km~s$^{-1}$ (BVZI of the broad component of H$\alpha$), this would then suggest that the outer radius of the CSM is $R_{\rm CSM}^{\rm out}$ $\gtrsim$ 4.4$\times$10$^{15}$~cm. Given that the interaction is already present at $\approx$ 15~days after explosion, the inner radius of the CSM is $R_{\rm CSM}^{\rm in}$ $<$ 0.4$\times$10$^{15}$~cm.

Following \citet{salamanca98}, a rough estimate of the mass-loss rate ($\dot{M}$) can be computed via: $L$(H$\alpha_{\rm broad}$) $\approx$ 0.25$\epsilon_{\rm H\alpha}\dot{M}v_s^3v_w^{-1}$, which is based on the assumption that the wind forming the CSM is steady.
Here we adopted the shock velocity $v_s$ $\approx$ 3000~km~s$^{-1}$, which is derived from the FWHM of the broad component,
and the wind velocity of $v_w$ $\approx$ 600~km~s$^{-1}$, which is obtained from the minimum of the narrow P-Cygni absorption.
The efficiency factor is assumed to be $\epsilon_{\rm H\alpha}$ $=$ 0.1, which is found in the literature \citep[see e.g][]{kiewe12}. 
The broad H$\alpha$ luminosity is measured to be 
$L$(H$\alpha_{\rm broad}$) $\approx$ 1.0$-$3.4$\times$10$^{40}$~erg~s$^{−1}$.
From these values, we obtain $\dot{M}$ $\approx$ 1.4$-$4.8$\times$10$^{-3}$ $M_{\odot}$~yr$^{-1}$.

\subsection{SN~2006aa}
SN~2006aa was discovered before maximum luminosity and its explosion date is well constrained (uncertainty of $\pm$8 days) thanks to 
pre-explosion images. We therefore assume the core collapse occurred 8 days before discovery (see Table~\ref{cbet}).

The spectra and the bolometric light curve suggest that the main source of energy is CSI, rather than the radioactive 
decay of $^{56}$Ni and $^{56}$Co. 
If we naively assume that the peak luminosity is powered solely due to radioactive decay, the combinination of the epoch of the peak and the peak bolometric luminosity with Arnett's rule \citep{arnett82,arnett85,branch92}
gives a $^{56}$Ni mass $=$ 0.3~$M_\odot$. This value is large compared to the typical $^{56}$Ni masses measured for SNe~II \citep[see Figure~9 in][]{smartt09}.

BB fits of the SEDs reveal that the peak temperature was reached close to maximum light --
i.e, around 60 days after explosion. This late-time peak and the initial rise of the temperature can be explained by a strengthened interaction due to the presence of a dense shell
of CSM.
Given that the maximum ejecta velocity is $v^{\rm max}_{\rm ejecta}$ $\approx$ 2500~km~s$^{-1}$ (BVZI of H$\alpha$) and $t_{\rm max}$ $\approx$ 60 days, the densest part of the CSM is likely located at 
$R_{\rm CSM}^{\rm sh}$ $\approx$ 1.3$\times$10$^{15}$~cm. Since the wind velocity, which is derived 
from the narrow P-Cygni absorption 
minimum, is $\approx$ 600~km~s$^{-1}$, the mass-loss episode that produces this dense shell likely occurred $\approx$ 8 months prior to core collapse.

As the spectra of SN~2006aa are quite similar to those of SN~2005kj, similar techniques can be adopted to estimate the electron density of the wind, the physical extension in radius from the progenitor of the CSM, and the progenitor's mass-loss rate. 
We obtain n$_{\rm H}$ $>$ 7.6$\times$10$^{8}$~cm$^{-3}$ by assuming v$_{3}$ $\approx$ 0.6 and r$_{15}$ $\approx$ 0.8.
 The inner CSM radius is $R_{\rm CSM}^{\rm in}$ $<$ 0.5$\times$10$^{15}$~cm, and the outer one is $R_{\rm CSM}^{\rm out}$ $\gtrsim$ 1.8$\times$10$^{15}$~cm, assuming the above mentioned value of $v^{\rm max}_{\rm ejecta}$ and the interaction  begins 24 and ends 85 days after explosion.
 
The mass-loss rate is $\dot{M}$ $\approx$ 5.4$-$16.2$\times$10$^{-3}$ $M_{\odot}$~yr$^{-1}$, where a wind velocity $v_w$ $\approx$ 600~km~s$^{-1}$ and a 
shock velocity $\approx$ 2000~km~s$^{-1}$ (FWHM$_{\rm b}$) were used.
The efficiency factor is again assumed to be $\epsilon_{\rm H\alpha}$ $=$ 0.1. The broad H$\alpha$ luminosity is found to be 
$L$(H$\alpha_{\rm broad}$) $\approx$ 1.1$-$3.4$\times$10$^{40}$~erg~s$^{-1}$.

\subsection{SN~2006bo}
The explosion date of SN~2006bo is not well constrained by pre-explosion images.
Since we observed a plateau in the light curve having a duration of about 70 days and then observed a sudden drop in luminosity, we classify SN~2006bo as a SN~1994W-like event. 
These transients are characterized by a 100 days plateau; therefore, it is possible that
SN~2006bo was discovered even a month after explosion. Because we did not
observe the SN between 70 and 145 days after discovery, it makes impossible to know whether 
the drop in luminosity occurred immediately after 70 days or later, and therefore, a proper
comparison of the plateau's length is difficult. On the other hand, the spectral comparison to SN~2009kn 
shown in Fig.~\ref{comp06bo} suggests that the SN could have been discovered at least 40 days 
after explosion.
The large uncertainty on the explosion date determines the uncertainty on the amount of $^{56}$Ni that is 
likely
to power the light curve tail, which was the case in SN~2009kn \citep{kankare12}. 
Assuming that the
explosion occurred on the day of discovery, the bolometric light curve tail then yields a $^{56}$ Ni mass of $M_{\rm ^{56}Ni}$ 
$=$ 0.007$M_{\odot}$. If the explosion occurred 40 days before, then a $M_{\rm ^{56}Ni}$ $=$ 0.011$M_{\odot}$ is obtained. In any case,
that value would be lower than the upper limit estimated for SN~2009kn of 
0.023~$M_{\odot}$ \citep{kankare12} and comparable
to the value 0.01 $M_{\odot}$, inferred for SN~2011ht \citep{94Wlike}.
Unfortunately, it is not possible to 
determine if the slope of the SN tail is consistent with the $^{56}$Co decay, since only a single photometric epoch was obtained at late epochs.

The Balmer lines appearing in the two spectra of SN~2006bo seem to be well reproduced by
the addition of a single Lorentzian function, with typical FWHM $\approx$ 1000~km~s$^{-1}$ and an absorption component that peaks at $-$600~km~s$^{-1}$. 
We believe that both components are generated in the CSM with wind velocity $v_w$ $\approx$ 600~km~s$^{-1}$.
The reason that the narrow emission component has slightly higher velocities than the narrow absorption might be due to the electron-scattering broadening of the narrow emission in the wind. 
The likely absence of a broad component might mean that ionized ejecta are not 
visible because of a CDS (cold-dense-shell). Given that we do not observe the broad component, we assume the shock velocity to have a lower limit equal to the FWHM of the narrow component, i.e. $v_s$ $\gtrsim$ 1000~km~s$^{-1}$.

The duration of the plateau is 70$-$110 days, which means the outer radius of the CSM surrounding 
SN~2006bo is $R_{\rm CSM}^{\rm out}$ $\approx$ 1.5$-$2.4$\times$10$^{15}$~cm. We assume here that the
emission after the plateau phase is mainly due to radioactive decay rather than to CSI and a maximum ejecta velocity of $v^{\rm max}_{\rm ejecta}$ $\approx$ 2500~km~s$^{-1}$ (BVZI of H$\alpha$). 
The limit on the inner CSM radius is computed from the
epoch of the first spectrum (11$-$41 days depending on the explosion date). 
We obtain $R_{\rm CSM}^{\rm in}$ $\approx$ 0.2$-$0.9$\times$10$^{15}$~cm.

The mass-loss rate estimate gives $\dot{M}$ $\lesssim$ 2.28$-$3.42$\times$10$^{-2}$ $M_{\odot}$~yr$^{-1}$. We assume here that the total H$\alpha$ luminosity is the upper limit value for the broad H$\alpha$ luminosity, $L$(H$\alpha_{\rm broad}$) $\lesssim$ 0.6$-$0.9$\times$10$^{40}$~erg~s$^{-1}$, using the mentioned values of $\epsilon_{\rm H\alpha}$, $v_s$ and $v_w$.
For a photospheric radius of 7.9$\times$10$^{14}$~cm (see Fig.~\ref{bolo}, bottom panel), the electron density is n$_{\rm H}$ $>$ 3.8$\times$10$^{8}$~cm$^{-3}$.

\subsection{SN~2006qq}
SN~2006qq was discovered relatively soon after explosion. Given the constraint from the pre-explosion images,
we assume the explosion occurred 16 days before discovery (see Table~\ref{cbet}). This appears consistent
with the temperature comparison to SN~2008fq (see Fig.~\ref{bolo}, middle panel), whose explosion date is better 
constrained,
although the temperature evolution is hard to standardize.

Given the bright emission lines, it is likely that CSI powers the emission of SN~2006qq, although the 
bolometric light curve 
seems to follow the $^{56}$Co radioactive decay at late times. If radioactive decay was the main powering 
source at late 
times, a $^{56}$Ni mass of $M_{\rm ^{56}Ni}$ $=$ 0.4~$M_{\odot}$ is derived.

As discussed in Sect.~\ref{sec:spectra}, the spectra reveal strong, broad H$\alpha$ and H$\beta$ emission from the ionized 
ejecta, whose luminosity reaches its maximum at 76 days after explosion. The broad emission from the ejecta is strongly asymmetric, where H$\alpha$ BVZI $\approx$ 
10$^4$~km~s$^{-1}$ and RVZI $\approx$ 3$\times$10$^3$~km~s$^{-1}$. The asymmetry can be explained in terms of
dust obscuration of the radiation coming from the back of the ejecta. Indeed, \citet{fox11} detected 
0.5$-$1.7$\times$10$^{-2}$~$M_{\odot}$ of (graphite) dust emitting in the MIR for this SN.
The strong asymmetric H$\alpha$ profile can also be produced through the occultation of the radiation that comes from the
receding ejecta by an opaque photosphere. The ratio between the photospheric radius and the maximum ejecta radius 
($R_{\rm p}/R_{\rm max}$) can be determined from H$\alpha$ BVZI and RVZI for each spectral epoch (see S12). 
In this case, we obtain $R_{\rm p}/R_{\rm max}$ $\approx$ 0.87 at 28 days and $R_{\rm p}/R_{\rm max}$ $\approx$ 0.95 at 76 days after explosion.

The epoch of the maximum broad emission and the maximum ejecta velocity ($v^{\rm max}_{\rm ejecta}$, from the H$\alpha$ BVZI) set the distance ($R_{\rm CSM}^{\rm sh}$) to the densest part of the CSM, which is $R_{\rm CSM}^{\rm sh}$ $\approx$ 6.6$\times$10$^{15}$~cm.
This dense shell was ejected during the end of the life of the progenitor star. If the wind velocity
is $v_w$ $\lesssim$ 200~km~s$^{-1}$, as measured from the FWHM of the narrow emission lines, 
then the mass-loss 
episodes that produced the dense shell occurred at least 10 years before core collapse.
With the assumed $v_w$ and $v_s$ $\approx$ 6000~km~s$^{-1}$ (that corresponds to the broad H$\alpha$ FWHM if we neglect the occultation effect), the broad H$\alpha$ luminosity gives
a mass-loss rate of $\dot{M}$ $\lesssim$ 0.3$-$0.7$\times$ 10$^{-3}$ $M_{\odot}$~yr$^{-1}$.
We derive an electron density lower limit of n$_{\rm H}$ $>$ 1.1$\times$10$^{9}$~cm$^{-3}$ from a photospheric radius of 1.6$\times$10$^{15}$~cm. 
The inner and the outer CSM radii that we can probe correspond to the phase of the first and the last spectral epoch (29 and 91 days since explosion, respectively) multiplied by $v^{\rm max}_{\rm ejecta}$. We obtain $R_{\rm CSM}^{\rm in}$ $\lesssim$ 2.5$\times$10$^{15}$~cm and
$R_{\rm CSM}^{\rm out}$ $\gtrsim$ 7.9$\times$10$^{15}$~cm.

\subsection{SN~2008fq}

The explosion date of SN~2008fq is well constrained by pre-explosion images ($\pm$4.5 days uncertainty).
The light curve comparison with SN~1998S (see bottom panel of Fig.~\ref{absmag}) suggests that the explosion occurred soon after 
the last non-detection, approximately 8 days before discovery.
The presence of broad H$\alpha$
P-Cygni features in the early time spectra allows us to estimate the ejecta velocity; we obtain 
$v_{min}=$7000$-$8000~km~s$^{-1}$. Given that the first spectrum was observed 1.7 days after discovery and that
the photospheric radius at that epoch was approximately 10$^{15}$~cm, we confirm that the core collapse occurred soon after the last non-detection by assuming free expansion.

The lack of a very early time spectrum prevents us from determining if SN~2008fq 
displayed strong CSI soon after core collapse, as was observed in SN~1998S 
(\citealp{leonard00,fassia01}, and see strong emission lines in the spectrum of SN~1998S at 15 days from its in Fig.~\ref{comp98S08fq}).
The presence of a wind moving at $v_w$ $\approx$ 500~km~s$^{-1}$ (from the narrow 
absorption minimum) is observed in SN~2008fq, as was also detected for SN~1998S at similar 
phases \citep{leonard00,fassia01}. We note, however, that the wind velocity of SN~1998S was lower ($\approx$ 150~km~s$^{-1}$).
From the extreme blue edge of the broad H$\alpha$ absorption, we determine
a maximum ejecta velocity $v^{\rm max}_{\rm ejecta}$ $\approx$ 9000~km~s$^{-1}$. 
If we assume that SN~2008fq had an inner CSM (ICSM) like that of SN~1998S \citep{leonard00,fassia01}, with the 
first spectrum observed approximately 10 
days after explosion and the above mentioned $v^{\rm max}_{\rm ejecta}$, the ICSM could not be more 
extended than $R_{\rm CSM}^{\rm in}$ 
$\approx$ 7.8$\times$10$^{14}$~cm. 

As shown in Sect.~\ref{sec:spectra}, a broad, prevalent Balmer emission component emerges about 20 days after maximum (45 days after explosion). Following the interpretation of \citet{leonard00,fassia01} for SN~1998S, this is likely due to the interaction with an outer, denser CSM (OCSM). Given the assumed $v^{\rm max}_{\rm ejecta}$, the inner part of the OCSM is placed at
$R_{\rm CSM}^{\rm sh}$ $\approx$ 3.5$\times$10$^{15}$~cm. A similar inner radius was found for the OCSM of SN~1998S \citep{fassia01}. The outer radius of the OCSM corresponds to the last spectral epoch (78 days after explosion); we obtain $R_{\rm CSM}^{\rm out}$ $\gtrsim$ 6.1$\times$10$^{15}$~cm.
 Given these radii and the assumed wind velocity, the OCSM was ejected during an episodic mass loss that started at least 3.9 years before core collapse and ended 2.2 years prior to core collapse. Assuming the same wind velocity, the ICSM was formed during a mass-loss that started less than 6 months before core collapse.

The most likely scenario is that the observed spectra up to 6 days after maximum are mainly powered by the ejecta interaction with a low-density, mid-CSM  (MCSM). As suggested by \citet{fassia01} for SN~1998S, this would explain the low contrast of the P-Cygni profiles (formed in the ejecta) on the blue continuum of the spectra (produced in a cold dense shell at the ejecta/MCSM interface). Only 20 days after maximum, the ejecta reach the OCSM, giving rise to a stronger interaction.

The mass-loss history of SN~2008fq is likely characterized by episodic outbursts that gave rise to a complex CSM. Therefore, an estimate of the mass-loss rate is difficult to obtain as the steady wind hypothesis is likely wrong. If we assume the shock velocity to be similar to the ejecta velocity $v_{min}$ (consistently with the FWHM of the broad H$\alpha$ component in the last spectra), the above-mentioned $v_w$ and $L$(H$\alpha$)$_{\rm broad}$ $\approx$ 1.5$\times$10$^{41}$~erg~s$^{-1}$, however, we obtain $\dot{M}$ $\approx$ 1.1$\times$ 10$^{-3}$ $M_{\odot}$~yr$^{-1}$.
We also obtain n$_{\rm H}$ $>$ 9.0$\times$10$^{8}$~cm$^{-3}$ (for a typical photospheric radius R$_{ph}$ $\approx$ 2.5$\times$10$^{15}$~cm).

\subsection{SN IIn subtypes and progenitor scenarios}

The seven CSP SNe~IIn confirm the diversity of observational properties from this subclass of CC explosions, while the case of SN~2006aa displays 
previously unobserved properties. Under close examination, these objects show how SNe~IIn can be separated out into different categories.

SN~2006bo belongs to the rare SN~1994W-like class \citep{94Wlike}.
Besides SN~1994W \citep{sollerman98}, only two additional objects are known to
belong to this group,
namely SNe~2009kn \citep{kankare12} and 2011ht \citep{94Wlike}.
These objects all have Type IIn spectral features but they also present a characteristic light curve plateau phase,
like that observed in SNe~IIP.

With a few other objects (e.g. SNe 2007pk and 1983K) SN~2008fq forms the SN~1998S-like category, which presents optical light curves that peak within 10$-$20 days at $-$18.5 $\lesssim$ $V_{\rm max}$ $\approx$ $\lesssim$ $-$19.3 mag, and which displays characteristic spectral evolution with broad emission components that arise after maximum light.
SN~2006qq, SNe~2005ip, and 2006jd (S12) resemble the
Type~IIn SNe~1988Z and 1995N \citep{turatto93, fransson02}.
Evidently, two common properties among these objects are an asymmetric 
H$\alpha$ profile and long-lasting dust emission that peaks at MIR wavelengths. 

SNe~2005kj is similar to the more common SNe~IIn, like those
presented in \citet{kiewe12} (e.g. SN~2005cp).
SN~2006aa shows an interesting and unprecedented initial plateau in the $u$ band and an uncommon late time increase of the photospheric temperature. Like SN~2005kj, its spectral features are however similar to those of typical SNe~IIn \citep{kiewe12}.

We note that the initial class of Type IIn SNe \citep{schlegel90} is likely to include a variety of phenomena with the common characteristics of CSI giving rise to narrow Balmer lines. Some subtypes have already been identified and singled out in the past, and we note that some of these rare interacting, core-collapse SN types are not represented in the CSP sample, like superluminous SNe-II \citep{galyam12} and SNe~Ibn \citep{pastorello08}.
A couple of SN 2002ic-like objects \citep{hamuy03} were observed by the CSP
and presented in \citet{prieto07} and \citet{taddia12}. These events show spectra resembling those of SNe~Ia,  although diluted by a blue continuum with the addition of Balmer emission lines. These properties can be interpreted as the result of the interaction between a thermonuclear SN and its dense CSM. 
Following this interpretation, we believe these transients do not belong to the family of core-collapse SNe~IIn, and therefore we do not include them in our sample. However, we note that alternative explanations support the idea that SN~2002ic-like objects might actually be core-collapse events \citep{benetti06}.

\citet{smith11rates} found that the SNe~IIn subclass constitutes about 9 percent of all CC~SNe. Although the CSP was 
not an unbiased survey, the number of SNe~IIn (7) and the total number of CC SNe (116) is consistent with the 
fraction found by \citet{smith11rates}. As noted by \citet{kiewe12}, who added 4 more SNe IIn, the current number of well 
studied SNe~IIn is still very low, and often publications are driven by more peculiar objects. The CSP sample presented here therefore 
contributes significantly to the current Type IIn sample.

Although our sample is still too small to allow further subclassifications, all the objects we have found do match up with similar objects in the literature, even if there is large diversity in 
the properties of SNe~IIn. It thus seems that the 
population of narrow-line CC~SNe separate out into discrete subcategories. This is
based on only a handful of objects, and further observations are clearly needed to map out the
diversity of SNe~IIn.

What determines the appearance of these objects is foremost the
mass-loss history, since this shapes the CSM with which the SN interacts.
For line-driven stellar winds, the main factors that affect the mass-loss are the progenitor mass and the metallicity.
If that was the main mass-loss mechanism, then we could expect a continuum of observational properties, which reflect
the continuous variation of progenitor mass and metallicity.
However, SN~1994W-like, SN~1998S-like, and long-lasting SNe~IIn look distinctively
different, and we have not yet found
any transitional or intermediate objects that connect these different subtypes.
This could imply that the final fate of these stars can only follow certain specific paths.

One overall aim of continued studies of SNe IIn is to identify the progenitor stars that end their lives following these paths. This is often done by comparing the derived mass-loss properties with those of massive stars. In this regard, we note some caveats.

The mass-loss rate estimates that we have presented assume steady winds. This makes them easy to calculate from the observables, and easy to compare to many similar estimates in the literature. It should be emphasized, however, that this is likely a much too simplistic assumption (e.g \citealp{dwarkadas11}). Much of the observations of SNe IIn seem to favor mass-loss histories of more complicated kinds, including clumped winds and ejected shells. The wind density parameters presented in Table~\ref{tabpar} are more closely tied to the observations, which are independent from
wind velocities.

We also remind the reader that the mass-loss period probed by observations of SNe~IIn is only a very tiny fraction of the lifetime of the progenitor stars. What we know observationally about the mass-loss properties of RSG or LBV stars is derived from stars that are not that close ($\lesssim$ 1000 years) to their final explosion.

If the mass-loss history in SNe~IIn was dominated by other mechanisms, like pulsations \citep{yoon10}, which drive episodic outbursts occurring only under specific conditions, we could more easily explain the existence of such different subtypes.
Our sample does imply that dense shells have likely been ejected soon before explosion by episodic outbursts, at least for some SNe (2006aa, 2006jd, 2006qq, and 2008fq).

The values derived for the mass-loss rates in this work (10$^{-4}$--10$^{-2}$~$M_{\odot}$~yr$^{-1}$) are similar to those found by \citet{kiewe12} for other SNe~IIn. Such high numbers may
imply that LBVs are the most likely progenitor channel, but it is unclear if RSG may not also produce some of the events when the steady wind assumption is not enforced. The wind density parameter of three SNe in our sample is higher than 10$^{16}$~g~cm$^{-1}$;  the other objects exhibit $w$ $\sim$ 10$^{15}$~g~cm$^{-1}$. These values are consistent with those found for other events
in the literature \citep[e.g.][]{chugai94,smith09_05ip}.

There are certainly objects like SN~2005gl \citep{galyam09} and SN~2009ip \citep{mauerhan12,pastorello12} with identified LBV progenitors; in these cases, the progenitor was directly detected before core collapse.
However, we caution that other channels (like RSGs) cannot be excluded on the basis of rough mass-loss rate estimates. 
Interestingly, \citet{anderson12} show
 that SNe~IIn are located in environments similar to those of SNe~IIP (whose progenitors are RSGs), indicating that the majority of events may not arise from very massive stars like LBVs.

The estimates of wind velocities from the narrow P-Cygni absorption features (for our sample, we find typical values of 500--600~km~s$^{-1}$) are also consistent
with LBV wind velocities. However, slower winds might have been accelerated at the shock breakout by the radiation pressure. In that
case, we would detect velocities of hundreds of km~s$^{-1}$, even though the wind had an original velocity of tenths of km~s$^{-1}$ (such as those measured for RSGs).

\section{Conclusions}
\label{sec:conclusions}
In this paper, we have presented the data for five of the seven CSP SNe~IIn, including extensive
optical and NIR photometry and visual-wavelength spectroscopy. We confirm the diversity of this class of SNe,
although a comparison to objects in the literature 
 provides counterparts for each object. 
 This implies that several subcategories
are present within the SNe~IIn class, including SN~1994W-like (SN~2006bo), SN~1998S-like (SN~2008fq) and SN~1988Z-like (SN~2006qq) events.
These subcategories appear significantly different from each other,
suggesting that SNe~IIn might be the result of different conditions in the progenitor mass/metallicity phases space.
 With a late-time maximum and an initial 
$u$-band plateau, the peculiar SN~2006aa was also studied.
Estimates for CSM parameters have been derived and compared to objects in the literature. 
Mass-loss rates and wind velocities derived from this sample are compatible with LBV progenitors, although caveats of such a connection are discussed.

\begin{acknowledgements}
We thank the referee, Nathan Smith, for providing comments that helped to improve the paper.
We are grateful to Eric Hsiao, David Osip and Yuri Beletsky for taking the $K$-band template of SN~2005kj. 
We also thank Erkki Kankare for sharing the spectra of SN~2009kn, Seppo Mattila for his inputs, Keiichi Maeda and Takashi J. Moriya for helpful discussions, and Miguel Roth for his important contribution to
the CSP.\\
F. Taddia acknowledges the Instrument Centre for Danish Astrophysics (IDA) for funding his visit to Aarhus University. 
This material is also based upon work supported by NSF under grants AST--0306969, AST--0607438 and AST--1008343. 
M. Stritzinger acknowledges generous support provided by the Danish Agency for Science and Technology and Innovation realized through a Sapere Aude Level 2 grant.
The Oskar Klein Centre is funded by the Swedish Research Council.
J.~P. Anderson and M. Hamuy acknowledge support by CONICYT through FONDECYT
grant 3110142, and by the Millennium Center for Supernova Science
(P10-064-F), with input from `Fondo de Innovaci\'{o}n para
la Competitividad, del Ministerio de Econom\'{i}a, Fomento y Turismo
de Chile'.
This research has made use of the NASA/IPAC Extragalactic Database (NED), which is operated by the Jet Propulsion Laboratory, California Institute of Technology, under contract with the National Aeronautics and Space Administration.
 
\end{acknowledgements}

\bibliographystyle{aa} 

\begin{thebibliography}{100}

\expandafter\ifx\csname natexlab\endcsname\relax\def\natexlab#1{#1}\fi




\bibitem[Aldering et al.(2006)]{aldering06} Aldering, G., 
Antilogus, P., Bailey, S., et al.\ 2006, \apj, 650, 510 

\bibitem[Anderson et al.(2012)]{anderson12} Anderson, J.~P., 
Habergham, S.~M., James, P.~A., \& Hamuy, M.\ 2012, \mnras, 424, 1372 

\bibitem[Andrews et al.(2010)]{andrews10} Andrews, J.~E., 
Gallagher, J.~S., Clayton, G.~C., et al.\ 2010, \apj, 715, 541 

\bibitem[Arnett(1982)]{arnett82} Arnett, W.~D.\ 1982, \apj, 253, 
785 

\bibitem[Arnett et al.(1985)]{arnett85} Arnett, W.~D., Branch, 
D., \& Wheeler, J.~C.\ 1985, \nat, 314, 337 

\bibitem[Benetti et al.(1999)]{benetti99} Benetti, S., Turatto, 
M., Cappellaro, E., Danziger, I.~J., 
\& Mazzali, P.~A.\ 1999, \mnras, 305, 811 

\bibitem[Benetti et al.(1998)]{benetti98} Benetti, S., 
Cappellaro, E., Danziger, I.~J., et al.\ 1998, \mnras, 294, 448 

\bibitem[Benetti et al.(2006)]{benetti06} Benetti, S., 
Cappellaro, E., Turatto, M., et al.\ 2006, \apjl, 653, L129 

\bibitem[Blanc et al.(2006)]{blanc06} Blanc, N., Copin, Y., 
Gangler, E., et al.\ 2006, The Astronomer's Telegram, 732, 1 

\bibitem[Blondin et al.(2006)]{blondin06_bo} Blondin, S., Modjaz, 
M., Kirshner, R., Challis, P., 
\& Hernandez, J.\ 2006, Central Bureau Electronic Telegrams, 481, 1 

\bibitem[Blondin et al.(2006)]{class06jd} Blondin, S., Modjaz, 
M., Kirshner, R., et al.\ 2006, Central Bureau Electronic Telegrams, 679, 1 

\bibitem[Blondin et al.(2009)]{blondin09}
Blondin, S., Prieto, J. L., Pata, F., et al. 2009, \apj, 693, 207

\bibitem[Boles et al.(2005)]{disco05ip} Boles, T., Nakano, S., 
\& Itagaki, K.\ 2005, Central Bureau Electronic Telegrams, 275, 1 

\bibitem[Boles 
\& Monard(2006)]{boles06} Boles, T., \& Monard, L.~A.~G.\ 2006, Central Bureau Electronic Telegrams, 468, 1 

\bibitem[Bonnaud et al.(2005)]{bonnaud05} Bonnaud, C., Pecontal, 
E., Blanc, N., et al. 2005, Central Bureau Electronic Telegrams, 296, 1 

\bibitem[Branch(1992)]{branch92} Branch, D.\ 1992, \apj, 392, 35 

\bibitem[{{Cardelli} {et~al.}(1989){Cardelli}, {Clayton}, \& {Mathis}}]{cardelli89}
{Cardelli}, J.~A., {Clayton}, G.~C., \& {Mathis}, J.~S. 1989, \apj, 345, 245 

\bibitem[Chandra et al.(2012)]{chandra12} Chandra, P., Chevalier, 
R.~A., Chugai, N., et al.\ 2012, \apj, 755, 110 

\bibitem[Chevalier 
\& Fransson(1994)]{chevalier94} Chevalier, R.~A., \& Fransson, C.\ 1994, \apj, 420, 268 

\bibitem[Chugai et al.(2004)]{chugai04} Chugai, N.~N., 
Blinnikov, S.~I., Cumming, R.~J., et al.\ 2004, \mnras, 352, 1213 

\bibitem[Chugai(2001)]{chugai01} Chugai, N.~N.\ 2001, \mnras, 
326, 1448 

\bibitem[Chugai 
\& Danziger(1994)]{chugai94} Chugai, N.~N., \& Danziger, I.~J.\ 1994, \mnras, 268, 173 

\bibitem[{{Contreras} {et~al.}(2010){Contreras}, {Hamuy}, {Phillips}, {et~al.}}]{contreras10} 
{Contreras} C., {Hamuy}, M., {Phillips}, M.~M., {et~al.} 2010, \aj, 139, 519 

\bibitem[Dilday et al.(2012)]{dilday12}
Dilday, B., Howell, D. A., Cenko, S. B., et al. 2012, Science, 337, 942

\bibitem[Leonard et al.(2000)]{leonard00} Leonard, D.~C., 
Filippenko, A.~V., Barth, A.~J., \& Matheson, T.\ 2000, \apj, 536, 239 

\bibitem[Dwarkadas \& Gruszko(2012)]{dwarkadas12} Dwarkadas, V.~V., \& Gruszko, J.\ 2012, \mnras, 419, 1515 

\bibitem[Dwarkadas(2011)]{dwarkadas11} Dwarkadas, V.~V.\ 2011, 
\mnras, 412, 1639         

\bibitem[Fassia et al.(2001)]{fassia01} Fassia, A., Meikle, 
W.~P.~S., Chugai, N., et al.\ 2001, \mnras, 325, 907 

\bibitem[Folatelli et al.(2006)]{folatelli06} Folatelli, G., 
Contreras, C., Phillips, M.~M., et al.\ 2006, \apj, 641, 1039 

\bibitem[Folatelli et al. (2010)]{folatelli10}
Folatelli, G., Phillips, M. M., Burns, C. R. et al. 2010, \aj, 139, 120

\bibitem[Fransson et al.(2002)]{fransson02}
Fransson, C., Chevalier, R. A., Filippenko, A. V., et al. 2002, \apj, 572, 350

\bibitem[Fox et al.(2011)]{fox11} Fox, O.~D., Chevalier, 
R.~A., Skrutskie, M.~F., et al.\ 2011, \apj, 741, 7 

\bibitem[Fransson et al.(2002)]{fransson02} Fransson, C., 
Chevalier, R.~A., Filippenko, A.~V., et al.\ 2002, \apj, 572, 350 

\bibitem[Gal-Yam et al.(2007)]{galyam07} Gal-Yam, A., Leonard, 
D.~C., Fox, D.~B., et al.\ 2007, \apj, 656, 372 

\bibitem[Gal-Yam \& Leonard(2009)]{galyam09}
Gal-Yam, A., \& Leonard, D.~C.\ 2009, \nat, 458, 865 

\bibitem[Gal-Yam(2012)]{galyam12} Gal-Yam, A.\ 2012, Science, 
337, 927 

\bibitem[{{Hamuy} {et~al.}(2006){Hamuy}, {Folatelli}, {Morrell}, \& {Phillips}}]{hamuy06}
{Hamuy}, M., {Folatelli}, G., {Morrell}, N., \& {Phillips}, M.~M. 2006, \pasp, 118, 2

\bibitem[Hamuy et al.(2003)]{hamuy03} Hamuy, M., Phillips, 
M.~M., Suntzeff, N.~B., et al.\ 2003, \nat, 424, 651 

\bibitem[Inserra et al.(2012)]{inserra12} Inserra, C., 
Pastorello, A., Turatto, M., et al.\ 2012, arXiv:1210.1411 

\bibitem[Kankare et al.(2012)]{kankare12} Kankare, E., Ergon, M., 
Bufano, F., et al.\ 2012, \mnras, 424, 855 

\bibitem[Kiewe et al.(2012)]{kiewe12} Kiewe, M., Gal-Yam, A., 
Arcavi, I., et al.\ 2012, \apj, 744, 10 

\bibitem[Komatsu et al.(2009)]{komatsu09} Komatsu, E., Dunkley, 
J., Nolta, M.~R., et al.\ 2009, \apjs, 180, 330 

\bibitem[Kudritzki 
\& Puls(2000)]{kudritzki00} Kudritzki, R.-P., \& Puls, J.\ 2000, \araa, 38, 613 

\bibitem[{{Landolt}(1992){Landolt}}]{landolt92}
{Landolt}, A.~U. 1992, \aj, 104, 340

\bibitem[Lee 
\& Li(2006)]{lee06} Lee, E., \& Li, W. 2006, Central Bureau Electronic Telegrams, 395, 1 

\bibitem[Leonard et al.(2000)]{leonard00} Leonard, D.~C., 
Filippenko, A.~V., Barth, A.~J., \& Matheson, T.\ 2000, \apj, 536, 239 

\bibitem[Liu et 
al.(2000)]{liu00} Liu, Q.-Z., Hu, J.-Y., Hang, H.-R., et al.\ 2000, \aaps, 144, 219 

\bibitem[Mauerhan et al.(2012a)]{94Wlike} Mauerhan, J.~C., 
Smith, N., Silverman, J.~M., et al.\ 2012, arXiv:1209.0821

\bibitem[Mauerhan et al.(2012b)]{mauerhan12} Mauerhan, J.~C., 
Smith, N., Filippenko, A., et al.\ 2012, arXiv:1209.6320 

\bibitem[Mihalas(1978)]{mihalas78} Mihalas, D.\ 1978, San 
Francisco, W.~H.~Freeman and Co., 1978.~650 p., 

\bibitem[Modjaz et al.(2005)]{class05ip} Modjaz, M., Kirshner, 
R., Challis, P., 
\& Calkins, M.\ 2005, Central Bureau Electronic Telegrams, 276, 1 

\bibitem[Niemela et al.(1985)]{niemela85} Niemela, V.~S., Ruiz, 
M.~T., \& Phillips, M.~M.\ 1985, \apj, 289, 52 

\bibitem[Nymark et al.(2006)]{nymark06} Nymark, T.~K., Fransson, C., \& Kozma, C.\ 2006, \aap, 449, 171 

\bibitem[Pastorello et al.(2005)]{pastorello05}
Pastorello, A., Aretxaga, I., Zampieri, L., et al. 2005, ASPC, 342, 285

\bibitem[Pastorello et al.(2002)]{pastorello02}
Pastorello, A., Turatto, M., Benetti, S., et al.\ 2002, \mnras, 333, 27 

 \bibitem[Pastorello et al.(2008)]{pastorello08} Pastorello, A., 
Mattila, S., Zampieri, L., et al.\ 2008, \mnras, 389, 113 %

\bibitem[Pastorello et al.(2012)]{pastorello12} Pastorello, A., 
Cappellaro, E., Inserra, C., et al.\ 2012, arXiv:1210.3568 

\bibitem[Phillips et al.(1990)]{phillips90} Phillips, M.~M., 
Hamuy, M., Maza, J., et al.\ 1990, \pasp, 102, 299

\bibitem[Poznanski et al.(2011)]{poznanski11}
Poznanski, D., Ganeshalingam, M., Silverman, J. M., Filippenko, A. V. 2011, \mnras, 415, L81

\bibitem[{{Persson} {et~al.}(1998){Persson}, {Murphy}, {Krzeminski}, {Roth}, \& {Rieke}}]{persson98}
{Persson}, S.~E., {Murphy}, D.~C., {Krzeminski}, W., {Roth}, M., \& {Rieke}, M.~J. 1998, \aj, 116, 2475

\bibitem[{{Poznanski} {et~al.}(2011){Poznanski}, {Ganeshalingam}, {Silverman}, \& {Filippenko}}]{poznanski11} 
{Poznanski}, D., {Ganeshalingam}, M., {Silverman}, J.~M., \& {Filippenko}, A.~V. 2011, \mnras, 415, L81 

\bibitem[Prasad 
\& Li(2006)]{disco06jd} Prasad, R.~R., \& Li, W.\ 2006, Central Bureau Electronic Telegrams, 673, 1 

\bibitem[Prieto et al.(2007)]{prieto07} Prieto, J.~L., 
Garnavich, P.~M., Phillips, M.~M., et al.\ 2007, arXiv:0706.4088

\bibitem[Quinn et al.(2008)]{class08fq} Quinn, J., Baade, D., 
Clocchiatti, A., et al.\ 2008, Central Bureau Electronic Telegrams, 1510, 1 


\bibitem[Roming et al.(2012)]{roming12} Roming, P.~W.~A., 
Pritchard, T.~A., Prieto, J.~L., et al.\ 2012, \apj, 751, 92

\bibitem[Salamanca et al.(1998)]{salamanca98} Salamanca, I., 
Cid-Fernandes, R., Tenorio-Tagle, G., et al.\ 1998, \mnras, 300, L17 

\bibitem[Schlafly 
\& Finkbeiner(2011)]{schlafly11} Schlafly, E.~F., \& Finkbeiner, D.~P.\ 2011, \apj, 737, 103 

\bibitem[Schlegel(1990)]{schlegel90} Schlegel, E.~M.\ 1990, 
\mnras, 244, 269 

\bibitem[Silverman et al.(2006)]{silverman06} Silverman, J.~M., 
Wong, D., Filippenko, A.~V., 
\& Chornock, R.\ 2006, Central Bureau Electronic Telegrams, 766, 2 

\bibitem[Smartt et al. (2009)]{smartt09} Smartt, S.~J., Eldridge, 
J.~J., Crockett, R.~M., \& Maund, J.~R.\ 2009, \mnras, 395, 1409 

\bibitem[Smith et al. (2012)]{smith12} Smith, N., Silverman, 
J.~M., Filippenko, A.~V., et al.\ 2012, \aj, 143, 17 

\bibitem[Smith et al. (2011)]{smith11rates} Smith, N., Li, W., 
Filippenko, A.~V., \& Chornock, R.\ 2011, \mnras, 412, 1522 

\bibitem[Smith et al. (2011)]{smith11} Smith, N., Li, W., 
Silverman, J.~M., Ganeshalingam, M., 
\& Filippenko, A.~V.\ 2011, \mnras, 415, 773  

\bibitem[Smith et al. (2009b)]{smith09} Smith, N., Hinkle, K.~H., 
\& Ryde, N.\ 2009, \aj, 137, 3558 

{\bibitem[Smith et al. (2009a)]{smith09_05ip} Smith, N., Silverman, 
J.~M., Chornock, R., et al.\ 2009, \apj, 695, 1334}

{\bibitem[Smith et al. (2008)]{smith08} Smith, N., Chornock, R., 
Li, W., et al.\ 2008, \apj, 686, 467}

\bibitem[Smith et al. (2007)]{smith07} Smith, N., Li, W., Foley, 
R.~J., et al.\ 2007, \apj, 666, 1116 

\bibitem[Smith 
\& Owocki(2006)]{smith06} Smith, N., \& Owocki, S.~P.\ 2006, \apjl, 645, L45 

\bibitem[Smith et al.(2002)]{smith02} Smith, J.~A., Tucker, 
D.~L., Kent, S., et al.\ 2002, \aj, 123, 2121 

\bibitem[Soker 
\& Kashi(2012)]{soker12} Soker, N., \& Kashi, A.\ 2012, arXiv:1211.5388 

\bibitem[Sollerman et al.(1998)]{sollerman98} Sollerman, J., 
Cumming, R.~J., \& Lundqvist, P.\ 1998, \apj, 493, 933 

\bibitem[Stathakis \& Sadler(1991)]{stathakis91}
Stathakis, R. A., \& Sadler, E. M. 1991, \mnras, 250, 786

\bibitem[Stritzinger et al.(2012)]{stritzinger12} Stritzinger, M., 
Taddia, F., Fransson, C., et al.\ 2012, \apj, 756, 173 

\bibitem[Stritzinger et al.(2011)]{stritzinger11} Stritzinger, M.~D., 
Phillips, M.~M., Boldt, L.~N., et al.\ 2011, \aj, 142, 156 

\bibitem[Taddia et 
al.(2012)]{taddia12} Taddia, F., Stritzinger, M.~D., Phillips, M.~M., et al.\ 2012, \aap, 545, L7 

\bibitem[Thrasher et al.(2008)]{disco08fq} Thrasher, P., Li, W., 
\& Filippenko, A.~V.\ 2008, Central Bureau Electronic Telegrams, 1507, 1 


\bibitem[{{Turatto} {et~al.}(2003){Turatto}, {Benetti}, \& {Cappellaro}}]{turatto03} 
{Turatto}, M., {Benetti}, S., \& {Cappellaro}, E. 2003, From Twilight to Highlight: The Physics of Supernovae, ed. W. Hillebrandt \& B. Leibundgut (Berlin: Springer), 200  


\bibitem[Turatto et al.(1993)]{turatto93} Turatto, M., 
Cappellaro, E., Danziger, I.~J., et al.\ 1993, \mnras, 262, 128 


\bibitem[Wegner \& Swanson(1996)]{wegner96}
Wegner, G., \& Swanson, S. R. 1996, \mnras, 278, 22

\bibitem[Yoon 
\& Cantiello(2010)]{yoon10} Yoon, S.-C., \& Cantiello, M.\ 2010, \apjl, 717, L62 




\end{thebibliography}

\onecolumn

\clearpage
\begin{deluxetable}{lcccccccccc}
\tabletypesize{\scriptsize}
\tablewidth{0pt}
\tablecaption{Full CSP sample of SNe~IIn.\label{sample}}
\tablehead{
\colhead{SN} &
\colhead{RA} &
\colhead{DEC} &
\colhead{Galaxy} &
\colhead{Galaxy } &
\colhead{Redshift} &
\colhead{Distance} &
\colhead{$E(B-V)_{\rm MW}$} &
\colhead{$E(B-V)_{\rm host}$} &
\colhead{EW$(\rm \ion{Na}{i}~D)_{\rm h}$}   \\
\colhead{} &
\colhead{} &
\colhead{} &
\colhead{} &
\colhead{Type} &
\colhead{} &
\colhead{(Mpc)} &
\colhead{(mag)} &
\colhead{(mag)} &
\colhead{(\AA)}}
\startdata
2005ip   & 09:32:06.42  & $+$08:26:44.4 & NGC 2906     &   Scd       &   0.0071    &    34.9$\pm$0.6                & 0.042        & \ldots              &\ldots              \\
2005kj   & 08:40:09.18&$-$05:36:02.2 & A084009-0536 &     $\cdots$       & 0.0170      &77.3$\pm$1.5               &0.025    &\ldots              &\ldots                \\     
2006aa  & 11:53:19.89&$+$20:45:18.2 & NGC 3947         &  SBb    & 0.0207      &97.5$\pm$1.8                     &0.036             & \ldots            &\ldots              \\  
2006bo  & 20:30:41.90&$+$09:11:40.8& UGC 11578        &  Sdm   & 0.0153    &70.0$\pm$1.3                     &0.091                 & \ldots          &\ldots              \\ 
2006jd   & 08:02:07.43 & $+$00:48:31.5 &UGC 4179       &   SBb     &  0.0186      &83.8$\pm$1.5                    & 0.048            &   \ldots        &\ldots               \\
2006qq  & 05:19:50.43&$-$20:58:06.4 & E553-G36          &   Sb    & 0.0290    &125.0$\pm$2.3     	&0.034                    &0.12$\pm$0.03          &0.76$\pm$0.21 \\ 
2008fq   & 20:25:06.20 & $-$24:48:28.00 &NGC 6907      & SB(s)bc & 0.0106  & 49.6$\pm$0.9        &0.056                       &0.46$\pm$0.03         & 2.90$\pm$0.20  \\
\enddata
\end{deluxetable}

\begin{deluxetable}{lllccc}
\tabletypesize{\scriptsize}
\tablewidth{0pt}
\tablecolumns{5}
\tablecaption{Discovery Telegrams.\label{cbet}}
\tablehead{
\colhead{SN} &
\colhead{Discovery\tablenotemark{a}} &
\colhead{Classification\tablenotemark{a}} &
\colhead{Last non detection} &
\colhead{Discovery} &
\colhead{Explosion date}  \\
\colhead{} &
\colhead{telegram} &
\colhead{telegram} &
\colhead{JD$-$2453000} &
\colhead{JD$-$2453000}&
\colhead{JD$-$2453000}}
\startdata
2005ip    & CBET 275       & CBET 276      &  391.50  &      679.66  &     535.58$\pm$144.08               \\
2005kj    & CBET 296      & CBET  296     & $\cdots$ &692.00               &        $\cdots$           \\
2006aa   &  CBET 395     &  ATel 732       &758.99    &774.98      & 766.99 $\pm$8.00     \\ 
2006bo   &  CBET 468 & CBET 481       &702.5 0 &  830.65      & 766.58 $\pm$64.07     \\
2006jd    & CBET 673    & CBET 679        &  846.7   &   1021.04          &       933.87$\pm$ 87.17            \\
2006qq   &  CBET 766  & CBET 766  &1035.90  &1069.87     & 1052.89$\pm$16.98   \\
2008fq    & CBET  1507 & CBET 1510 &   1715.80  &   1724.79   &  1720.29$\pm$4.49\\    
\enddata
\tablenotetext{a}{References: 
CBET 275: \citet{disco05ip}; 
CBET 276: \citet{class05ip}; 
CBET 296: \citet{bonnaud05}; 
CBET 395: \citet{lee06}; 
Atel 732: \citet{blanc06}; 
CBET 468: \citet{boles06}; 
CBET 481: \citet{blondin06_bo};  
CBET 673: \citet{disco06jd}; 
CBET 679: \citet{class06jd}; 
CBET 766: \citet{silverman06}; 
CBET 1507: \citet{disco08fq}; 
CBET 1510: \citet{class08fq}.} 
\end{deluxetable}

\onltab{3}{
\begin{deluxetable} {lllcccccccccc}
\rotate
\tabletypesize{\tiny}
\tablecolumns{13}
\tablewidth{0pt}
\tablecaption{Optical and near-infrared photometry of the local sequences in the standard system\label{tabstars}}
\tablehead{
\colhead{} &
\colhead{}  &
\colhead{}  &
\colhead{$u'$}   &
\colhead{$g'$}   &
\colhead{$r'$}   &
\colhead{$i'$}   &
\colhead{$B$}   &
\colhead{$V$}  &
\colhead{$Y$}   &
\colhead{$J$}   &
\colhead{$H$}  &
\colhead{$K_s$}\\
\colhead{STAR} &
\colhead{$\alpha(2000)$}  &
\colhead{$\delta(2000)$}  &
\colhead{(mag)}   &
\colhead{(mag)}   &
\colhead{(mag)}   &
\colhead{(mag)}   &
\colhead{(mag)}   &
\colhead{(mag)}  &
\colhead{(mag)}   &
\colhead{(mag)}   &
\colhead{(mag)}  &
\colhead{(mag)}}
\startdata
\multicolumn{13}{c}{\bf SN 2005kj}\\
01	& 08:40:14.48	& $-$05:35:21.98	& $ \cdots $	& $ \cdots $	& $ \cdots $	& $ \cdots $	&  14.142(016)	& $ \cdots $	& $ \cdots $	& $ \cdots $	& $ \cdots $	& $ \cdots $	\\
02	& 08:40:06.53	& $-$05:37:07.63	& $ \cdots $	& $ \cdots $	& $ \cdots $	& $ \cdots $	& $ \cdots $	& $ \cdots $	& $ \cdots $	& $ \cdots $	& $ \cdots $	& $ \cdots $	\\
03	& 08:40:05.50	& $-$05:37:05.35	&  15.772(015)	& $ \cdots $	& $ \cdots $	& $ \cdots $	&  14.555(009)	&  13.753(014)	&  12.616(005)	&  12.348(016)	&  11.932(005)	& $ \cdots $	\\
04	& 08:39:59.04	& $-$05:36:21.37	& $ \cdots $	& $ \cdots $	& $ \cdots $	& $ \cdots $	& $ \cdots $	& $ \cdots $	& $ \cdots $	& $ \cdots $	& $ \cdots $	& $ \cdots $	\\
05	& 08:40:16.65	& $-$05:37:38.33	&  15.540(015)	&  14.434(015)	&  14.189(015)	&  14.133(015)	&  14.710(008)	&  14.270(008)	&  13.639(005)	&  13.457(005)	&  13.239(005)	& $ \cdots $	\\
06	& 08:39:59.30	& $-$05:33:06.59	&  16.148(010)	&  14.931(007)	&  14.534(007)	&  14.394(008)	&  15.282(008)	&  14.685(007)	& $ \cdots $	& $ \cdots $	& $ \cdots $	& $ \cdots $	\\
07	& 08:40:09.91	& $-$05:32:33.87	&  16.671(008)	&  15.326(007)	&  14.847(007)	&  14.673(009)	&  15.715(008)	&  15.039(007)	&  14.047(007)	&  13.775(007)	&  13.419(010)	& $ \cdots $	\\
\enddata
\tablecomments{Uncertainties given in parentheses in thousandths of a
  magnitude correspond to an rms of the magnitudes obtained on
  photometric nights.}
\end{deluxetable}}

\onltab{4}{
\begin{deluxetable}{cccccccccc}
\tablewidth{0pt}
\tabletypesize{\scriptsize}
\tablecaption{Optical Photometry of five CSP SNe~IIn in the natural system.\label{tabphot}}
\tablehead{
\colhead{JD$-$2453000} &
\colhead{Phase\tablenotemark{a}} &
\colhead{$u$ (mag)}&
\colhead{$g$ (mag)}&
\colhead{$r$ (mag)}&
\colhead{$i$ (mag)}&
\colhead{$B$ (mag)}&
\colhead{$V$ (mag)}&
\colhead{Telescope\tablenotemark{b}}}
\startdata
\multicolumn{9}{c}{\bf SN~2005kj}\\
692.00  &+0.00 &       \ldots &  \ldots      &   16.3\tablenotemark{c}    &\ldots &\ldots  &  \ldots     &  $\cdots$ \\
695.83 &+3.83 &  16.196(012) &  16.067(007) &  16.066(008) &  16.192(011) &  16.209(009) &  16.074(008)  &  SWO \\ 
698.83 &+6.83 &  16.224(011) &  16.102(005) &  16.084(004) &  16.195(007) &  16.222(007) &  16.089(005)  &  SWO \\ 
699.80 &+7.80 &  16.234(012) &  16.103(006) &  16.082(008) &  16.187(008) &  16.231(009) &  16.097(009)  &  SWO \\ 
702.79 &+10.79 &  16.284(012) &  16.122(005) &  16.106(004) &  16.201(007) &     \ldots   &  16.116(007) &  SWO \\ 
703.84 &+11.84 &  16.313(013) &  16.134(006) &  16.104(008) &  16.212(009) &  16.264(007) &  16.106(007) &  SWO \\ 
706.79 &+14.79 &  16.356(013) &  16.152(007) &  16.099(009) &  16.217(009) &  16.300(008) &  16.137(008) &  SWO \\
\enddata 
\tablecomments{Values in parentheses are 1$\sigma$ measurement uncertainties in
millimag. The quoted uncertainties include the error on the zero point as determined from the flux measure of the local sequence stars and the uncertainty on the measured SN flux.}
\tablenotetext{a}{Days since discovery.}
\tablenotetext{b}{DUP and SWO correspond to the du Pont and Swope telescopes, respectively.}
\tablenotetext{c}{Unfiltered magnitudes from discovery and confirmation images (see  Table~\ref{cbet} for  CBET references).}
\end{deluxetable}}

\onltab{5}{
\begin{deluxetable}{ccccccc}
\tablewidth{0pt}
\tabletypesize{\scriptsize}
\tablecaption{NIR Photometry of five CSP SNe~IIn.\label{tabphotIR}}
\tablehead{
\colhead{JD$-$2453000} &
\colhead{Phase\tablenotemark{a}} &
\colhead{$Y$ (mag)}&
\colhead{$J$ (mag)}&
\colhead{$H$ (mag)}&
\colhead{$K_s$ (mag)}&
\colhead{Telescope\tablenotemark{b}}}
\startdata
\multicolumn{7}{c}{\bf SN~2005kj}\\
696.83 & $+$4.83    &  15.875(014)& 15.764(019)& 15.545(019)& \ldots      &SWO\\
701.82 & $+$9.82    &  15.890(020)& 15.748(020)& 15.567(023)& \ldots      &SWO\\
705.86 & $+$13.86   &  15.898(020)& 15.782(023)& 15.564(017)& \ldots      &SWO\\
710.86 & $+$18.86   &  15.871(015)& 15.837(026)& 15.566(025)& \ldots      &SWO\\
714.79 & $+$22.79   &  15.918(019)& 15.791(026)& 15.579(024)& \ldots      &SWO\\
717.81 & $+$25.81   &  15.913(015)& 15.779(023)& 15.581(020)& \ldots      &SWO\\
719.78 & $+$27.78   &  15.948(020)& 15.808(025)& 15.596(018)& \ldots      &SWO\\
\enddata 
\tablecomments{Values in parentheses are 1$\sigma$ measurement uncertainties in
  millimag. The quoted uncertainties include the error on the zero point as determined from the flux measure of the local sequence stars and the uncertainty on the measured SN flux.}
\tablenotetext{a}{Days since discovery.}
\tablenotetext{b}{DUP, SWO, and BAA correspond to the du Pont, Swope and Baade telescopes, respectively.}
\end{deluxetable}}

\onltab{6}{
\begin{deluxetable}{lllllllllll}
\tablewidth{0pt}
\tabletypesize{\scriptsize}
\tablecaption{UT dates of optical and NIR template observations.\label{tabtemplate}}
\tablehead{
\colhead{SN/filter}&
\colhead{$u$}&
\colhead{$B$}&
\colhead{$g$}&
\colhead{$V$}&
\colhead{$r$}&
\colhead{$i$}&
\colhead{$Y$}&
\colhead{$J$}&
\colhead{$H$}&
\colhead{$K_s$}}
\startdata
        2005kj & 2008-03-06 & 2008-03-14 & 2008-03-06 & 2008-03-06 & 2008-03-14 & 2008-03-06 & 2011-12-20 & 2011-12-20 & 2011-12-20 & 2013-02-02 \\ 
        2005ip & \ldots     & \ldots     & \ldots     & \ldots     & \ldots     & \ldots     & \ldots     & \ldots     & \ldots     & \ldots     \\ 
        2006aa & 2008-03-14 & 2008-03-14 & 2008-03-14 & 2008-03-14 & 2008-03-14 & 2008-03-14 & 2007-06-28 & 2007-06-28 & 2011-02-21 & \ldots     \\ 
        2006bo & 2007-09-16 & 2007-09-16 & 2007-09-16 & 2007-09-16 & 2007-09-16 & 2007-09-16 & 2007-06-29 & 2007-06-29 & 2007-06-29 & \ldots     \\ 
        2006jd & \ldots     & \ldots     & \ldots     & \ldots     & \ldots     & \ldots     & \ldots     & \ldots     & \ldots     & \ldots     \\ 
        2006qq & 2008-03-14 & 2008-03-14 & 2008-03-14 & 2008-03-14 & 2008-03-14 & 2008-03-14 & 2009-04-16 & 2009-04-16 & 2009-05-03 & \ldots     \\ 
        2008fq & 2012-04-22 & 2012-11-07 & 2012-09-15 & 2012-09-15 & 2012-09-15 & 2012-09-15 & 2010-04-29 & 2010-04-29 & 2011-07-21 & 2010-09-25 \\ 
    \enddata 
\tablecomments{SNe~2005ip and 2006jd are still visible given their extraordinary long-lasting CSM-interaction; therefore, templates are not available yet.}
\end{deluxetable}}

\onltab{7}{
\begin{deluxetable}{cccccccc}
\tabletypesize{\scriptsize}
\tablewidth{0pt}
\tablecolumns{8}
\tablecaption{Spectroscopic Observations of five CSP SNe~IIn.\label{tabspectra}}
\tablehead{
\colhead{Date} &
\colhead{Julian Date} &
\colhead{Epoch\tablenotemark{a}} &
\colhead{Telescope} &
\colhead{Instrument} &
\colhead{Spectral Range} &
\colhead{Resolution} &
\colhead{Integration} \\ 
\colhead{(UT)} &
\colhead{JD$-$2453000} &
\colhead{(days)} &
\colhead{} &
\colhead{} &
\colhead{(\AA)} &
\colhead{(FWHM \AA)} &
\colhead{(s)}}
\startdata

\multicolumn{8}{c}{\bf SN~2005kj}\\
2005 Nov. 24 & 698.77 & +6.8 & Du~Pont    & MS     &  3780 -- 7289   & 5 & 600x3  \\ 
2005 Nov. 25 & 699.76 & +7.8 & Du~Pont    & MS     &  3780 -- 7289   & 5 & 600x3  \\ 
2005 Dec. 08 & 712.76 & +20.8 & T60        & CS     &  3200 -- 9606   & 14 & 900x3  \\ 
2005 Dec. 20 & 724.81 & +32.8 & Du~Pont    & WFCCD  &  3800 -- 9235     & 8 & 400x3  \\ 
2005 Dec. 21 & 725.76 & +33.8 & Du~Pont    & WFCCD  &  3800 -- 9235     & 8 & 400x3  \\  
2005 Dec. 23 & 727.76 & +35.8 & Du~Pont    & WFCCD  &  3800 -- 9235     & 8 & 400x3  \\ 
2006 Jan. 16 & 751.76 & +59.8 & NTT        & EMMI   &  3200 -- 10200    & 6--9 & 200x3  \\
\enddata
\tablenotetext{a}{Days since discovery.} 
\end{deluxetable}}

\begin{deluxetable}{lccccccccc}
\tabletypesize{\scriptsize}
\tablewidth{0pt}
\tablecolumns{10}
\tablecaption{Decay Rate at 100 Days Since Discovery for five CSP SNe~IIn.\label{tabslope}}
\tablehead{
\colhead{SN} &
\colhead{$B$} &
\colhead{$g$} &
\colhead{$V$} &
\colhead{$r$} &
\colhead{$i$} &
\colhead{UVOIR} \\
\colhead{      } &
\colhead{(mag/10 days)} &
\colhead{(mag/10 days)} &
\colhead{(mag/10 days)} &
\colhead{(mag/10 days)} &
\colhead{(mag/10 days)} &
\colhead{(mag/10 days)} }
\startdata
2005ip &     0.28     &  0.27      &   0.32       &   0.29    &  0.29 & 0.11   \\
2005kj  & 0.28 &0.24 &0.20 &0.18 &0.15 &0.16 \\
2006aa  & 0.4 &0.33 &0.27 &0.20 &0.20&0.23 \\
2006jd  & 0.07 &0.06 &0.06 &0.07 &0.08 &  0.03     \\
2006qq  & 0.06 &0.10 &0.08 &0.05 &0.12 &0.09 \\
\enddata
\end{deluxetable}

\begin{deluxetable}{lrrrrrrrrr}
\tabletypesize{\scriptsize}
\tablewidth{0pt}
\tablecolumns{10}
\tablecaption{Peak Absolute Magnitudes  for the Full CSP SN~IIn Sample.\label{max}}
\tablehead{
\colhead{SN} &
\colhead{$u$ (mag)} &
\colhead{$B$ (mag)} &
\colhead{$g$ (mag)} &
\colhead{$V$ (mag)} &
\colhead{$r$  (mag)} &
\colhead{$i$  (mag)} &
\colhead{$Y$ (mag)} &
\colhead{$J$  (mag)} &
\colhead{$H$ (mag)}}
\startdata
2005ip & $<$ $-$16.99  &  $<$ $-$17.28  &  $<$ $-$17.50  &  $<$ $-$17.63  &  $<$ $-$17.78  &  $<$ $-$17.77  &  $<$ $-$18.18  &  $<$ $-$18.34  &  $<$ $-$18.56\\
2005kj&$<$ $-$18.40 & $<$ $-$18.38 & $<$ $-$18.49 &  $<$ $-$18.48 &  $<$ $-$18.47 &  $<$ $-$18.33 &     $<$ $-$18.60 &    $<$ $-$18.70 &    $<$ $-$18.91 \\ 
2006aa&    $-$17.41 &     $-$17.49 &     $-$17.63 &      $-$17.60 &      $-$17.67 &      $-$17.58 &      $-$17.94 &      $-$18.03 &      $-$18.14 \\ 
2006bo&$<$ $-$15.97 & $<$ $-$16.54 & $<$ $-$16.75 &  $<$ $-$16.84 &  $<$ $-$16.95 &  $<$ $-$17.00 &  $<$ $-$17.26 &  $<$ $-$17.48 &  $<$ $-$17.67 \\ 
2006jd &$<$ $-$16.26 & $<$ $-$16.69 & $<$ $-$16.92 & $<$ $-$17.03 & $<$ $-$17.29 & $<$ $-$17.17 & $<$ $-$-17.96 &  $-$18.14 &  $-$19.16\\
2006qq&$<$ $-$17.84 & $<$ $-$18.10 & $<$ $-$18.25 &  $<$ $-$18.32 &  $<$ $-$18.36 &  $<$ $-$18.33 &  $<$ $-$18.75 &  $<$ $-$18.96 &  $<$ $-$19.21 \\ 
2008fq & $<$ $-$18.73 &      $-$19.09 &      $-$19.26 &      $-$19.33 &      $-$19.31 &      $-$19.19 &       $-$19.43 &      $-$19.43 &      $-$19.58 \\
\enddata
\tablecomments{Typical uncertainties are 0.1 mag and are dominated by the errors on distance and extinction. 
Upper limits are given when the SN was observed after maximum.}
\end{deluxetable}

\onltab{10}{
\begin{deluxetable}{cccccccc}
\tabletypesize{\scriptsize}
\tablewidth{0pt}
\tablecaption{Balmer line fluxes in units of 10$^{-14}~$erg~s$^{-1}$~cm$^{-2}$ for SN~2005kj.\label{tab_flux05kj}}
\tablehead{
\colhead{Days}&
\colhead{F(H$\alpha_{\rm b}$)}&
\colhead{F(H$\alpha_{\rm n}$)}&  
\colhead{F(H$\alpha_{\rm tot}$)}&
\colhead{F(H$\beta_{\rm b}$)}&  
\colhead{F(H$\beta_{\rm n}$)}&  
\colhead{F(H$\beta_{\rm P-Cyg}$)}&  
\colhead{F(H$\beta_{\rm tot}$)}\\
\colhead{since discovery}&
\colhead{~}&
\colhead{~}&
\colhead{~}&
\colhead{~}&
\colhead{~}&
\colhead{~}&
\colhead{~}}
\startdata
$+$6.77   & 4.76(0.50) & 2.86(0.47)  & 7.63(0.68) &   3.18(0.27) & 0.58(0.28)  &    \ldots        &   3.76(0.39) \\ 
$+$7.76   & 4.14(0.90) & 1.25(0.97)  & 5.39(1.32) &   3.37(0.18) & 0.37(0.17)  &    \ldots        &   3.74(0.25) \\ 
$+$20.76  & 4.03(0.50) & 1.98(0.50)  & 6.02(0.71) &   2.94(0.50) & 1.67(0.50)  &    \ldots        &   4.61(0.71) \\ 
$+$32.81  & 3.31(0.34) & 1.75(0.36)  & 5.06(0.50) &   0.80(0.73) & 1.19(0.66)  &    \ldots        &   1.99(0.98) \\ 
$+$33.76  & 2.70(0.30) & 1.76(0.30)  & 4.46(0.43) &   1.75(0.31) & 0.35(0.31)  &    \ldots        &   2.10(0.44) \\ 
$+$35.76  & 2.85(0.29) & 1.80(0.31)  & 4.65(0.43) &   1.96(0.33) & 1.03(0.34)  &    \ldots        &   2.98(0.47) \\ 
$+$59.76  & 2.45(0.28) & 1.63(0.27)  & 4.08(0.39) &   \ldots     & 0.85(0.14)  &    \ldots        &   0.85(0.14) \\ 
$+$67.66  & 2.00(0.17) & 1.31(0.18)  & 3.31(0.25) &   \ldots     & 0.50(0.05)  &    $-$0.08(0.03)   &   0.42(0.08) \\
\enddata
\tablecomments{Values in parentheses are 1$\sigma$ uncertainties.} 
\end{deluxetable}}

\onltab{11}{
\begin{deluxetable}{cccccc}
\tabletypesize{\scriptsize}
\tablewidth{0pt}
\tablecaption{Balmer line velocities for SN~2005kj.\label{tab_vel05kj}}
\tablehead{
\colhead{Phase}&
\colhead{FWHM(H$\alpha_{\rm b}$)}&
\colhead{FWHM(H$\alpha_{\rm n}$)}&  
\colhead{FWHM(H$\beta_{\rm b}$)}&  
\colhead{FWHM(H$\beta_{\rm n}$)}&  
\colhead{v$_{\rm min}$(H$\beta_{\rm P-Cyg}$)}\\
\colhead{(days)}&
\colhead{km~s$^{-1}$}&
\colhead{km~s$^{-1}$}&
\colhead{km~s$^{-1}$}&
\colhead{km~s$^{-1}$}&
\colhead{km~s$^{-1}$}}
\startdata
$+$6.77   & 3610(391) &  839(96)    & 2432(313) &  490(170) & \ldots     \\ 
$+$7.76   & 2175(499) &  641(269)   & 2460(219) &  443(153) & \ldots     \\ 
$+$20.76  & 2518(584) &  1041(360)  & 2969(500) &  1455(500)& \ldots     \\ 
$+$32.81  & 2972(381) &  770(95)    & 1385(515) &  808(234) & \ldots     \\ 
$+$33.76  & 2758(267) &  833(81)    & 1597(254) &  387(264) & \ldots     \\ 
$+$35.76  & 2837(297) &  815(80)    & 3487(780) &  805(176) & \ldots     \\ 
$+$59.76  & 3345(485) &  710(85)    & \ldots    &  758(177) & \ldots     \\ 
$+$67.66  & 2256(179) &  661(51)    & \ldots    &  425(56)  & $-$582(21) \\ 
$+$87.72  & 1933(211) &  694(100)   & \ldots    &  907(554) & $-$805(228) \\ 
$+$107.64 & \ldots    &  978(41)    & \ldots    &  531(268) & $-$892(331) \\ 
$+$116.66 & \ldots    &  1020(15)   & \ldots    &  653(118) & $-$722(94) \\ 
$+$124.63 & \ldots    &  1036(49)   & \ldots    &  802(429) & $-$652(103) \\ 
$+$126.57 & \ldots    &  975(28)    & \ldots    &  542(163) & $-$819(103) \\ 
$+$135.56 & \ldots    &  858(33)    & \ldots    &  726(430) & $-$929(467) \\ 
$+$159.49 & \ldots    &  1028(24)   & \ldots    &  889(368) & $-$911(296) \\ 
\enddata
\tablecomments{Values in parentheses are 1$\sigma$ uncertainties.} 
\end{deluxetable}}

\onltab{12}{
\begin{deluxetable}{ccccccccc}
\tabletypesize{\scriptsize}
\tablewidth{0pt}
\tablecaption{Balmer line fluxes in units of 10$^{-14}~$erg~s$^{-1}$~cm$^{-2}$ for SN~2006aa.\label{tab_flux06aa}}
\tablehead{
\colhead{Days}&
\colhead{F(H$\alpha_{\rm b}$)}&
\colhead{F(H$\alpha_{\rm n}$)}&  
\colhead{F(H$\alpha_{\rm P-Cyg}$)}&  
\colhead{F(H$\alpha_{\rm tot}$)}&
\colhead{F(H$\beta_{\rm b}$)}&  
\colhead{F(H$\beta_{\rm n}$)}&  
\colhead{F(H$\beta_{\rm P-Cyg}$)}&  
\colhead{F(H$\beta_{\rm tot}$)}\\
\colhead{since discovery}&
\colhead{~}&
\colhead{~}&
\colhead{~}&
\colhead{~}&
\colhead{~}&
\colhead{~}&
\colhead{~}&
\colhead{~}}
\startdata
$+$18.82 & 1.13(0.17)& 0.96(0.18) &  \ldots      & 2.10(0.24) & 0.50(0.37) & 0.41(0.38)  &  $-$0.09(0.35)& 0.82(0.64)  \\   
$+$24.77 & 1.86(0.13)& 0.76(0.14) &  \ldots      & 2.62(0.19) & 1.08(0.29) & 0.34(0.26)  &  $-$0.09(0.09)& 1.33(0.39)  \\   
$+$34.78 & 2.32(0.14)& 1.04(0.15) &  $-$0.06(0.01) & 3.29(0.21) & 1.71(0.10) & 0.38(0.10)  &  $-$0.09(0.02)& 2.00(0.15)  \\  
$+$34.82 & 2.51(0.28)& 0.82(0.29) &  \ldots      & 3.33(0.41) & 2.14(0.12) & 0.07(0.08)  &  \ldots               & 2.21(0.14) \\   
$+$43.71 & 3.02(0.14)& 0.76(0.12) &  $-$0.13(0.04) & 3.65(0.19) & 1.84(0.19) & 0.37(0.19)  &  $-$0.09(0.03)& 2.12(0.27)  \\  
$+$49.75 & 2.96(0.13)& 0.78(0.14) &  \ldots      & 3.74(0.19) & 2.02(0.24) & 0.35(0.21)  &  $-$0.09(0.05)& 2.29(0.33)  \\  
$+$76.62 & 2.18(0.21)& 1.00(0.23) &  \ldots      & 3.18(0.31) & 0.90(0.50) & 0.22(0.50)  &  \ldots               & 1.12(0.71) \\  
\enddata
\tablecomments{Values in parentheses are 1$\sigma$ uncertainties.} 
\end{deluxetable}}

\onltab{13}{
\begin{deluxetable}{ccccccc}
\tabletypesize{\scriptsize}
\tablewidth{0pt}
\tablecaption{Balmer line velocities for SN~2006aa.\label{tab_vel06aa}}
\tablehead{
\colhead{Phase}&
\colhead{FWHM(H$\alpha_{\rm b}$)}&
\colhead{FWHM(H$\alpha_{\rm n}$)}& 
\colhead{v$_{\rm min}$(H$\alpha_{\rm P-Cyg}$)}& 
\colhead{FWHM(H$\beta_{\rm b}$)}&  
\colhead{FWHM(H$\beta_{\rm n}$)}&  
\colhead{v$_{\rm min}$(H$\beta_{\rm P-Cyg}$)}\\
\colhead{(days)}&
\colhead{km~s$^{-1}$}&
\colhead{km~s$^{-1}$}&
\colhead{km~s$^{-1}$}&
\colhead{km~s$^{-1}$}&
\colhead{km~s$^{-1}$}&
\colhead{km~s$^{-1}$}}
\startdata
$+$18.82 & 2444(460)&  595(70)  & \ldots    & 1735(823)&  593(313)  & $-$780(204) \\ 
$+$24.77 & 2077(191)&  490(58)  & \ldots    & 1744(342)&  518(229)  & $-$619(65)  \\ 
$+$34.78 & 2401(154)&  674(58)  & $-$759(17)   & 2332(192)&  513(94)   & $-$702(17)  \\  
$+$34.82 & 2221(240)&  657(133) & \ldots    & \ldots   &  \ldots    & \ldots   \\ 
$+$43.71 & 2046(120)&  429(45)  & $-$648(37)   & 2190(319)&  457(164)  & $-$596(125) \\  
$+$49.75 & 2029(143)&  410(53)  & \ldots    & 2158(347)&  461(177)  & $-$600(120) \\ 
$+$76.62 & 2070(240)&  547(74)  & \ldots    & 1301(479)&  441(544)  & \ldots   \\ 
\enddata
\tablecomments{Values in parentheses are 1$\sigma$ uncertainties.} 
\end{deluxetable}}

\onltab{14}{
\begin{deluxetable}{ccccccc}
\tabletypesize{\scriptsize}
\tablewidth{0pt}
\tablecaption{Balmer line fluxes in units of 10$^{-14}~$erg~s$^{-1}$~cm$^{-2}$ for SN~2006bo.\label{tab_flux06bo}}
\tablehead{
\colhead{Days}&
\colhead{F(H$\alpha_{\rm n}$)}&  
\colhead{F(H$\alpha_{\rm P-Cyg}$)}&  
\colhead{F(H$\alpha_{\rm total}$)}&
\colhead{F(H$\beta_{\rm n}$)}&  
\colhead{F(H$\beta_{\rm P-Cyg}$)}&  
\colhead{F(H$\beta_{\rm tot}$)}\\
\colhead{since discovery}&
\colhead{~}&
\colhead{~}&
\colhead{~}&
\colhead{~}&
\colhead{~}&
\colhead{~}}
\startdata
 $+$11.21 & 1.62(0.03) & $-$0.11(0.01) & 1.51(0.03)  & 0.79(0.16) & $-$0.25(0.15) & 0.54(0.22)  \\ 
$+$21.23 & 1.28(0.08) & $-$0.20(0.05) & 1.08(0.10)  & 0.52(0.21) & $-$0.27(0.19) & 0.25(0.28)  \\ 
\enddata
\tablecomments{Values in parentheses are 1$\sigma$ uncertainties.} 
\end{deluxetable}}

\onltab{15}{
\begin{deluxetable}{ccccc}
\tabletypesize{\scriptsize}
\tablewidth{0pt}
\tablecaption{Balmer line velocities for SN~2006bo.\label{tab_vel06bo}}
\tablehead{
\colhead{Phase}&
\colhead{FWHM(H$\alpha_{\rm n}$)}& 
\colhead{v$_{\rm min}$(H$\alpha_{\rm P-Cyg}$)}& 
\colhead{FWHM(H$\beta_{\rm n}$)}&  
\colhead{v$_{\rm min}$(H$\beta_{\rm P-Cyg}$)}\\
\colhead{(days)}&
\colhead{km~s$^{-1}$}&
\colhead{km~s$^{-1}$}&
\colhead{km~s$^{-1}$}&
\colhead{km~s$^{-1}$}}
\startdata
$+$11.21 &  1007(20) & $-$661(14)&  933(82) & $-$555(90)    \\ 
$+$21.23 &  1008(47) & $-$607(39)&  876(233) & $-$561(149)  \\ 
\enddata
\tablecomments{Values in parentheses are 1$\sigma$ uncertainties.} 
\end{deluxetable}}

\onltab{16}{
\begin{deluxetable}{ccccccc}
\tabletypesize{\scriptsize}
\tablewidth{0pt}
\tablecaption{Balmer line fluxes in units of 10$^{-14}~$erg~s$^{-1}$~cm$^{-2}$ for SN~2006qq.\label{tab_flux06qq}}
\tablehead{
\colhead{Days}&
\colhead{F(H$\alpha_{\rm b}$)}&
\colhead{F(H$\alpha_{\rm n}$)}&  
\colhead{F(H$\alpha_{\rm tot}$)}&
\colhead{F(H$\beta_{\rm b}$)}&  
\colhead{F(H$\beta_{\rm n}$)}&  
\colhead{F(H$\beta_{\rm tot}$)}\\
\colhead{since discovery}&
\colhead{~}&
\colhead{~}&
\colhead{~}&
\colhead{~}&
\colhead{~}&
\colhead{~}}
\startdata
$+$12.84 & 0.42(0.24)& 1.27(0.19)  & 1.69(0.31)  & 0.18(0.22) & 0.26(0.24) & 0.44(0.33)  \\ 
$+$17.90 & 0.82(0.08)& 1.24(0.04)  & 2.06(0.09)  & 0.53(0.06) & 0.15(0.05) & 0.69(0.07)  \\ 
$+$31.86 & 3.24(0.08)& 0.77(0.02)  & 4.00(0.08)  & 1.05(0.07) & 0.05(0.01) & 1.10(0.07)  \\ 
$+$36.88 & 3.13(0.16)& 1.22(0.07)  & 4.35(0.17)  & 1.41(0.23) & 0.08(0.07) & 1.48(0.24)  \\ 
$+$43.91 & 2.92(0.15)& 1.05(0.06)  & 3.97(0.16)  & 0.94(0.27) & 0.25(0.21) & 1.20(0.34)  \\ 
$+$48.72 & 3.06(0.09)& 0.66(0.03)  & 3.72(0.10)  & 0.73(0.04) & 0.07(0.01) & 0.81(0.05)  \\ 
$+$59.71 & 6.62(1.92)& 0.36(0.10)  & 6.99(2.00)  & 1.27(0.08) & 0.04(0.03) & 1.32(0.09)  \\ 
$+$73.72 & 2.22(0.19)& 0.76(0.05)  & 2.98(0.20)  & \ldots     & \ldots     &\ldots       \\
$+$74.77 & 1.38(0.11)& 0.43(0.03)  & 1.81(0.11)  & \ldots     & \ldots     &\ldots       \\
\enddata
\tablecomments{Values in parentheses are 1$\sigma$ uncertainties.} 
\end{deluxetable}}

\onltab{17}{
\begin{deluxetable}{ccccc}
\tabletypesize{\scriptsize}
\tablewidth{0pt}
\tablecaption{Balmer line velocities for SN~2006qq.\label{tab_vel06qq}}
\tablehead{
\colhead{Phase}&
\colhead{BVZI(H$\alpha_{\rm b}$)}& 
\colhead{RVZI(H$\alpha_{\rm b}$)}& 
\colhead{FWHM(H$\alpha_{\rm n}$)}&  
\colhead{FWHM(H$\beta_{\rm n}$)}\\
\colhead{(days)}&
\colhead{km~s$^{-1}$}&
\colhead{km~s$^{-1}$}&
\colhead{km~s$^{-1}$}&
\colhead{km~s$^{-1}$}}
\startdata
$+$12.84  &6523(200)  & 2613(200)&  450(39) &  706(343)  \\ 
$+$17.90  &7437(200)  & 2613(200)&  328(9)  &  346(115)  \\ 
$+$31.86  &12462(200) & 4897(200)&  208(5)  &  85(25)    \\ 
$+$36.88  &12005(200) & 3070(200)&  579(34) &  254(100)  \\ 
$+$43.91  &9721(200)  & 3526(200)&  572(36) &  1516(929) \\ 
$+$48.72  &10178(200) & 3983(200)&  381(20) &  370(78)   \\ 
$+$59.71  &8807(200)  &   \ldots &  228(70) &  370(293)  \\ 
$+$73.72  &9721(200)  & 1699(200)&  296(24) &  \ldots    \\ 
$+$74.77  &10178(200) & 1699(200)&  267(22) &   \ldots   \\
\enddata
\tablecomments{Values in parentheses are 1$\sigma$ uncertainties.} 
\end{deluxetable}}

\onltab{18}{
\begin{deluxetable}{cccccccc}
\tabletypesize{\scriptsize}
\tablewidth{0pt}
\tablecaption{Balmer line fluxes in units of 10$^{-14}~$erg~s$^{-1}$~cm$^{-2}$ for SN~2008fq.\label{tab_flux08fq}}
\tablehead{
\colhead{Days}&
\colhead{F(H$\alpha_{\rm b}$)}&
\colhead{F(H$\alpha_{\rm n}$)}& 
 \colhead{F(H$\alpha_{\rm P-Cyg^b}$)}&  
 \colhead{F(H$\alpha_{\rm P-Cygni^n}$)}&  
\colhead{F(H$\alpha_{\rm tot}$)}&
\colhead{F(H$\beta_{\rm n}$)}& 
  \colhead{F(H$\alpha_{\rm P-Cyg^n}$)}\\
\colhead{since discovery}&
\colhead{~}&
\colhead{~}&
\colhead{~}&
\colhead{~}&
\colhead{~}&
\colhead{~}&
\colhead{~}}
\startdata
$+$1.75 & 3.65(0.51) & 2.85(0.25)  & -0.37(0.12)  & -0.96(0.26) & 5.17(0.63)  & 1.00(0.30) & -0.22(0.23)  \\  
$+$6.77 & 6.58(0.73) & 3.00(0.41)  & -1.19(0.32)  & -2.25(0.48) & 6.13(1.01)  & 1.00(0.84) & -1.69(1.04)  \\  
$+$13.78& 10.00(0.74) & 1.30(0.30) & -1.31(0.34)  & -0.87(0.42) & 9.12(0.97)  & \ldots     & \ldots       \\ 
$+$28.75& 70.00(7.80) & \ldots     & -20.11(8.40) & \ldots      & 49.89(11.47)& \ldots     & \ldots       \\ 
$+$70.74& 60.00(1.41) & 1.64(0.18) & -10.19(1.31) & \ldots      & 51.44(1.94) & \ldots     & \ldots       \\ 
\enddata
\tablecomments{Values in parentheses are 1$\sigma$ uncertainties.} 
\end{deluxetable}}

\onltab{19}{
\begin{deluxetable}{ccccccc}
\tabletypesize{\scriptsize}
\tablewidth{0pt}
\tablecaption{Balmer line velocities for SN~2008fq.\label{tab_vel08fq}}
\tablehead{
\colhead{Phase}&
\colhead{FWHM(H$\alpha_{\rm b}$)}&
\colhead{FWHM(H$\alpha_{\rm n}$)}& 
\colhead{v$_{\rm min}$(H$\alpha_{\rm P-Cyg^b}$)}&
\colhead{v$_{\rm min}$(H$\alpha_{\rm P-Cyg^n}$)}&  
\colhead{FWHM(H$\beta_{\rm n}$)}&  
\colhead{v$_{\rm min}$(H$\beta_{\rm P-Cyg^n}$)}\\
\colhead{(days)}&
\colhead{km~s$^{-1}$}&
\colhead{km~s$^{-1}$}&
\colhead{km~s$^{-1}$}&
\colhead{km~s$^{-1}$}&
\colhead{km~s$^{-1}$}&
\colhead{km~s$^{-1}$}}
\startdata
$+$1.75  & 4166(581)&  446(36)  &  $-$7387(239)  & $-$531(63)  &  435(177) & 770(100)   \\ 
$+$6.77  & 4568(545)&  366(38)  &  $-$7236(450)  & $-$414(52)  &  410(284) & 482(232)   \\ 
$+$13.78 & 4568(711)&  229(63)  &  $-$7581(307)  & $-$441(104) & \ldots    & \ldots     \\ 
$+$28.75 & 7063(371)&  \ldots   &  $-$7049(1947) & \ldots      & \ldots    & \ldots      \\ 
$+$70.74 & 6463(216)&  183(46)  &  $-$7049(364)  & \ldots      & \ldots    & \ldots      \\ 
\enddata
\tablecomments{Values in parentheses are 1$\sigma$ uncertainties.} 
\end{deluxetable}}

\onltab{20}{
\begin{deluxetable}{ccc}
\tabletypesize{\scriptsize}
\tablewidth{0pt}
\tablecaption{H$\alpha$ EWs for 5 CSP SNe~IIn.\label{tab_EW}}
\tablehead{
\colhead{Phase}&
\colhead{EW(H$\alpha$)}&
\colhead{$\Delta_{EW(H\alpha)}$}\\
\colhead{(days since discovery)}&
\colhead{(\AA)}&
\colhead{(\AA)}}
\startdata
\multicolumn{3}{c}{\bf SN~2005kj}\\
$+$6.77   &64 &7\\
$+$7.76   &56 &15\\
$+$20.76  &67 &26\\
$+$32.81  &62 &6\\
$+$33.76  &55 &6\\
$+$35.76  &58 &5\\
$+$59.76  &65 &6\\
$+$67.66  &56 &4\\
$+$87.72  &54 &10\\
$+$107.64 &41 &1\\
$+$116.66 &55 &1\\
$+$124.63 &61 &2\\
$+$126.57 &64 &1\\
$+$135.56 &65 &2\\
$+$159.49 &112&2\\
\multicolumn{3}{c}{\bf SN~2006aa}\\
$+$18.82 &99  &8\\
$+$24.77 &108 &5\\
$+$34.78 &118 &5\\
$+$34.82 &119 &11\\
$+$43.71 &128 &4\\
$+$49.75 &130 &5\\
$+$76.62 &146 &9\\
\multicolumn{3}{c}{\bf SN~2006bo}\\
$+$11.21 &53 &10\\
$+$21.23 &43 &10\\
\multicolumn{3}{c}{\bf SN~2006qq}\\
$+$12.84 &57  &10\\
$+$17.90 &72  &3\\
$+$31.86 &146 &3\\
$+$36.88 &169 &8\\
$+$43.91 &165 &6\\
$+$48.72 &157 &4\\
$+$59.71 &271 &50\\
$+$73.72 &163 &9\\
$+$74.77 &100 &6\\
\multicolumn{3}{c}{\bf SN~2008fq}\\
$+$1.75 &11  &1\\
$+$6.77 &11  &2\\
$+$13.78& 16 &2\\
$+$28.75& 143&30\\
$+$70.74& 238&9\\
\enddata
\end{deluxetable}}

\clearpage

\begin{deluxetable}{cccccccccccc}
\rotate
\tabletypesize{\scriptsize}
\tablewidth{0pt}
\tablecaption{CSI Parameters for the full CSP SN~IIn Sample.\label{tabpar}}
\tablehead{
\colhead{SN} &
\colhead{$L(H\alpha_{\rm broad}$)}&
\colhead{$v_s$} &
\colhead{$v_w$} &
\colhead{$v^{\rm max}_{\rm ejecta}$} &
\colhead{$\dot{M}$}&
\colhead{$w$}&
\colhead{$n_{\rm H}$}&
\colhead{$R_{\rm CSM}^{\rm in}$}&
\colhead{$R_{\rm CSM}^{\rm out}$}&
\colhead{$R_{\rm CSM}^{\rm sh}$}&
\colhead{$t_{\rm sh}$}\\
\colhead{      } &
\colhead{(10$^{40}$~erg~s$^{-1})$} &
\colhead{(10$^3$~km~s$^{-1}$)} &
\colhead{(10$^2$~km~s$^{-1}$)} &
\colhead{(10$^3$~km~s$^{-1}$)} &
\colhead{(10$^{-3}$~M$_{\odot}$~yr$^{-1}$)}&
\colhead{(10$^{15}$~g~cm$^{-1}$)}&
\colhead{(10$^8$~cm$^{-3}$)} &
\colhead{(10$^{15}$~cm)} &
\colhead{(10$^{15}$~cm)} &
\colhead{(10$^{15}$~cm)} &
\colhead{(yr before core collapse)}}
\startdata
05ip &4  &1.0 &1.2& 16.0      & 0.2$-$0.4&     1.0$-$2.0  &                       $>$3$\times$10$^{-3}$ &$<$0.1 &$>$325.1 &$-$ &$-$\\
05kj  &1.0$-$3.4 &3.0 &6.0&     3.0       & 1.4$-$4.8 &     1.5$-$5.0&      $>$1.8 & $<$0.4 &$>$4.4 & $-$& $-$ \\
06aa  & 1.1$-$3.4 &2.0 &6.0 &   2.5         & 5.0$-$16.0 &      5.7$-$17.1&     $>$7.6 & $<$0.5 & $>$1.8 & 1.3 & 0.7\\
06bo & $<$0.6$-$0.9  &$>$1.0 & 6.0 & 2.5  & $<$ 23.0$-$34.0 &   $<$24.0$-$36.0&       $>$3.8 & $<$0.2--0.9 & $>$1.5$-$2.4& $-$ & $-$ \\
06jd &  3.0$-$13.0&1.6 &$<$1.6 &    16.0             &$<$7.0$-$32.0&   29.3$-$126.9 &        $>$10$^{-2}$& $<$3.0&$>$213.1 & 55.3& $>$110\\
06qq  &5.6$-$12.1  & 6.0  & $<$2.0 & 10.0    &$<$0.3$-$0.7 &   1.0$-$2.2  &       $>$11.4 & $<$2.5 & $>$7.9&6.6&$>$10  \\
08fq &  15.0 &7.5 &5.0& 9.0 &1.1&               1.4&     $>$9.0 &$<$0.8& $>$6.1& 3.5&  0.5$-$0.0~/~$>$3.9$-$2.2\\
\enddata
\tablecomments{Luminosity values for H$\alpha$ of SNe~2005ip and 2006jd are from S12 (here, we use the intermediate component, which is associated with post-shock gas). $v_s$, $v_w$, $v^{\rm max}_{\rm ejecta}$, and n$_{\rm H}$ for SNe~2005ip and 2006jd are from \citet{smith09_05ip} and S12, respectively. The mass-loss rate {and the wind density ($w$)} of SN~2005ip are
from \citet{smith09_05ip}. 
The limits for $R_{\rm CSM}^{\rm in}$ and $R_{\rm CSM}^{\rm out}$ of SNe~2005ip and 2006jd are obtained from the first and the last spectral epochs \citep[][, S12]{smith09_05ip}, and from the assumed $v^{\rm max}_{\rm ejecta}$.
The CSM of SN~2006jd is characterized by the presence of a dense structure, as indicated by the re-brightening of the $r$-band light curve approximately 400 days after explosion. Its radius ($R_{\rm CSM}^{\rm sh}$) and epoch of the outburst are computed from the epoch of the $r$-band maximum and from the assumed $v^{\rm max}_{\rm ejecta}$ and $v_w$. In the case of SN~2008fq, we present the epochs of two outbursts that formed the inner and the outer CSM, as described in Section~\ref{sec:discussion}. Here, $R_{\rm CSM}^{\rm in}$ is the extent of the inner CSM,
$R_{\rm CSM}^{\rm sh}$ the inner radius of the outer CSM, and $R_{\rm CSM}^{\rm out}$ the outer radius of the outer CSM.}
\end{deluxetable}

\clearpage

\onlfig{1}{
\begin{figure}[h]
\begin{center}$
\begin{array}{cc}
\includegraphics[width=6cm]{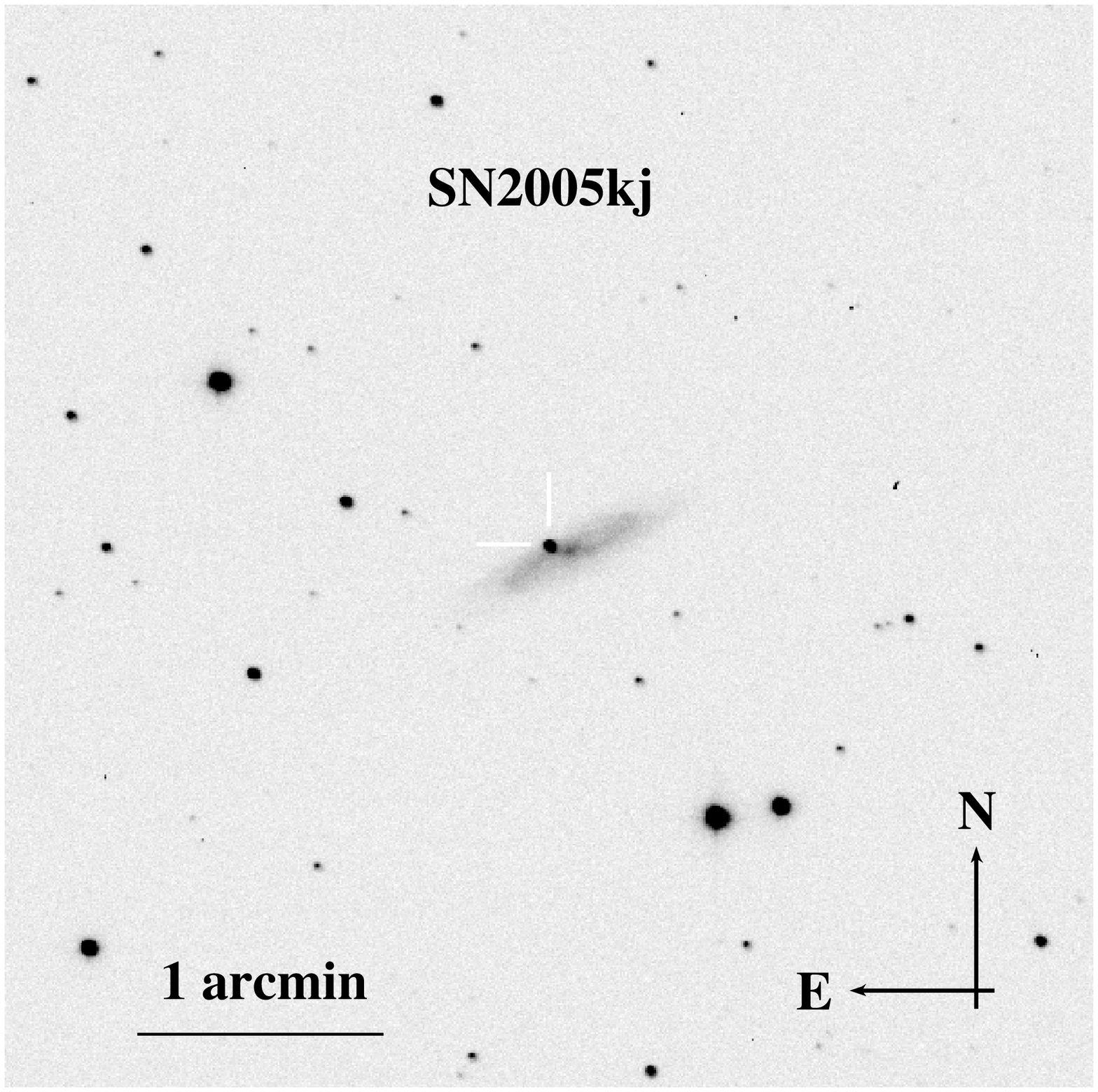} &
\includegraphics[width=6cm]{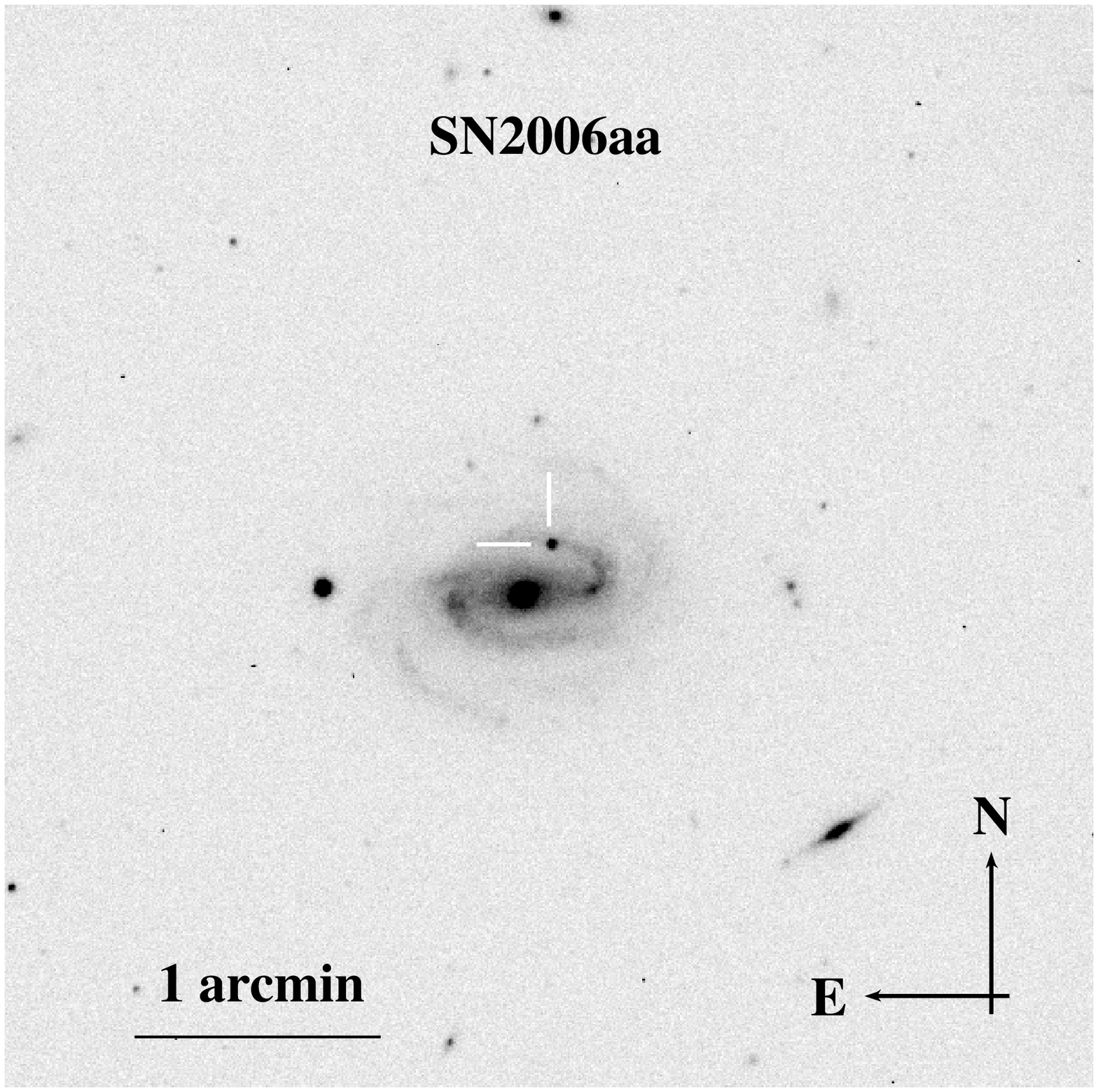}\\
\includegraphics[width=6cm]{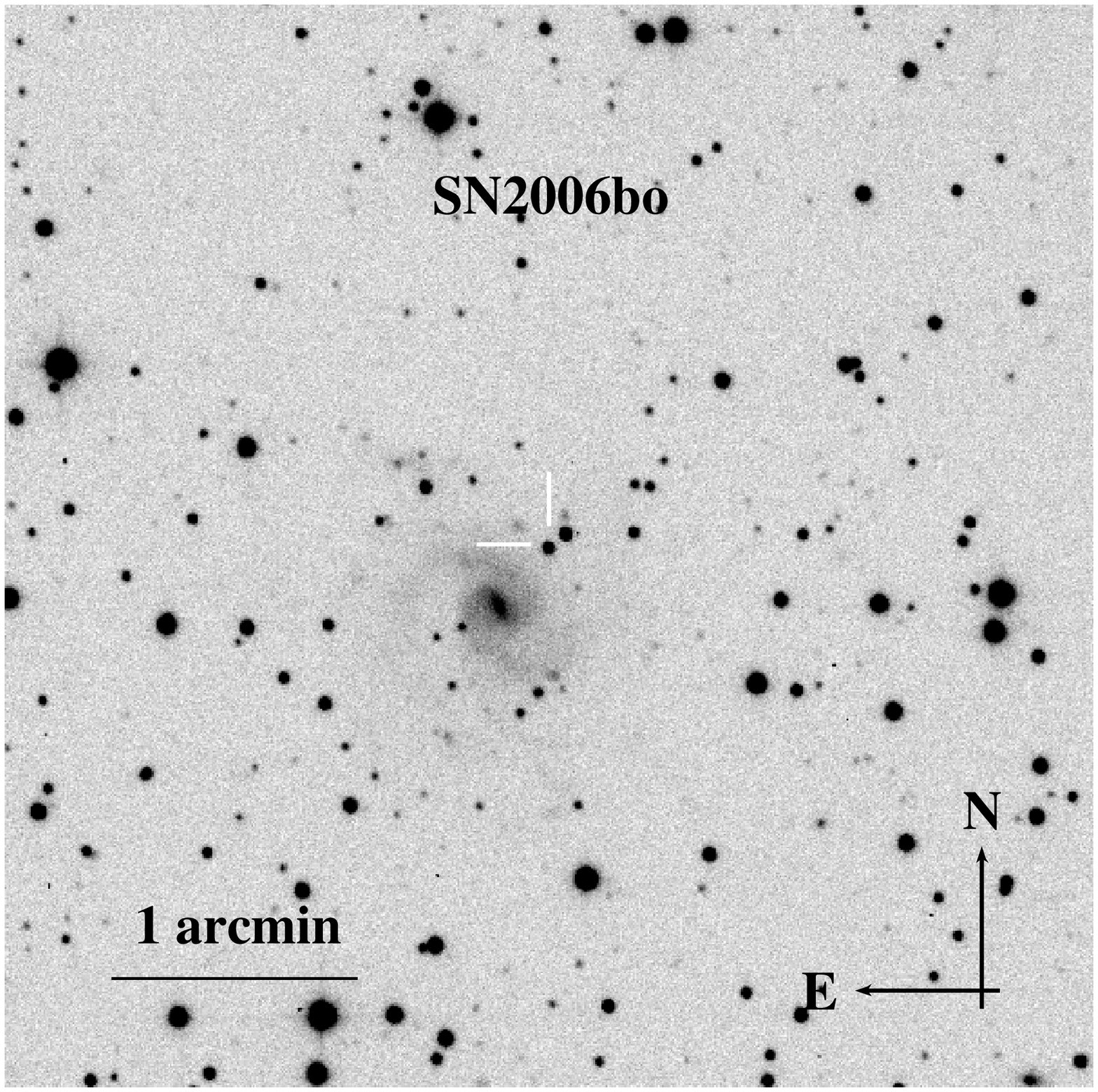} &
\includegraphics[width=6cm]{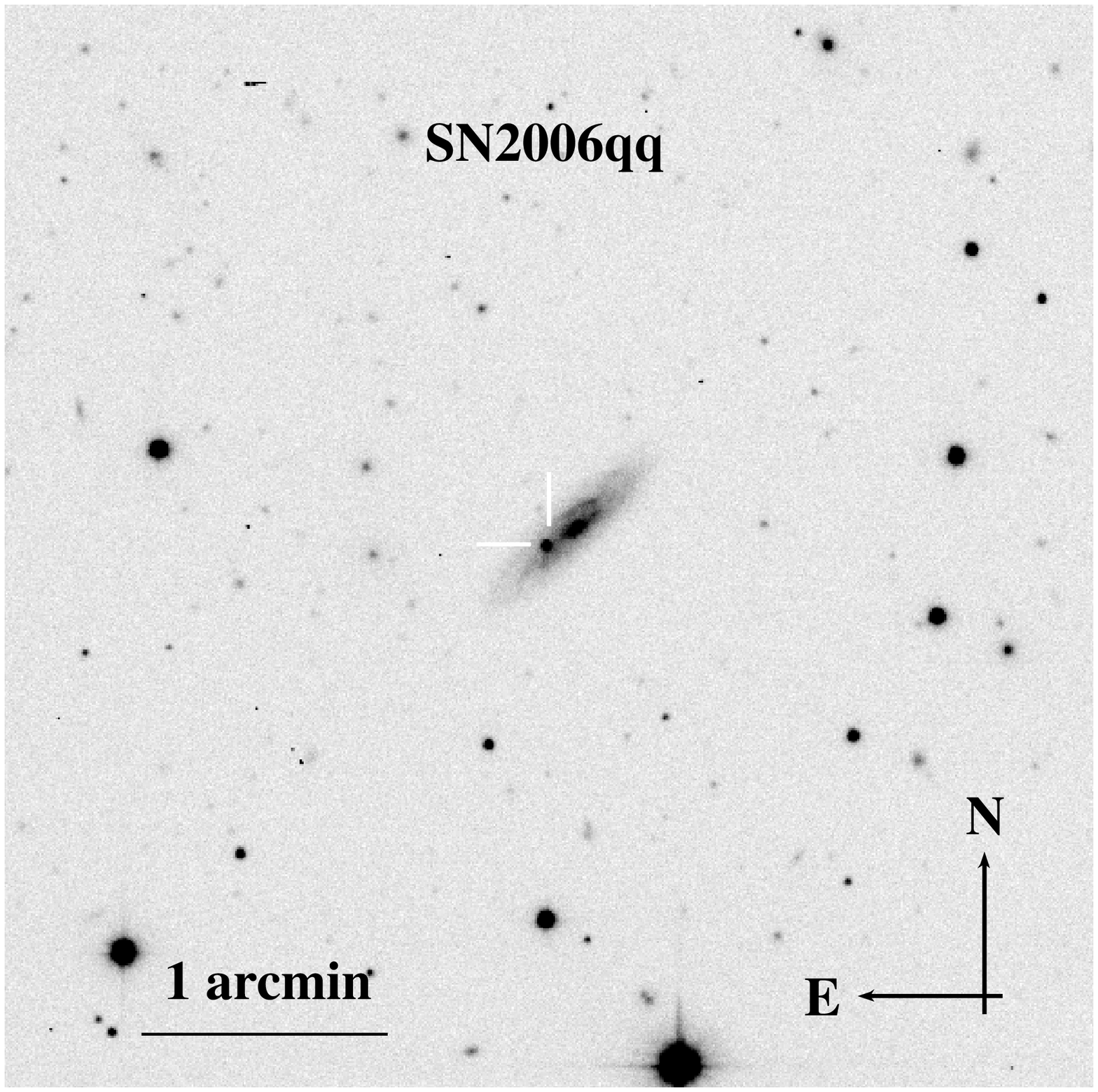} \\
\includegraphics[width=6cm]{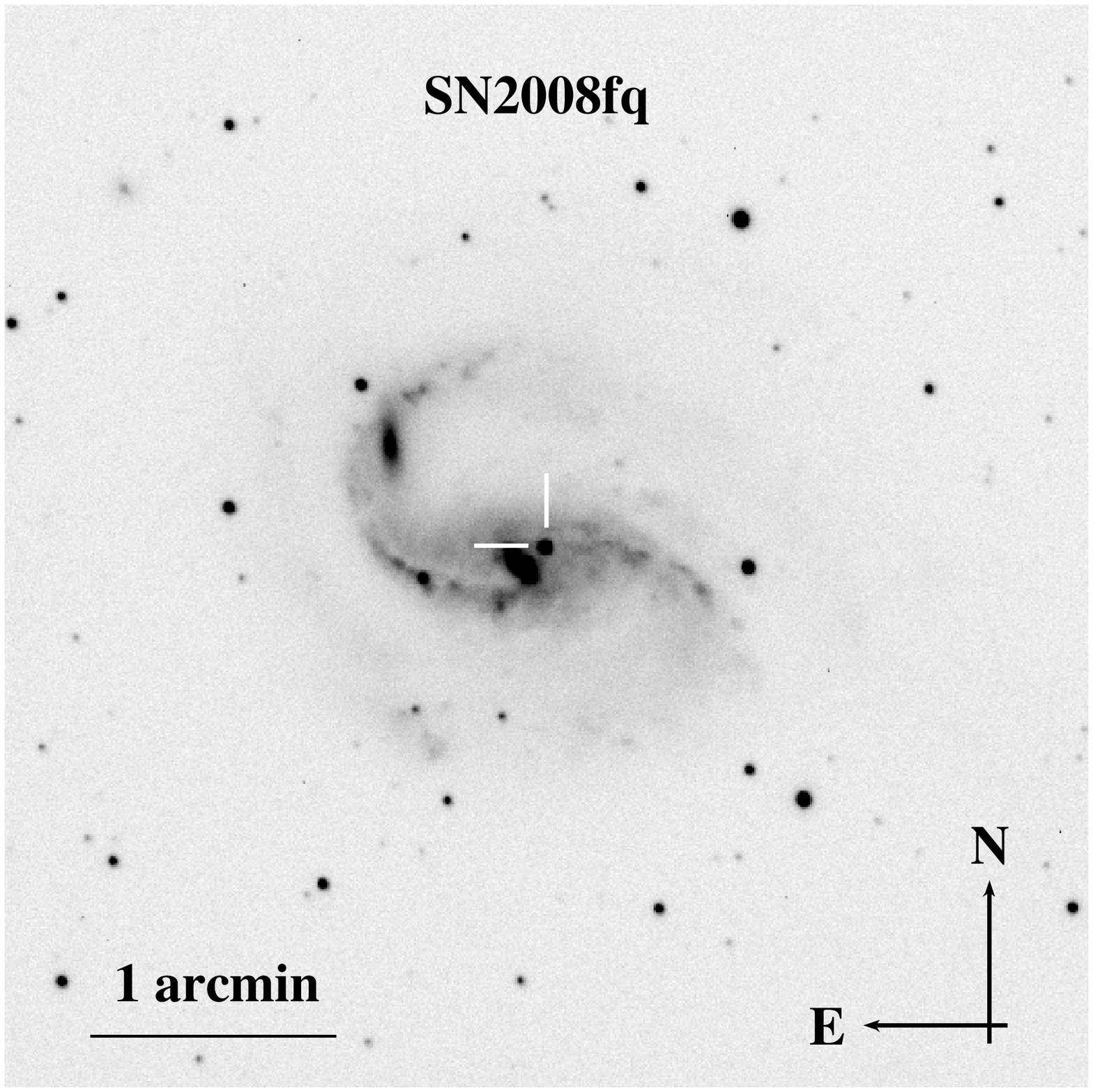} &
\end{array}$
\end{center}
\caption{\label{fc}Swope $V$-band images of 5 CSP SNe~IIn.}
\end{figure}}

 \clearpage
\begin{figure}[h]
\begin{center}$
\begin{array}{cc}
\includegraphics[width=3.6in]{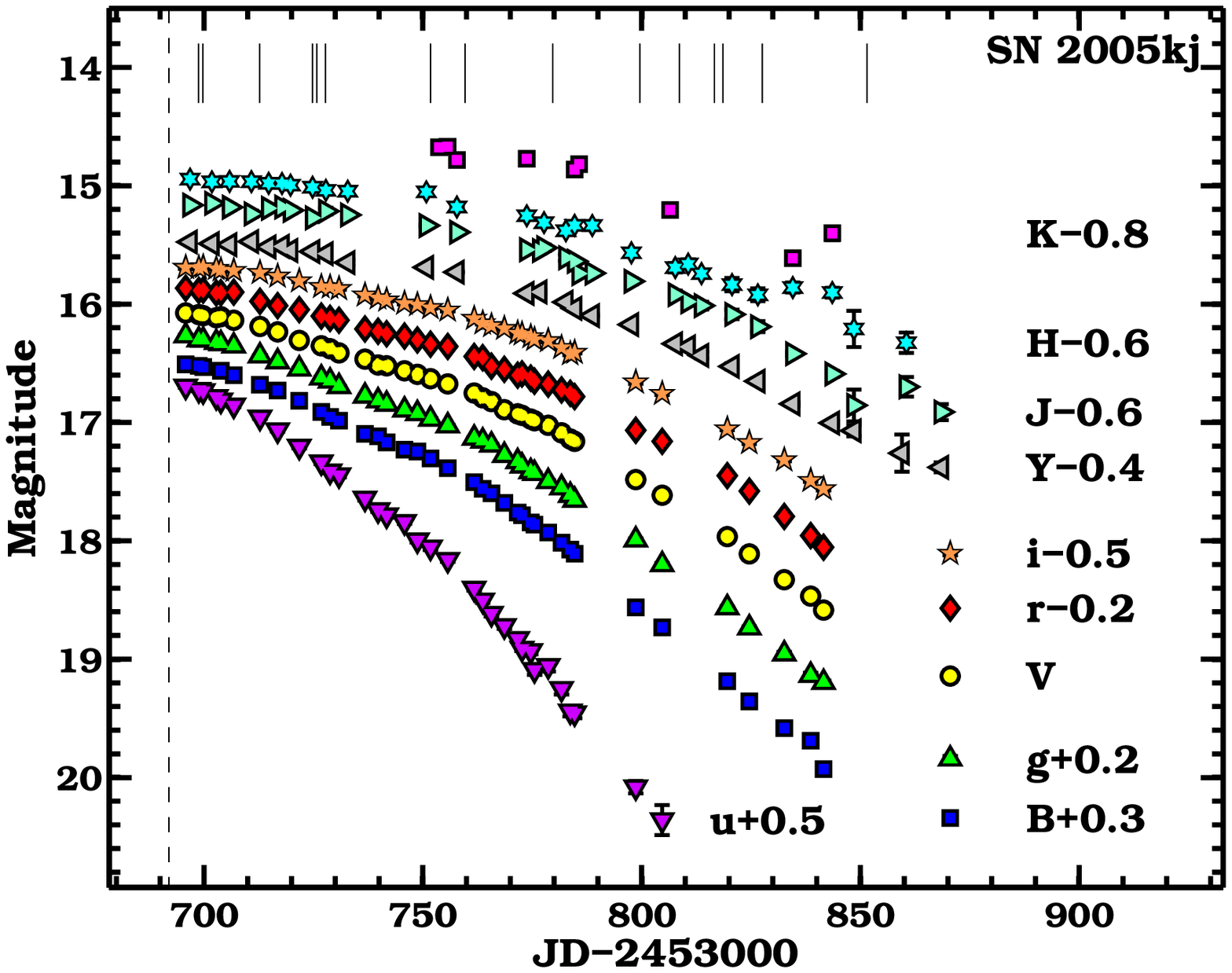} &
\includegraphics[width=3.6in]{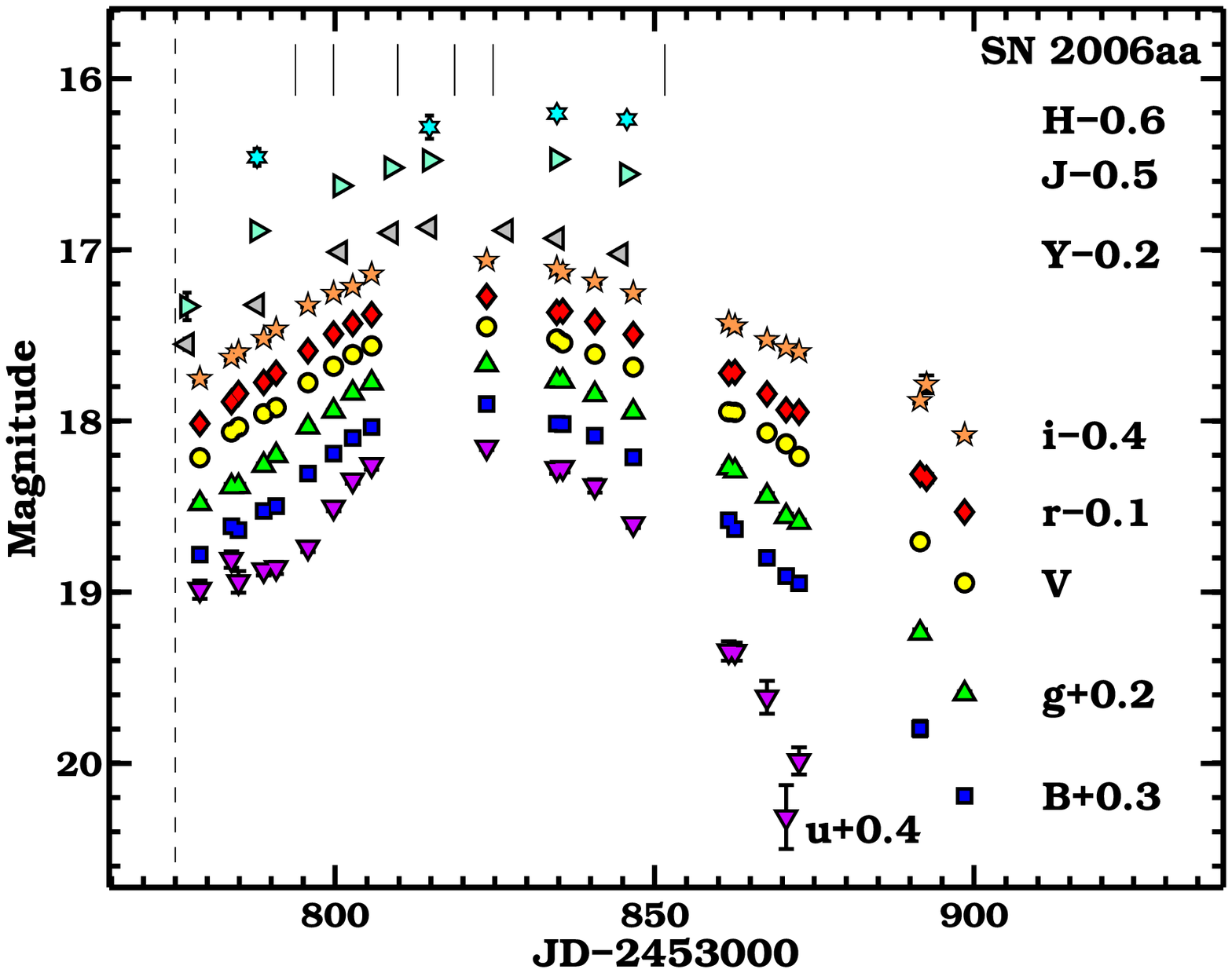} \\
\includegraphics[width=3.6in]{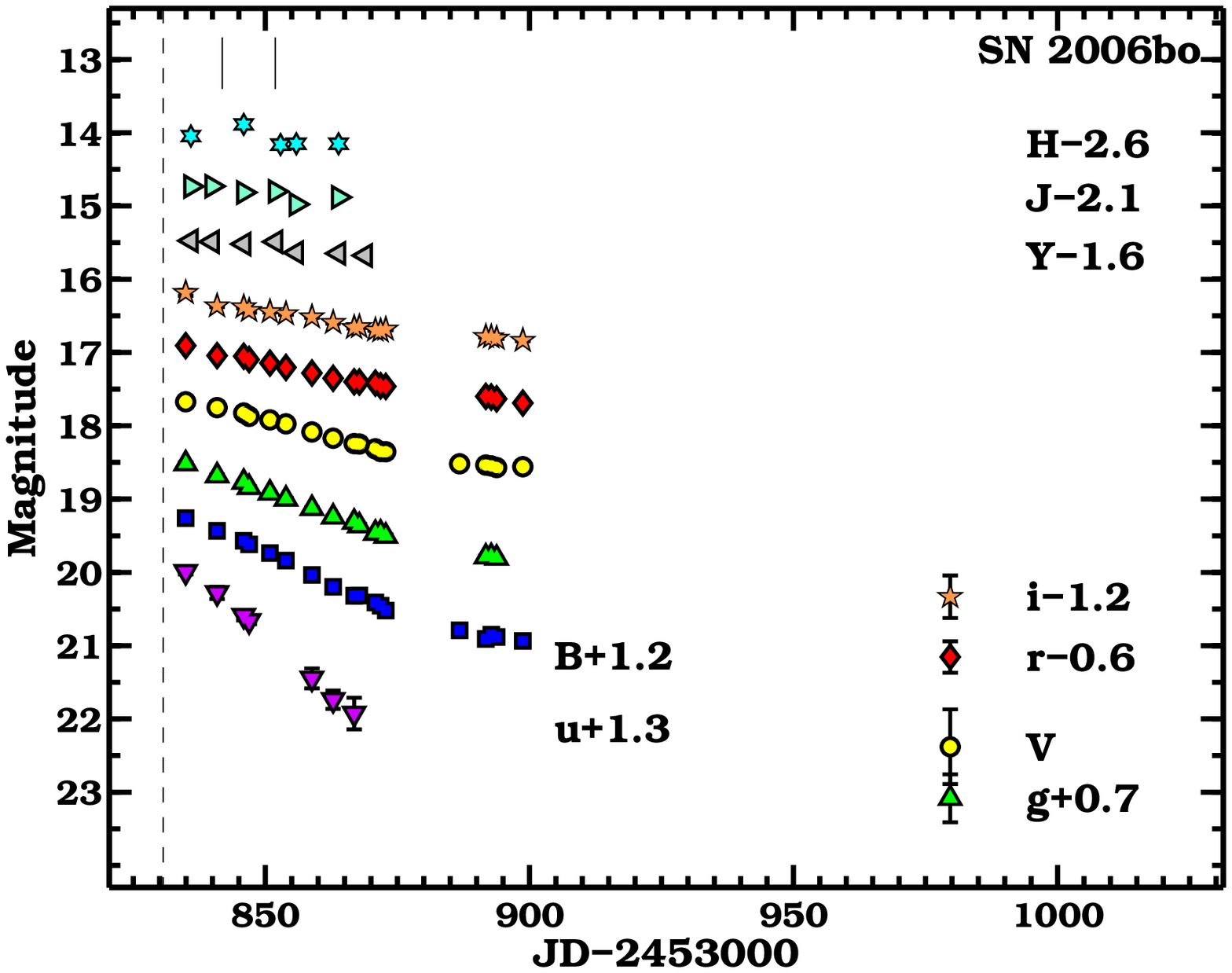} &
\includegraphics[width=3.6in]{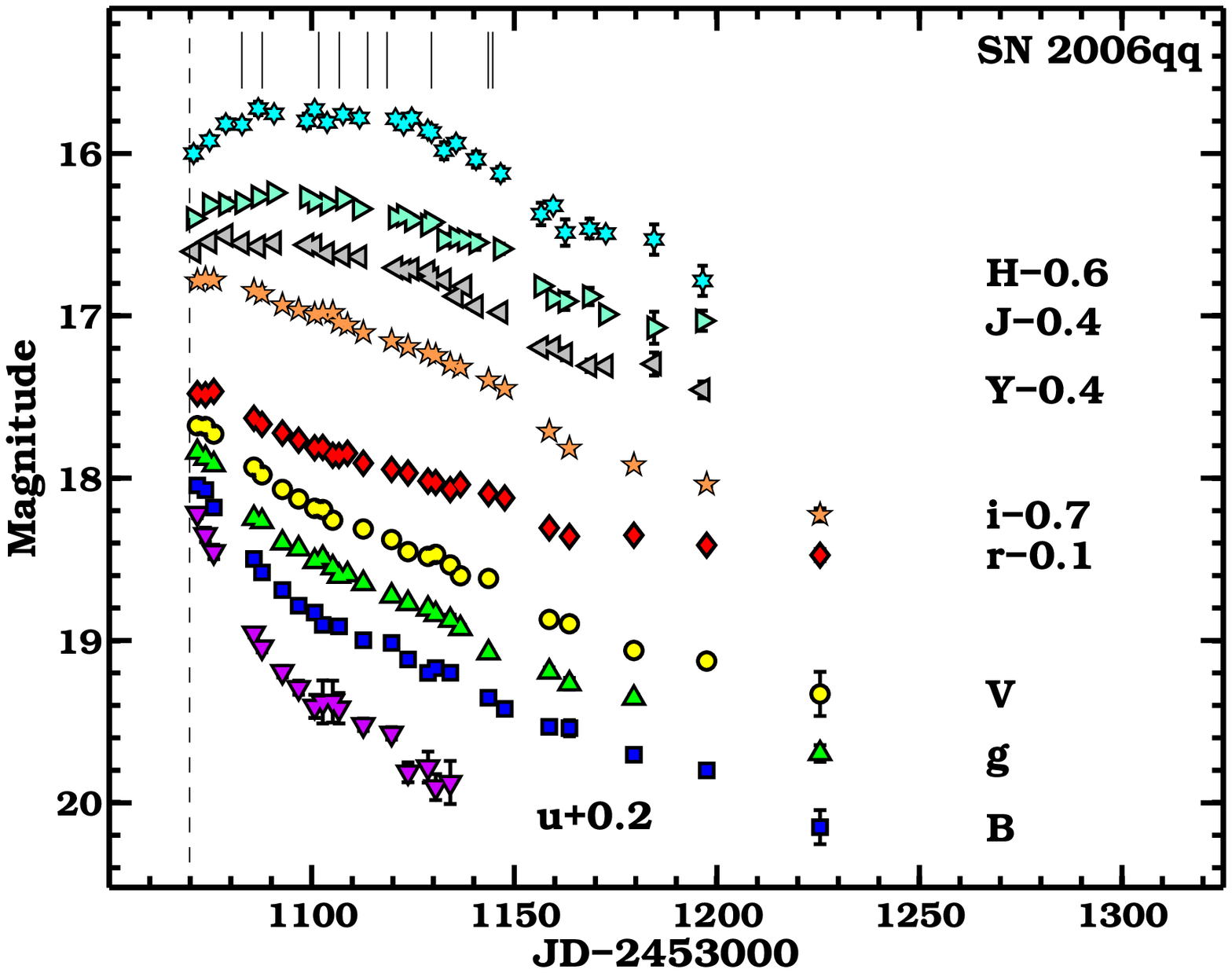} \\
\includegraphics[width=3.6in]{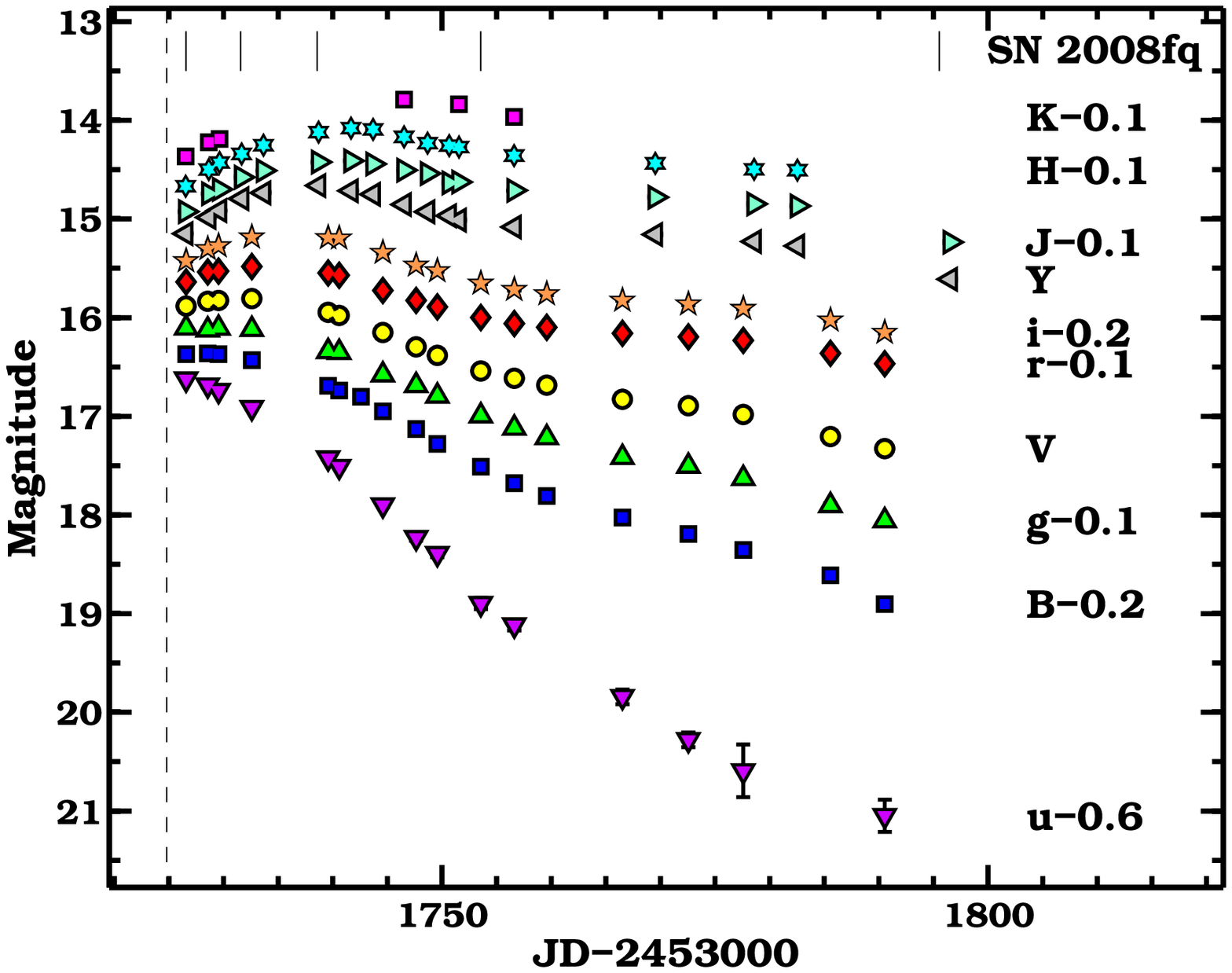} &
\end{array}$
\end{center}
\caption{\label{lc}Optical and NIR light curves of 5  CSP SNe~IIn.
For presentation the light curves have been offset by arbitrary values.  Epochs coinciding with  spectral observations are indicated with vertical lines at the top of each panel. The discovery epoch is indicated by a vertical dashed line.}
\end{figure}

 \clearpage
\begin{figure}[h]
\begin{center}
\includegraphics[width=7.0in]{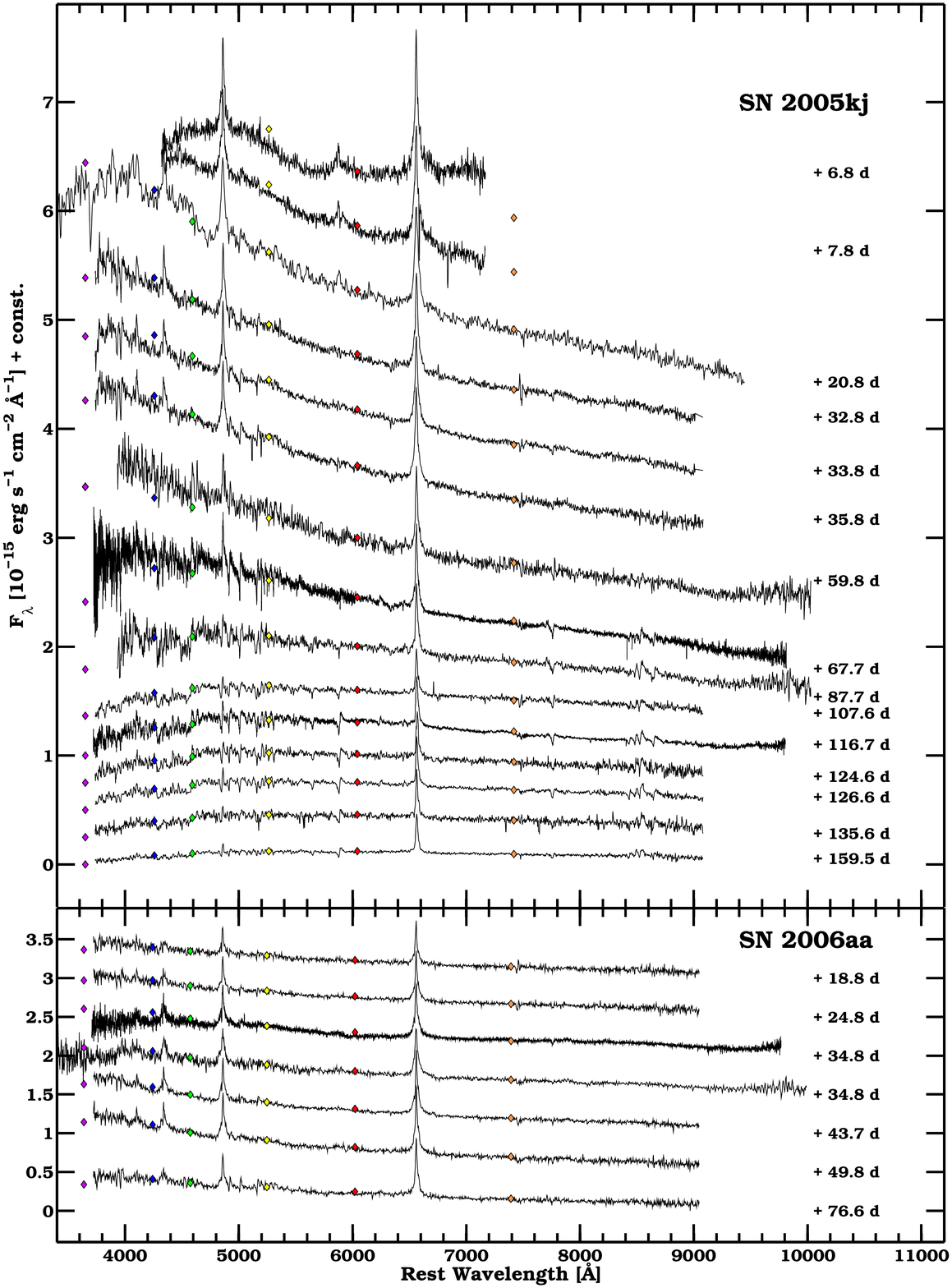} \\
\end{center}
\caption{Spectroscopic sequences of  SNe~2005kj and 2006aa. 
Days relative to the discovery epoch are reported next to each spectrum. Fluxes obtained from interpolated magnitudes at the epoch of each spectrum in each optical filter are marked by diamonds.\label{spec1}}
\end{figure}

  \clearpage
\begin{figure}[h]
\begin{center}
\includegraphics[width=7.0in]{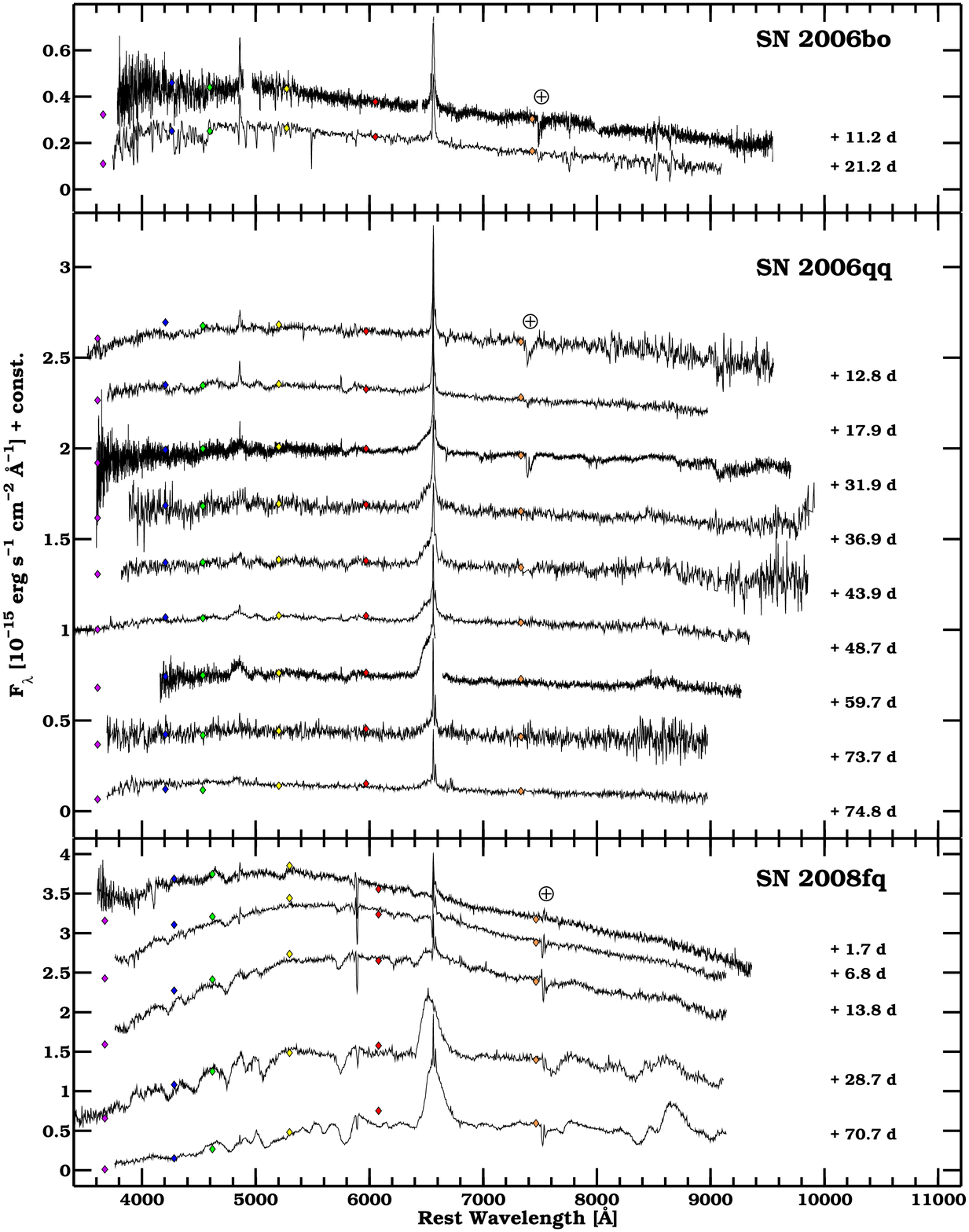} \\
\end{center}
\caption{Spectroscopic  sequences of SNe~2006bo, 2006qq and 2008fq. Days relative to the discovery epoch are reported next to each spectrum. Fluxes obtained from interpolated magnitudes at the epoch of each spectrum in each optical filter are marked by diamonds.\label{spec2}}
\end{figure}

\clearpage
\begin{figure}
 \centering
 $\begin{array}{cc}
\includegraphics[width=13cm]{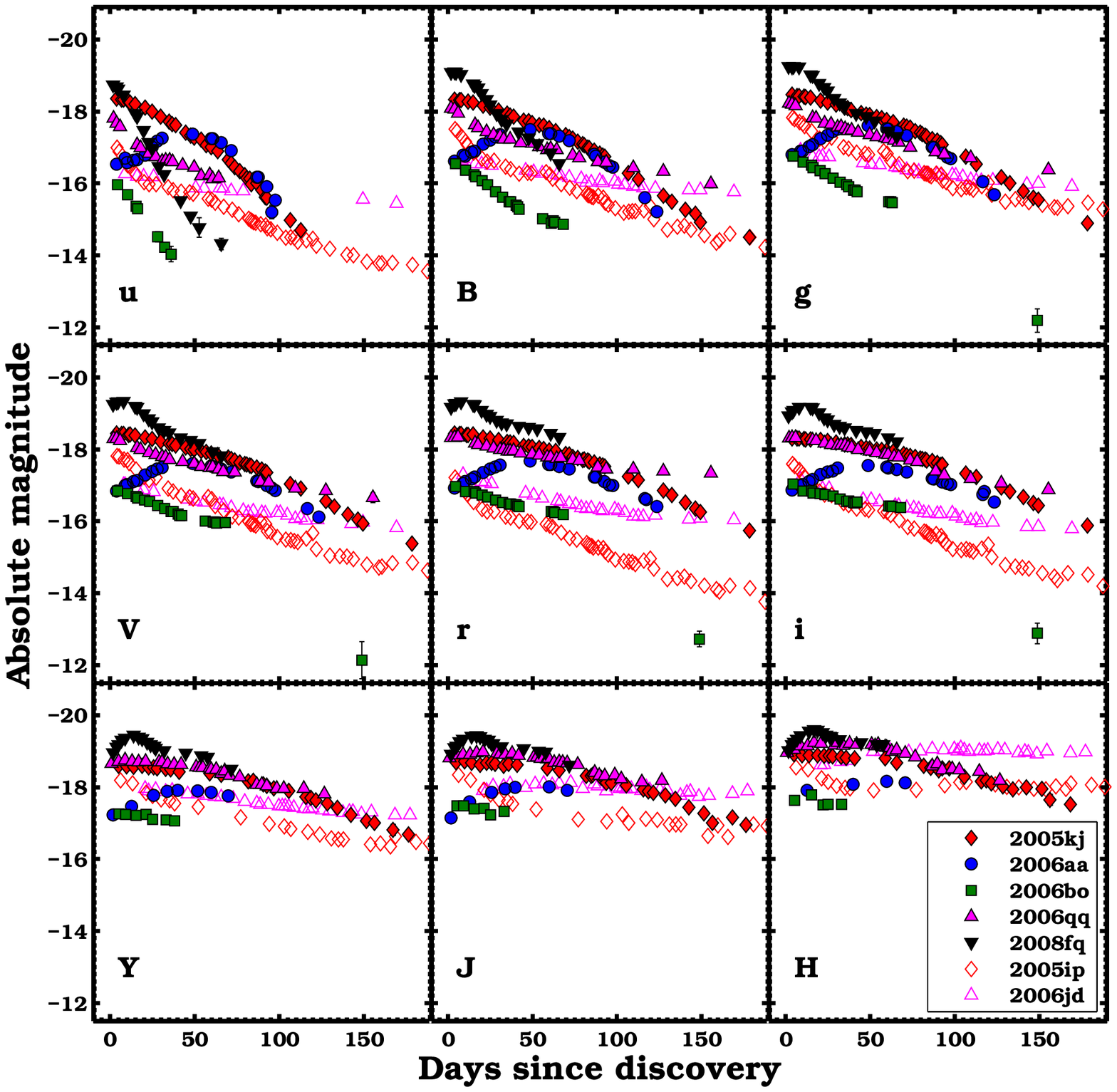} \\
\includegraphics[width=13cm]{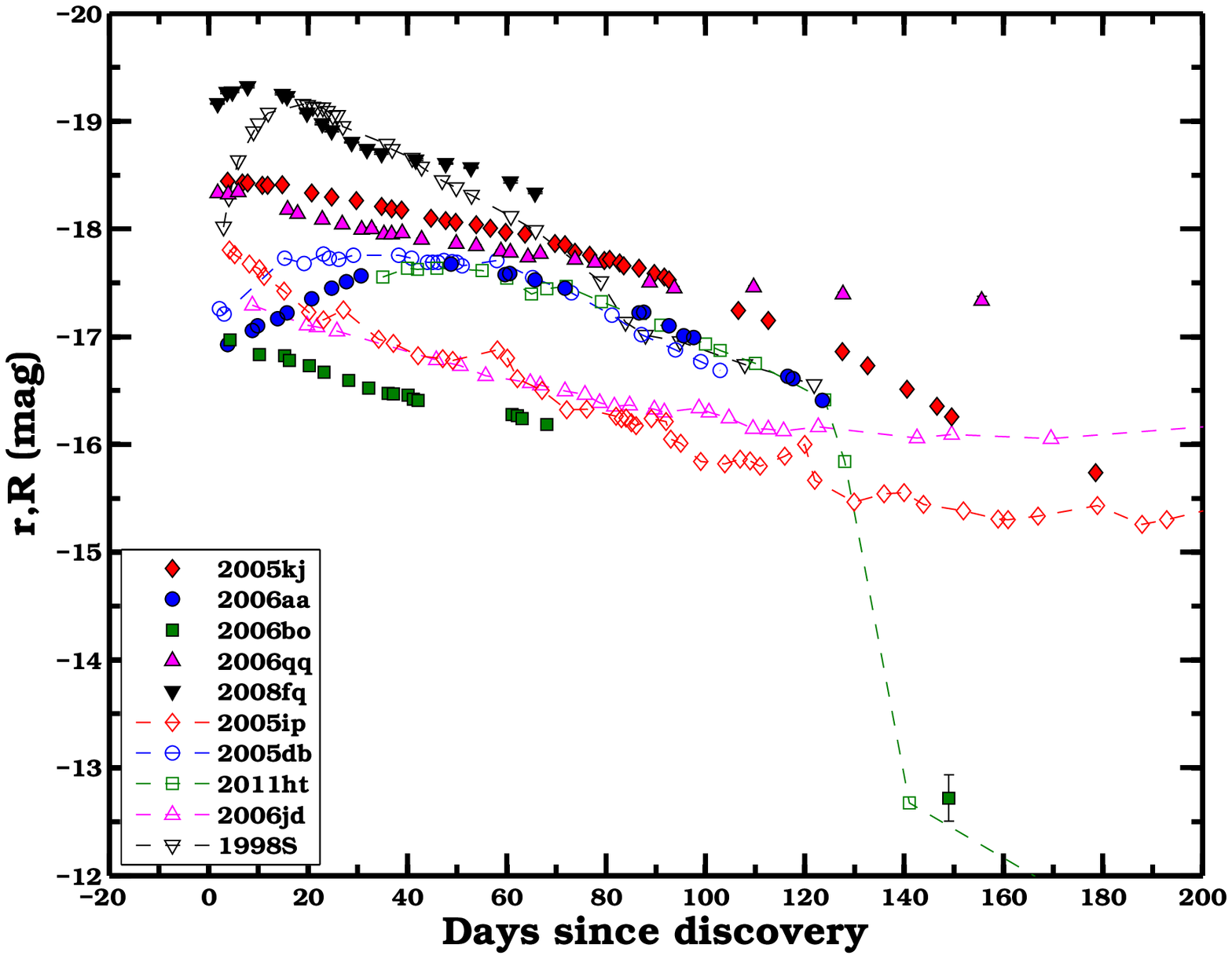}  
\end{array}$
  \caption{\label{absmag}{\em (Top panel)} Optical and NIR absolute magnitude light curves of the full CSP SN~IIn sample plotted vs. days relative to the epoch of discovery. {\em (Bottom panel)} The $r$-band absolute magnitude light curves for the full CSP SN~IIn sample are compared to those of other SNe~IIn in the literature. Photometry and values of  $E(B-V)_{host}$ of SNe~1998S, 2005db, 2005ip, 2006jd,
 and 2011ht are taken from  \citet{liu00}, \citet{kiewe12}, S12 and \citet{94Wlike}. To set the absolute flux scale, luminosity distances from NED were adopted, using WMAP5  cosmological parameters and correcting for peculiar motions.} 
 \end{figure}

\clearpage
\begin{figure}
 \centering
\includegraphics[width=16cm]{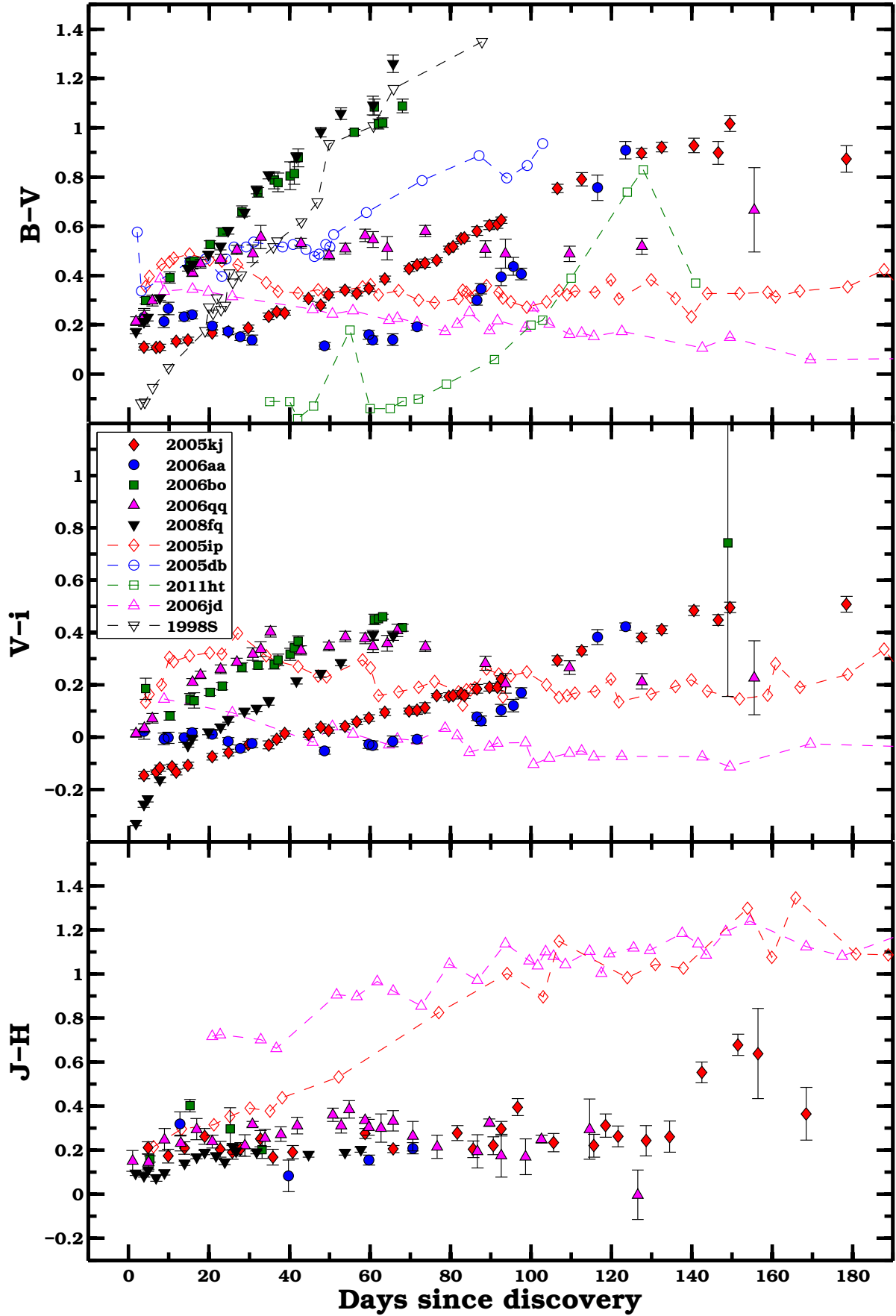}  
  \caption{\label{color}Intrinsic optical and NIR color-curve evolution of the CSP SNe~IIn
  sample. 
  The top panel $B-V$ color curves 
  also include several objects from the literature (see caption of Fig.~\ref{absmag} for references).}
 \end{figure}

\clearpage
\begin{figure}
 \centering
\includegraphics[width=16cm]{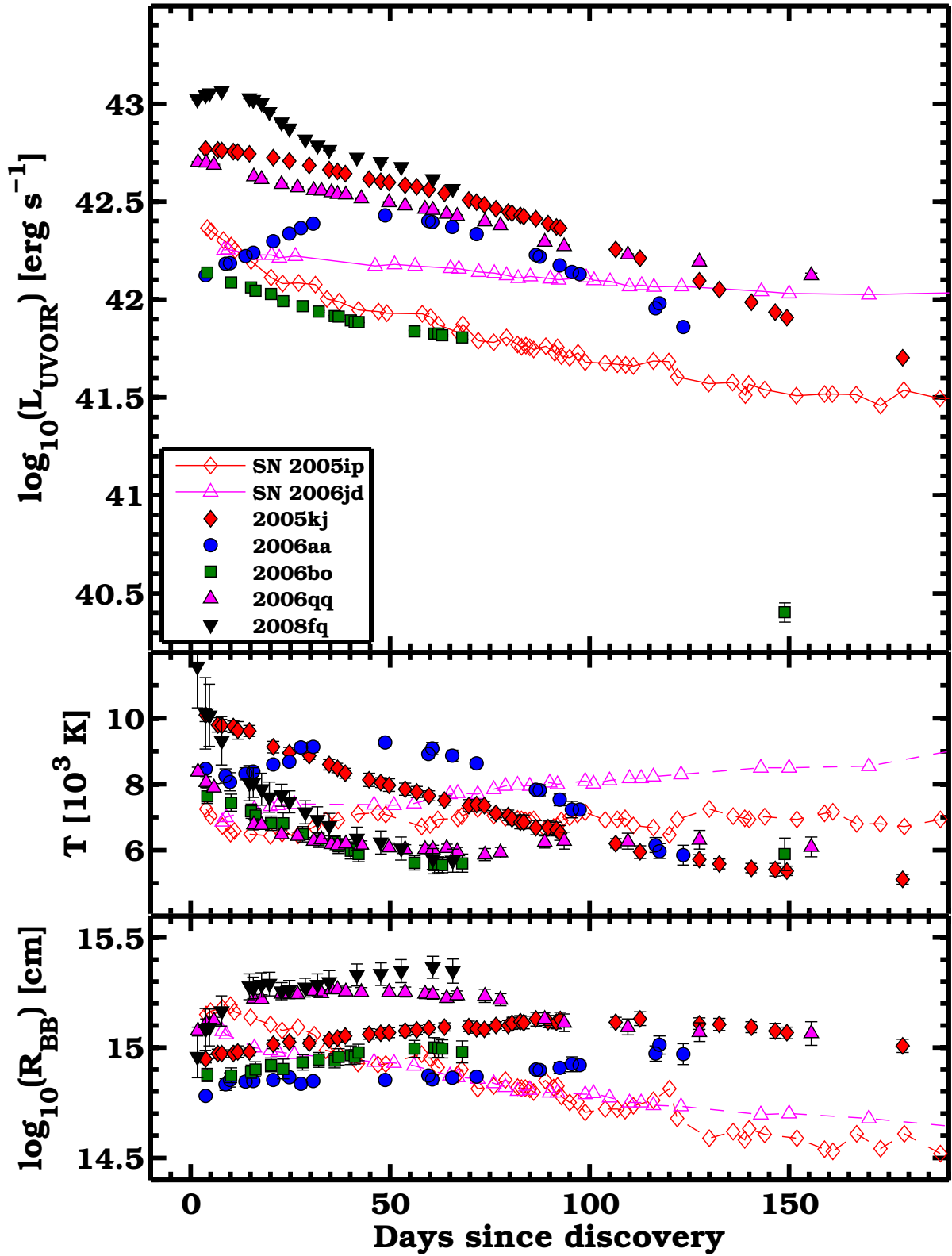}  
  \caption{\label{bolo}{\em (Top panel)} UVOIR light curves for the full CSP SN~IIn 
  sample. Theoretical light curve tails powered by radioactive decay for different amounts of $^{56}$Ni are included as dashed lines. {\em (Middle panel)} Temperature evolution
  derived from BB fits to the time-series of SEDs. 
  For SNe~2005ip and 2006jd we plot the temperature associated with the optical emission (see S12).
{\em (Bottom panel)} Radius evolution was computed assuming BB emission. For SNe~2005ip and 2006jd, we plot the radius associated with the optical emission (see S12).}
 \end{figure}

\clearpage
\begin{figure}[h]
\begin{center}$
\begin{array}{cc}
\includegraphics[height=8.5cm]{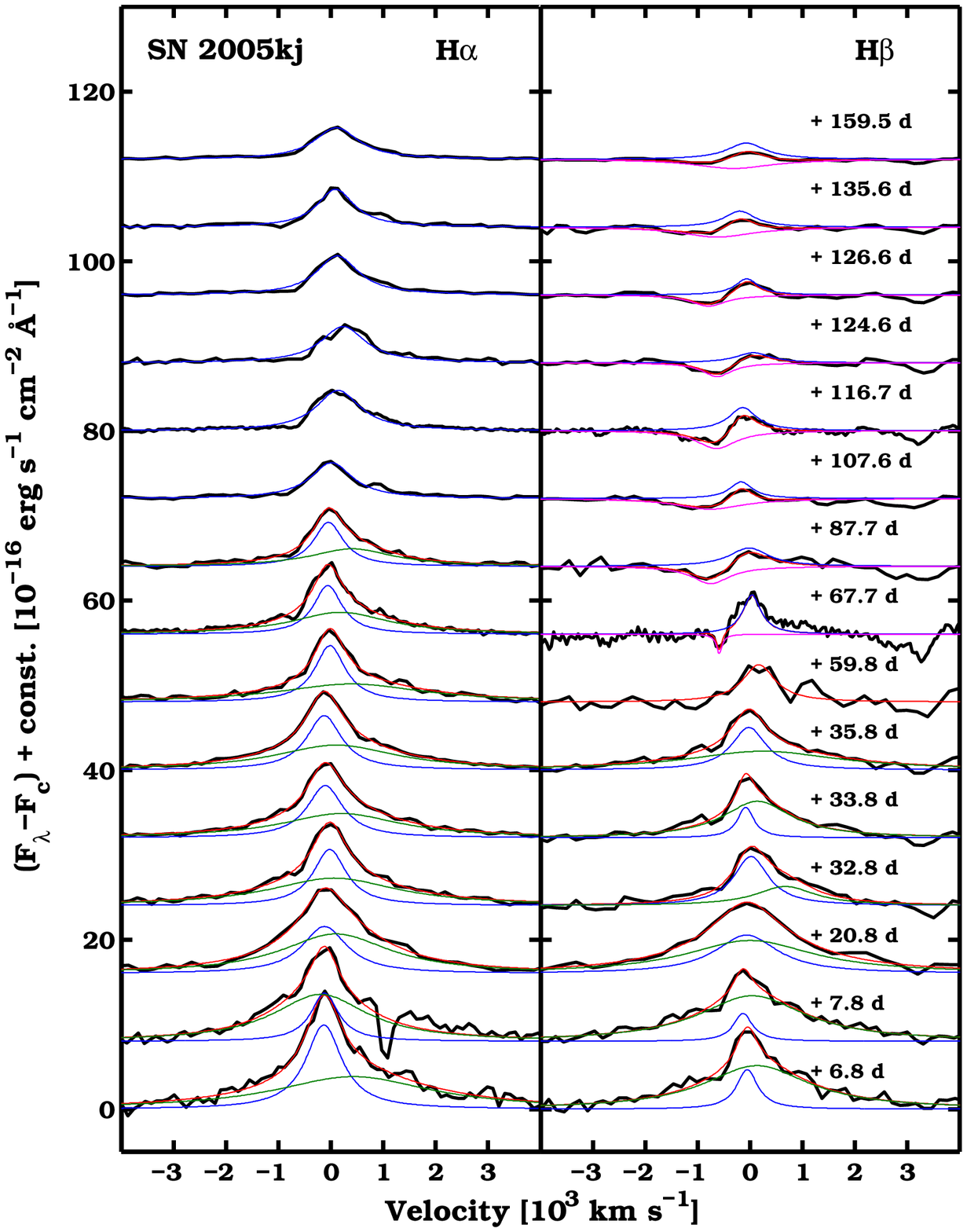} &
\includegraphics[height=8.5cm]{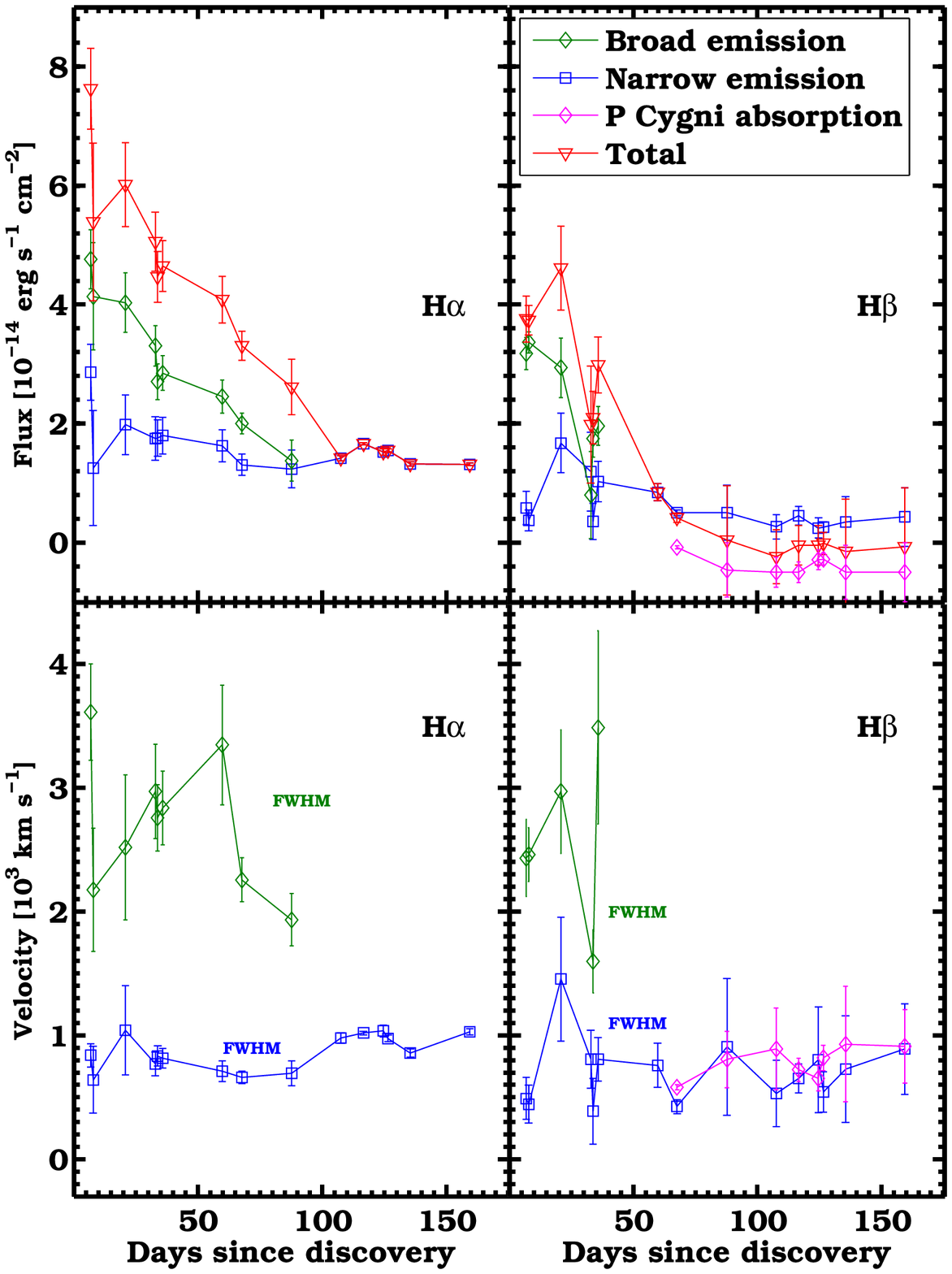} \\
\end{array}$
\end{center}
\caption{\label{ha_5kj}{\em (Left panel)} H$\alpha$ and H$\beta$ profiles of SN~2005kj after low-order polynomial continuum subtraction and reddening correction. A combination of Lorentzians has been used to fit the profiles. {\em (Right panel)} 
Fluxes and velocities for the different Lorentzian components.}
\end{figure}

\begin{figure}[h]
\begin{center}$
\begin{array}{cc}
\includegraphics[height=8.5cm]{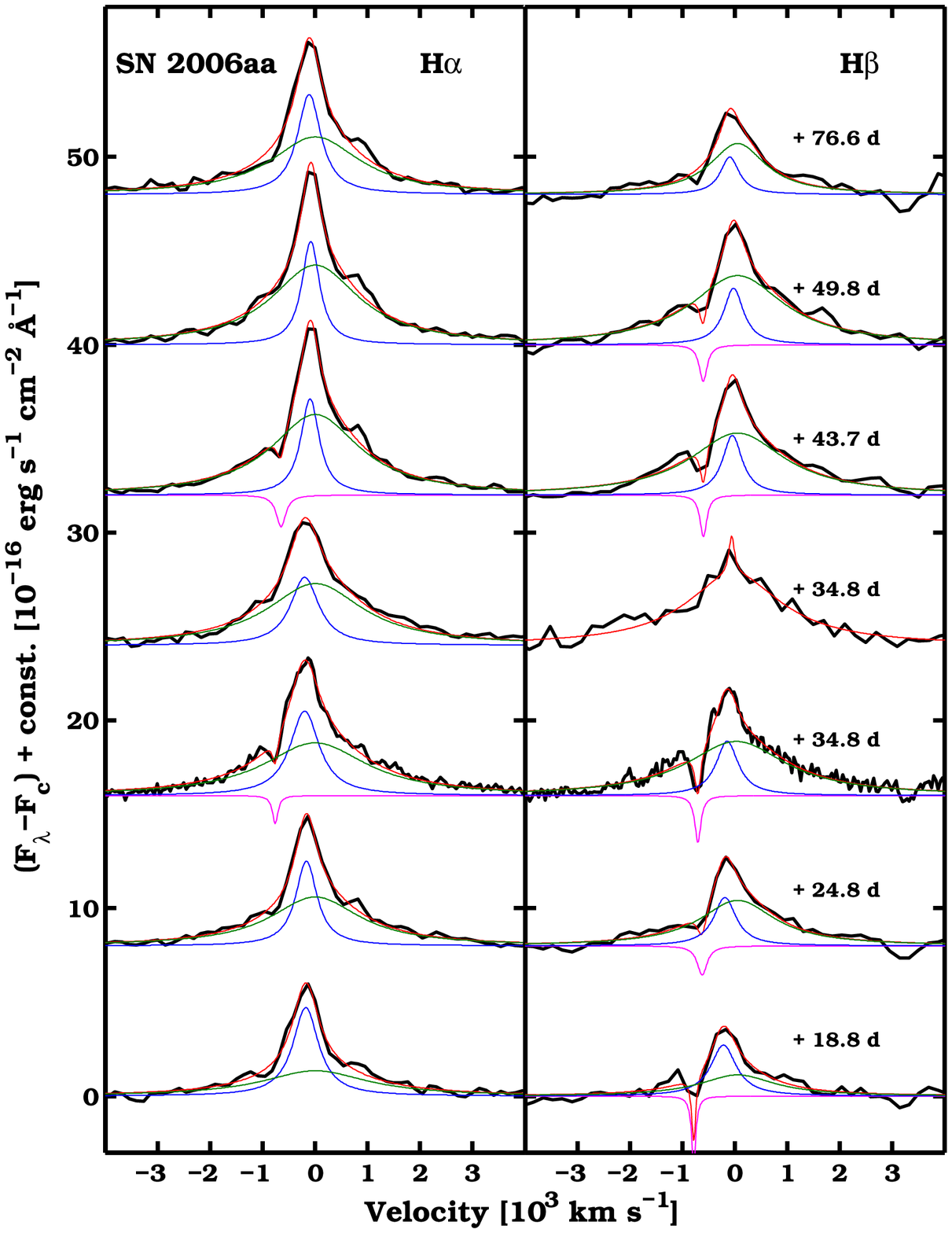} &
\includegraphics[height=8.5cm]{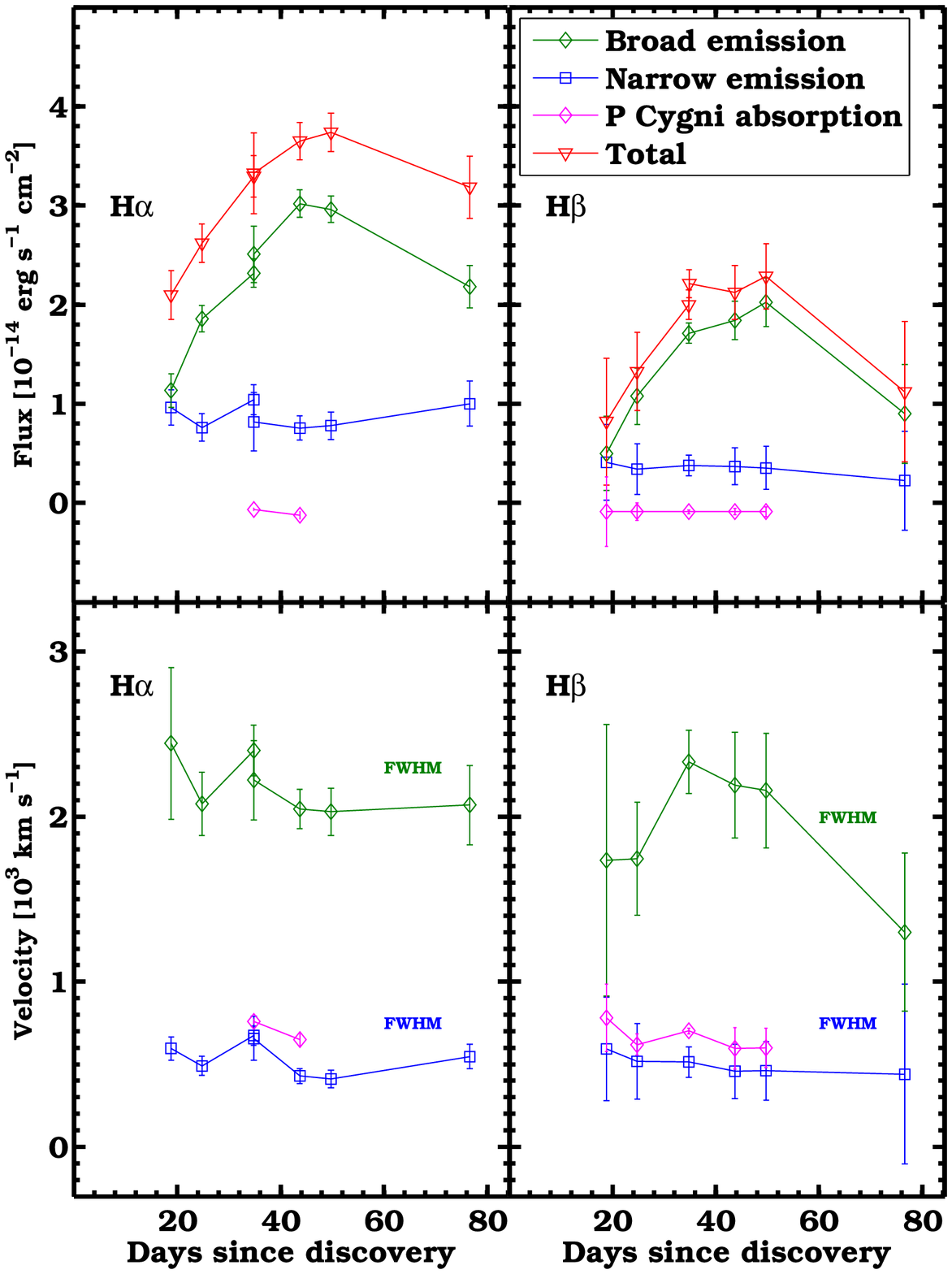} \\
\end{array}$
\end{center}
\caption{\label{ha_6aa}{\em (Left panel)} H$\alpha$ and H$\beta$ profiles of SN~2006aa after low-order polynomial continuum subtraction and reddening correction. A combination of Lorentzians has been used to fit the profiles. {\em (Right panel)} Fluxes and velocities for the different Lorentzian components.}
\end{figure}

\clearpage
\begin{figure}[h]
\begin{center}$
\begin{array}{cc}
\includegraphics[height=8.5cm]{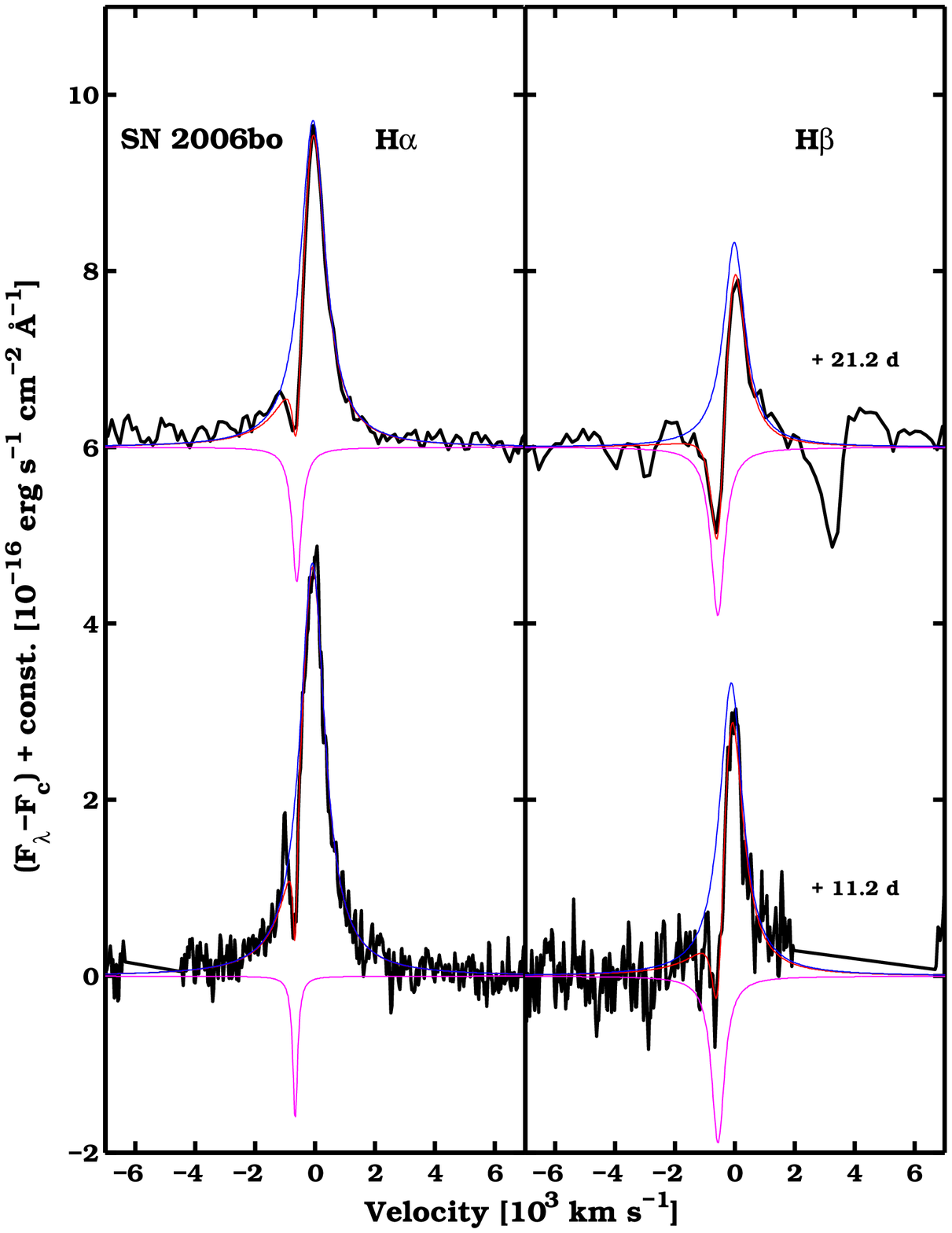} &
\includegraphics[height=8.5cm]{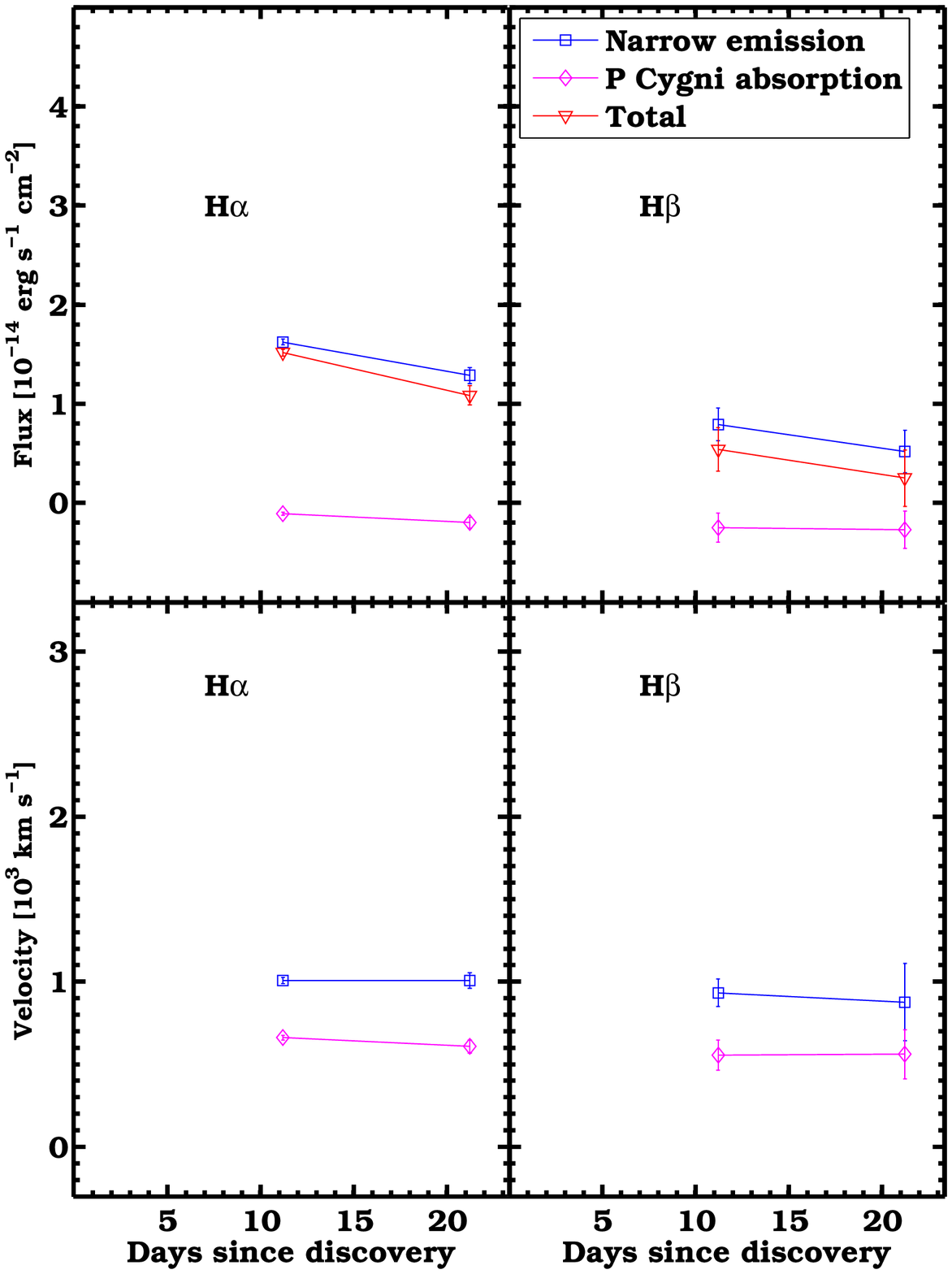} \\
\end{array}$
\end{center}
\caption{\label{ha_6bo}{\em (Left panel)}  H$\alpha$ and H$\beta$ profiles of SN~2006bo after low-order polynomial continuum subtraction and reddening correction. A combination of Lorentzians has been used to fit the profiles. {\em (Right panel)}  Fluxes and velocities for the different Lorentzian components.}
\end{figure}  
  
\begin{figure}[h]
\begin{center}$
\begin{array}{cc}
\includegraphics[height=8.5cm]{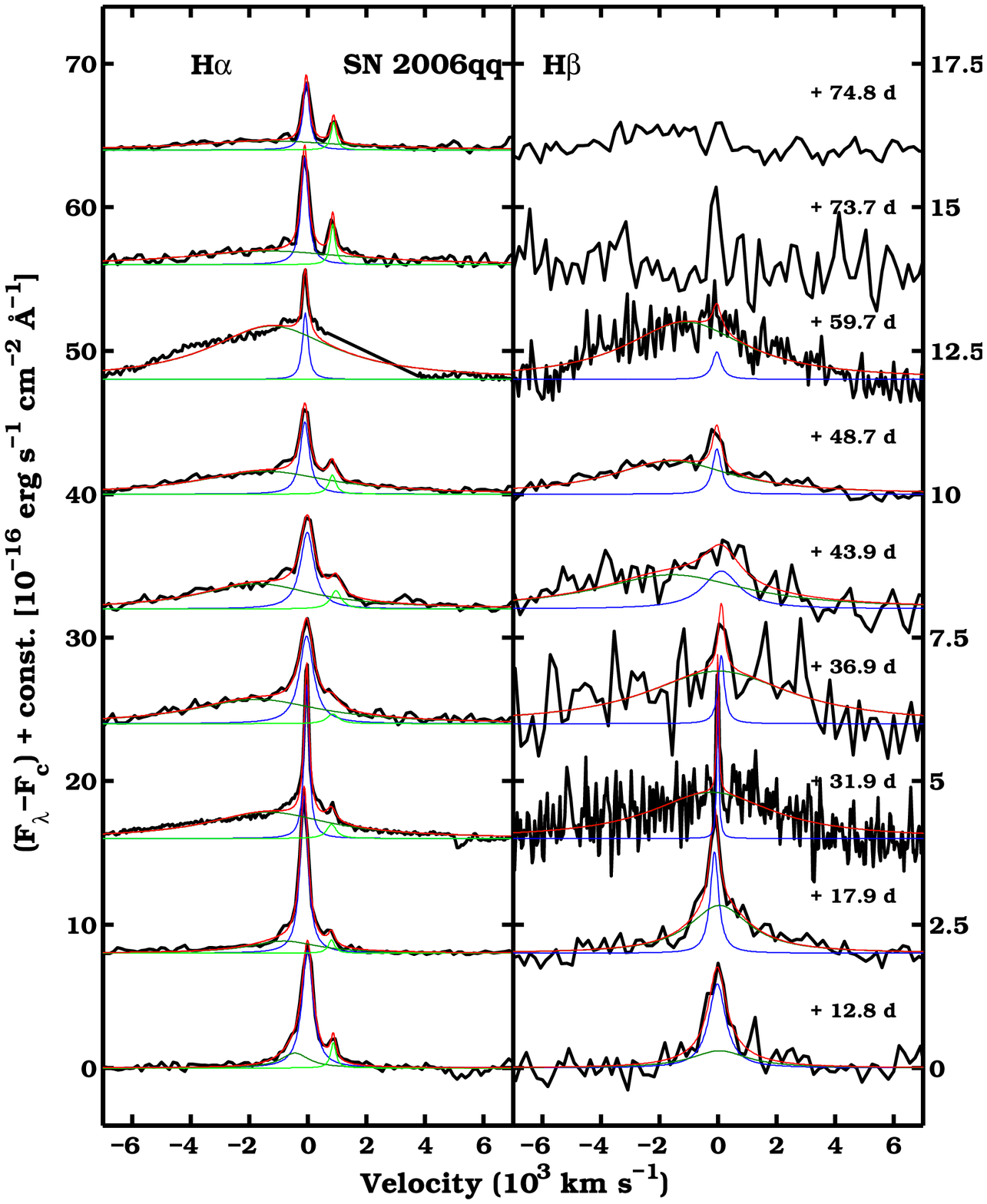} &
\includegraphics[height=8.5cm]{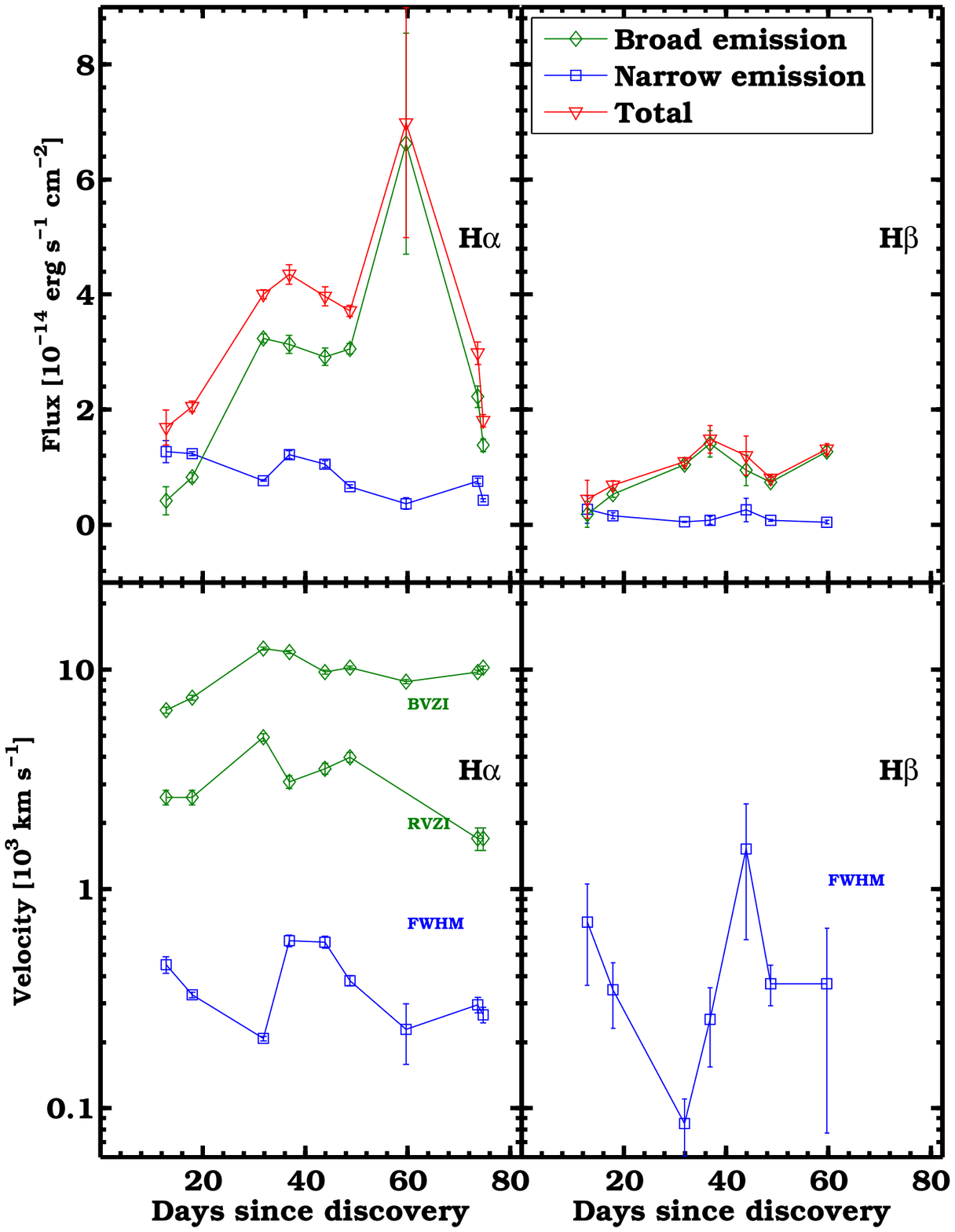} \\
\end{array}$
\end{center}
\caption{\label{ha_6qq}{\em (Left panel)}  H$\alpha$ and H$\beta$ of SN~2006qq profiles after low-order polynomial continuum subtraction and reddening correction. A combination of Lorentzians has been used to fit the profiles. {\em (Right panel})  Fluxes and velocities for the different Lorentzian components.}
\end{figure}

\clearpage
\begin{figure}[h]
\begin{center}$
\begin{array}{cc}
\includegraphics[height=8.5cm]{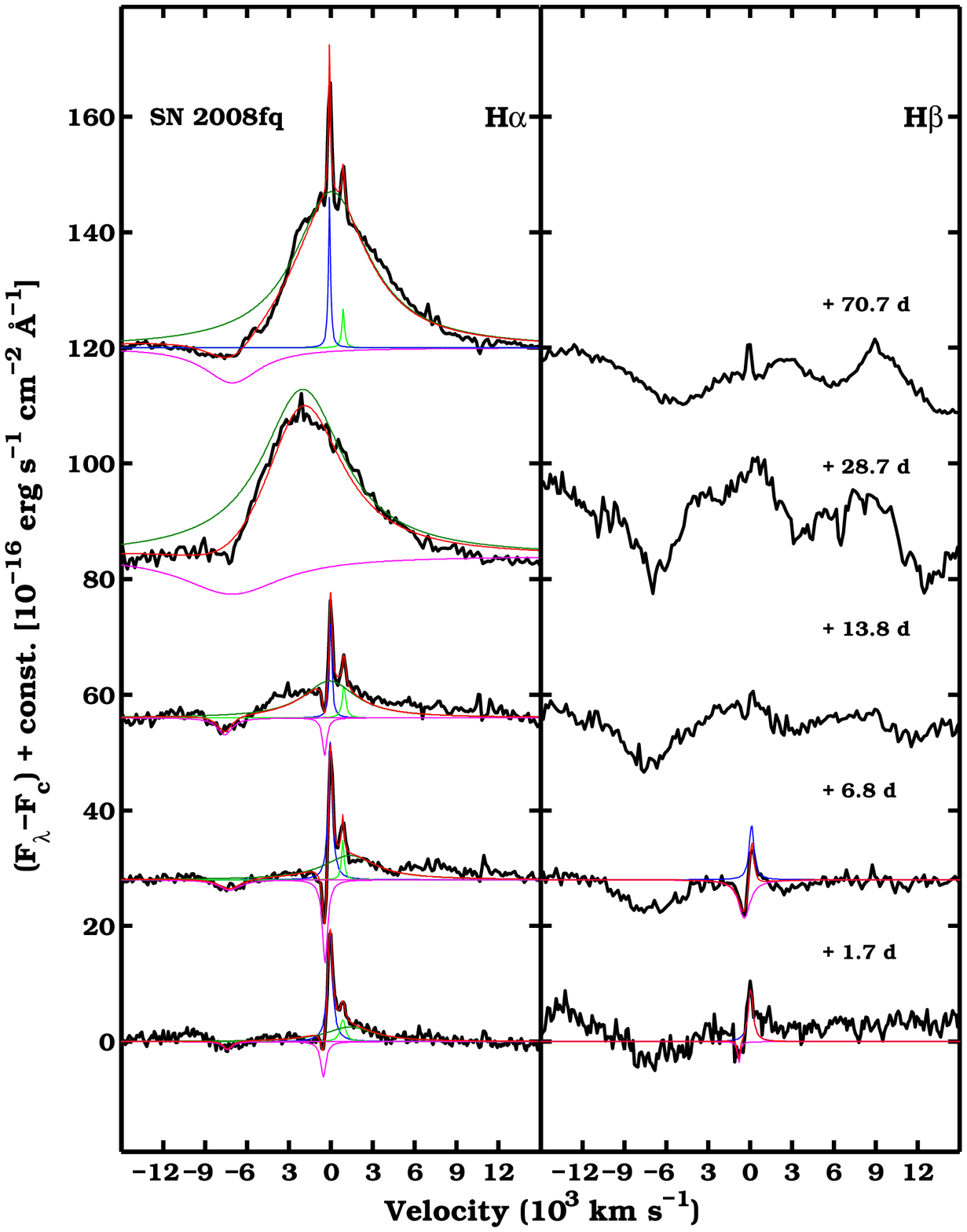} &
\includegraphics[height=8.5cm]{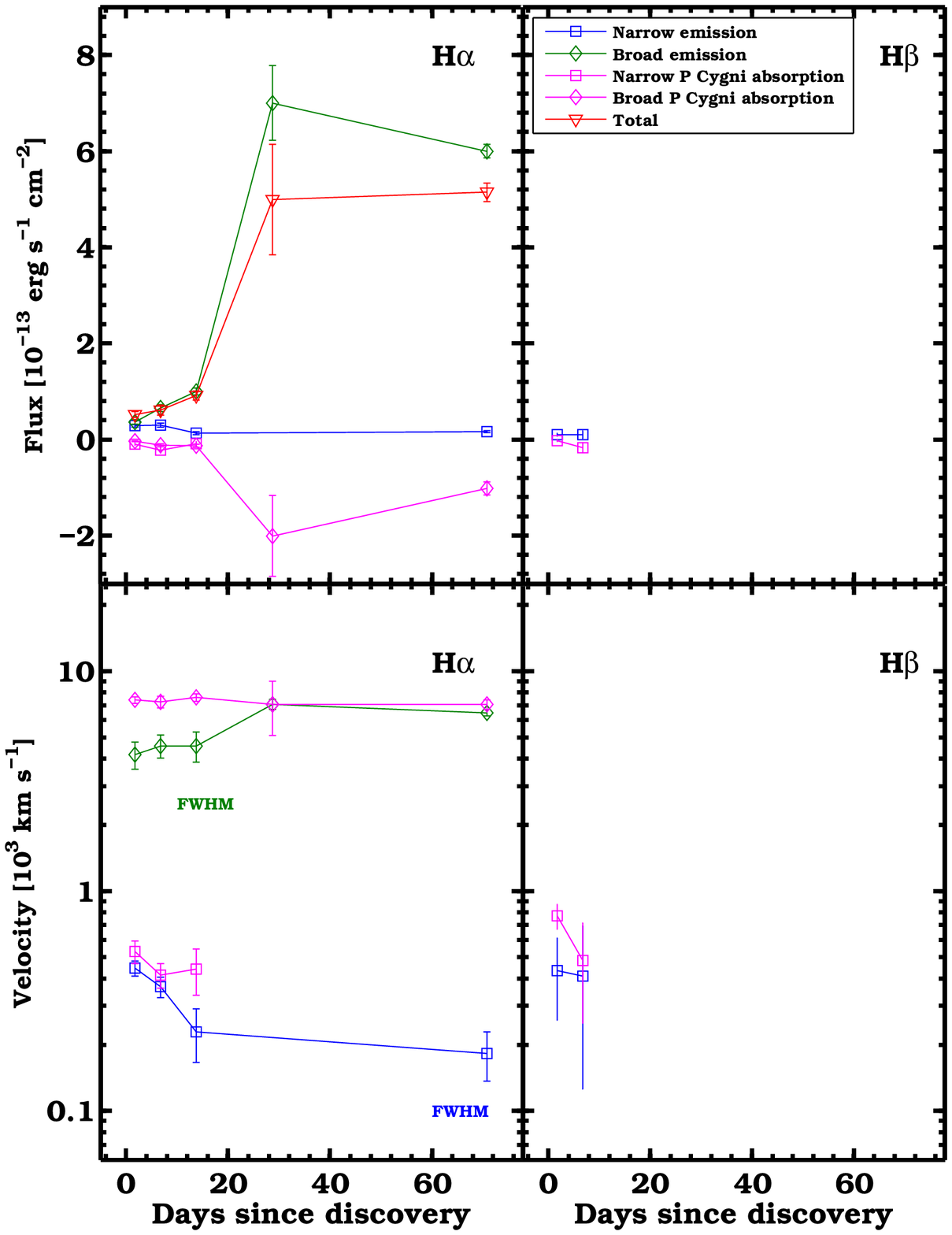} \\
\end{array}$
\end{center}
\caption{\label{ha_8fq}{\em (Left panel)} H$\alpha$ and H$\beta$ profiles of SN~2008fq after low-order polynomial continuum subtraction and reddening correction. A combination of Lorentzians has been used to fit the profiles. {\em (Right panel)}  Fluxes and velocities for the different Lorentzian components.}
\end{figure}

\begin{figure}
 \centering
\includegraphics[width=16cm]{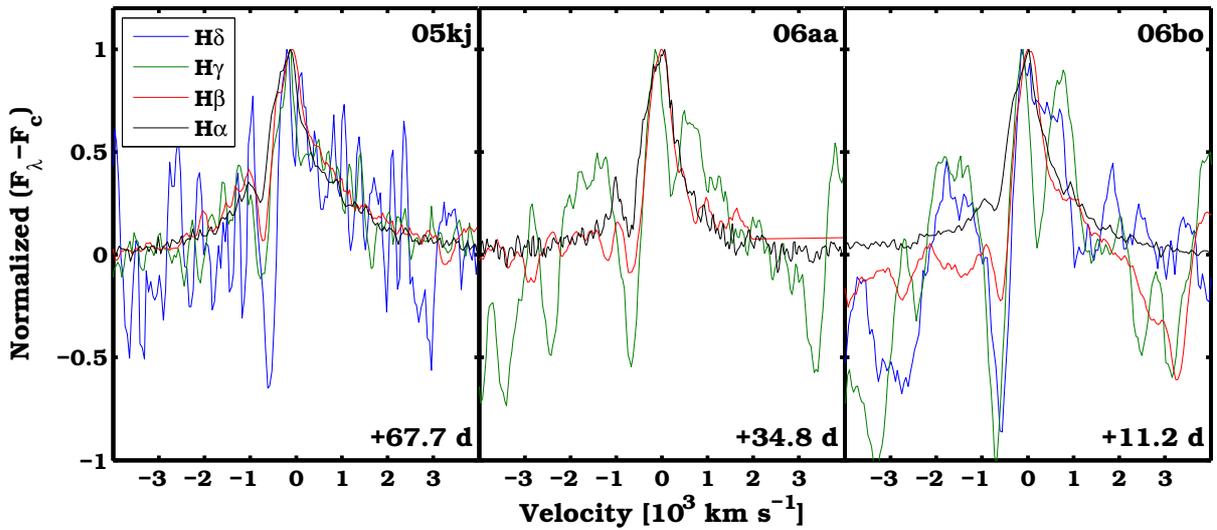}  
  \caption{\label{wingsH}Over-imposed Balmer line profiles after continuum subtraction and peak normalization for
  SNe~2005kj, 2006aa, and 2006bo.}
 \end{figure}

\clearpage
\begin{figure}[h]
\begin{center}$
\begin{array}{cc}
\includegraphics[height=5.9cm]{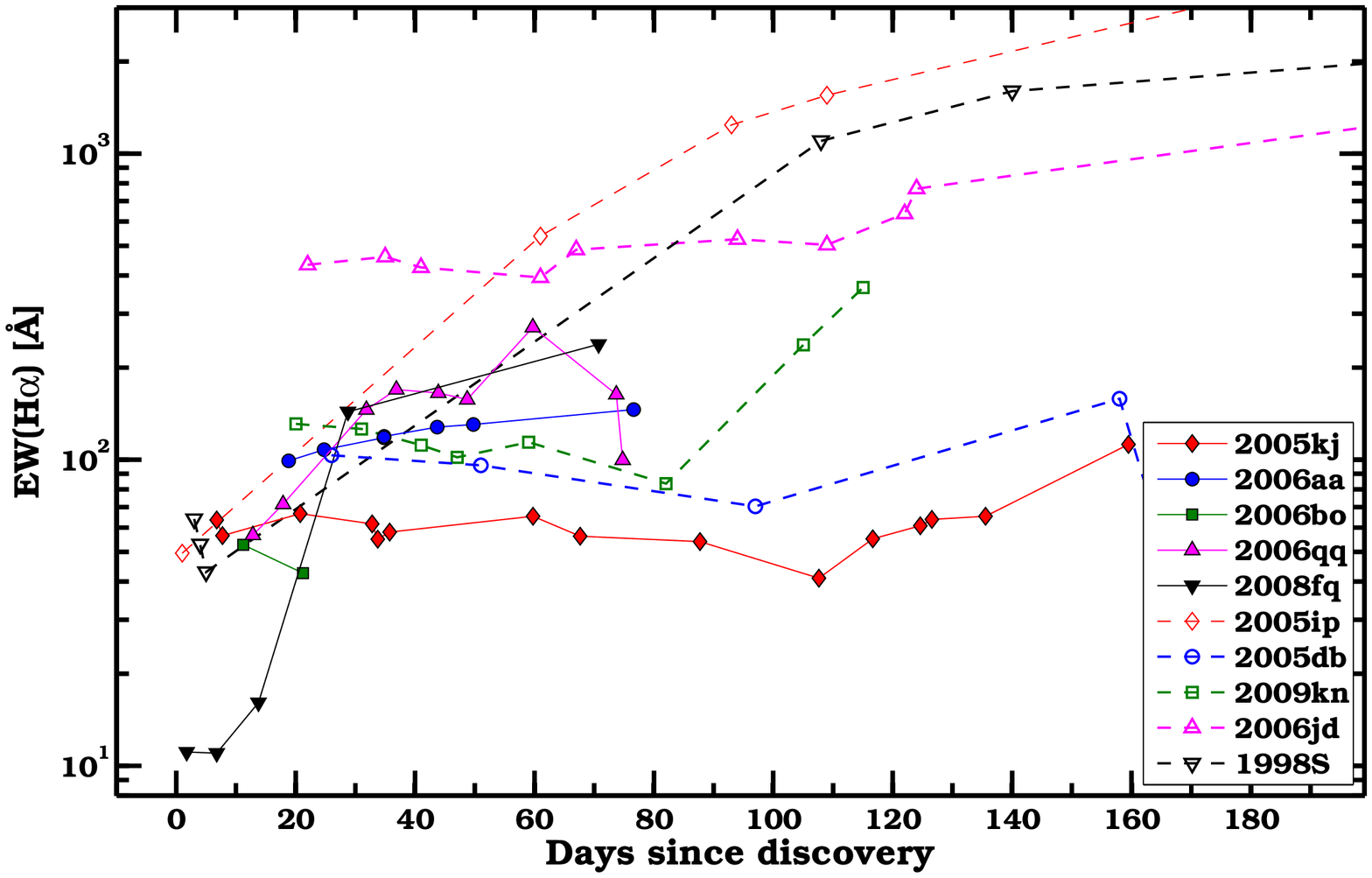} &
\includegraphics[height=5.9cm]{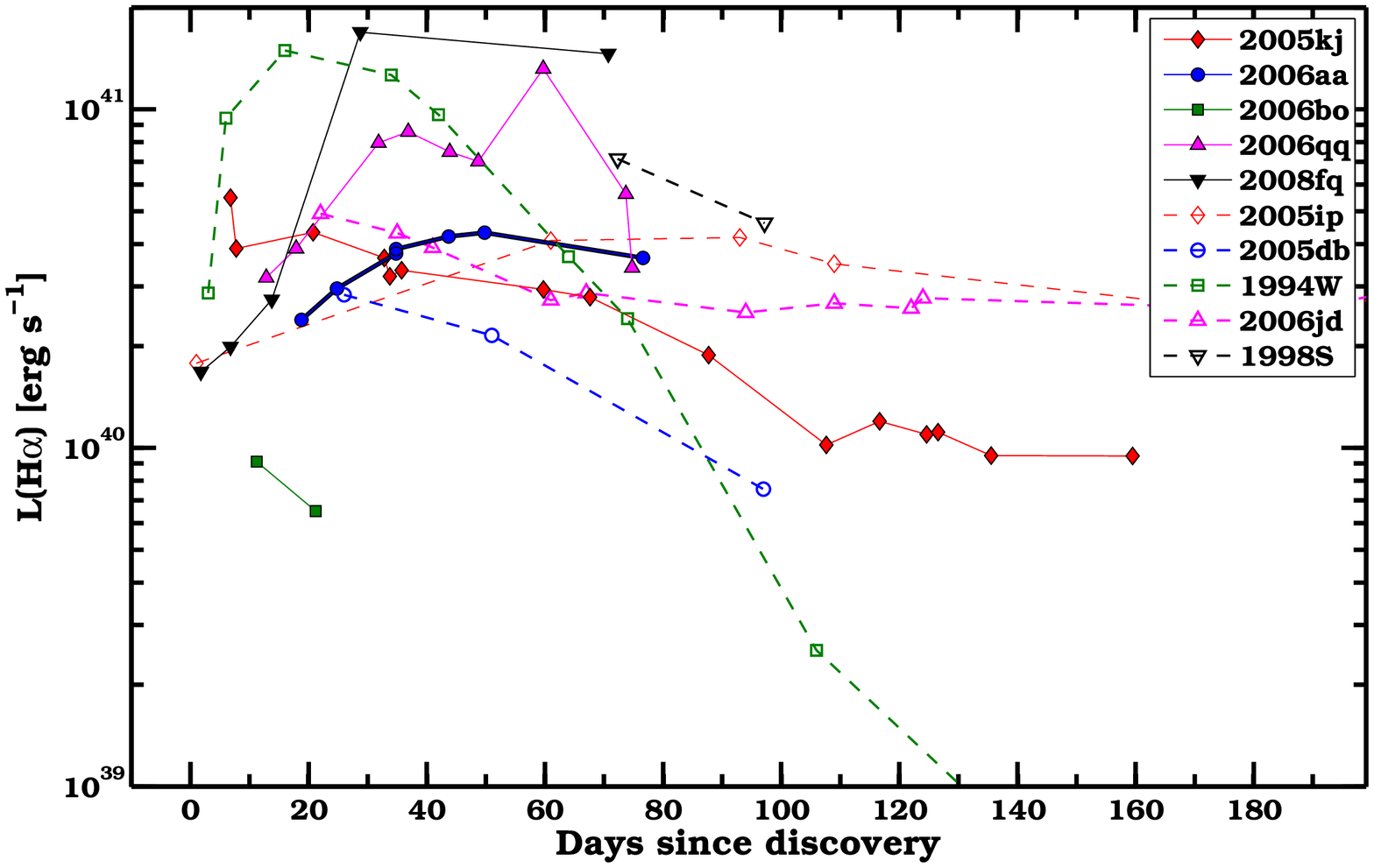} \\
\end{array}$
\end{center}
\caption{\label{ew}  {\em (Left panel)} H$\alpha$ EW time evolution for the full CSP SN~IIn sample and several objects  in the 
literature including:  SNe~1998S \citep{leonard00}, 
  2005db \citep{kiewe12}, and 2009kn \citep{kankare12}.
  {\em (Right panel)}  H$\alpha$ luminosities for the full CSP SN~IIn sample compared to those of some objects in the literature.
Data for SNe 2005db and 1998S are from the spectra published by \citet{kiewe12}, and \citet{fassia01}. The H$\alpha$ fluxes of SN~1994W were published by \citet{chugai04}.}
\end{figure} 

\begin{figure}[h]
\begin{center}$
\begin{array}{cc}
\includegraphics[height=6.8cm]{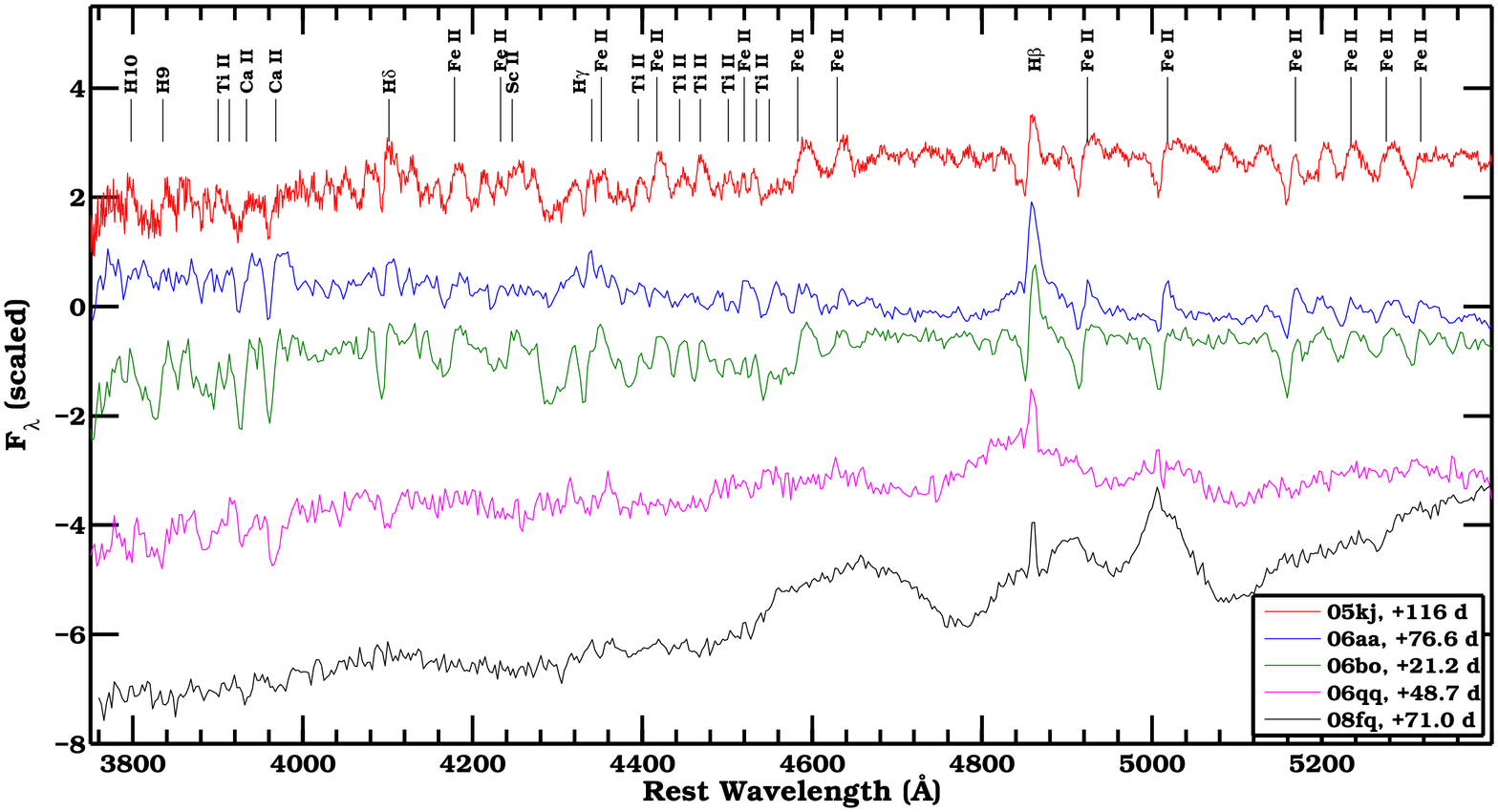} &
\includegraphics[height=6.8cm]{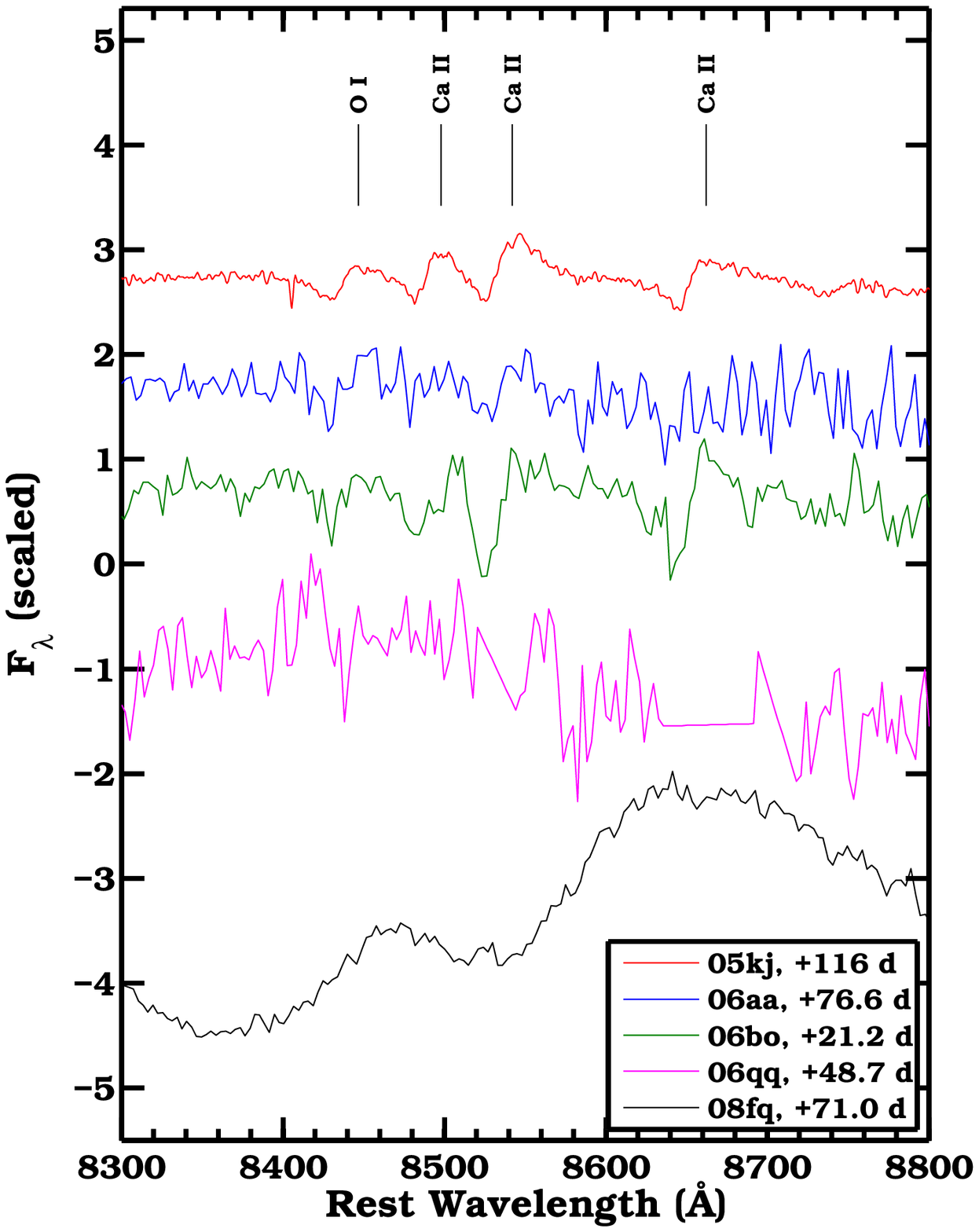}\\
\end{array}$ 
\end{center}
\caption{\label{FeCaII}{\em (Left panel)} Spectral comparison of the five new CSP SNe~IIn between wavelength 3800 and 5300~\AA. Line identification is based on  \citet{kankare12}. 
In addition to Balmer lines, each spectrum is  dominated by \ion{Fe}{ii} and \ion{Ti}{ii} features. {\em (Right panel)} Spectral comparison of the same objects in the wavelength region of the \ion{Ca}{ii} NIR triplet.}
\end{figure}

\clearpage
\begin{figure}
 \centering
\includegraphics[width=15cm]{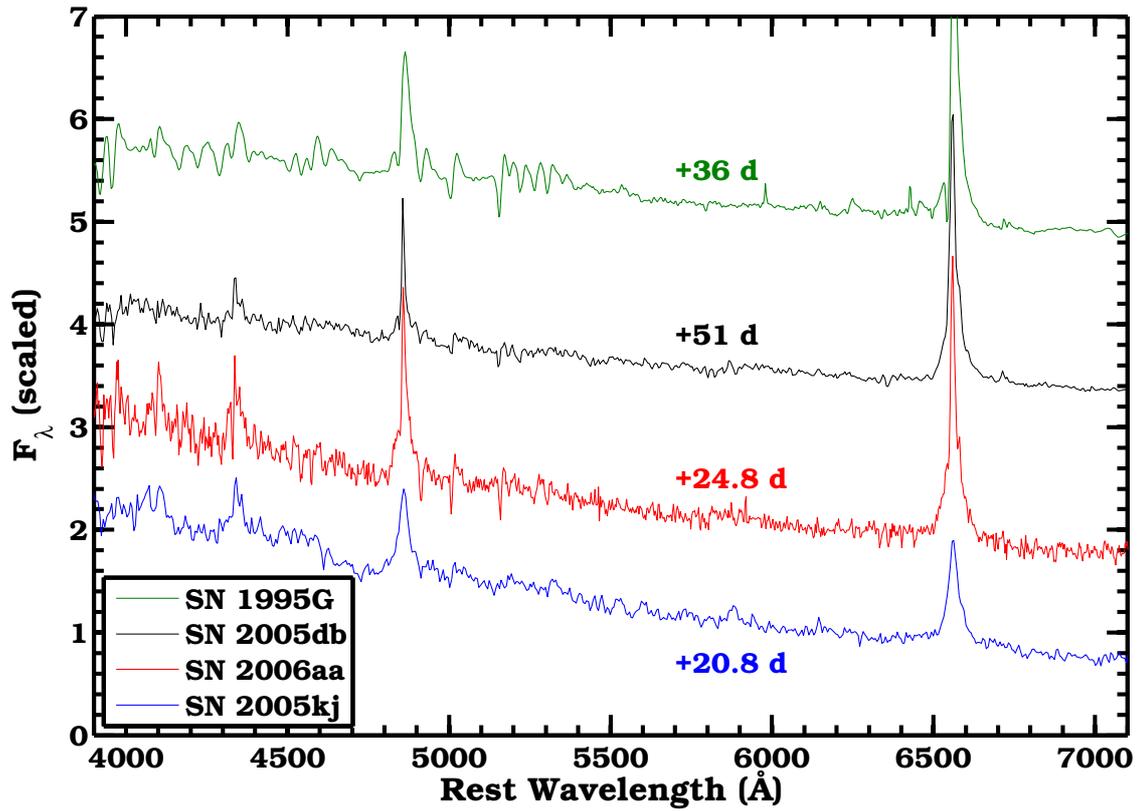}  
  \caption{\label{compnormalIIn}Spectral comparison between SNe~2005kj and 2006aa to SNe~2005db and 1995G.}
 \end{figure}

\begin{figure}
 \centering
\includegraphics[width=15cm]{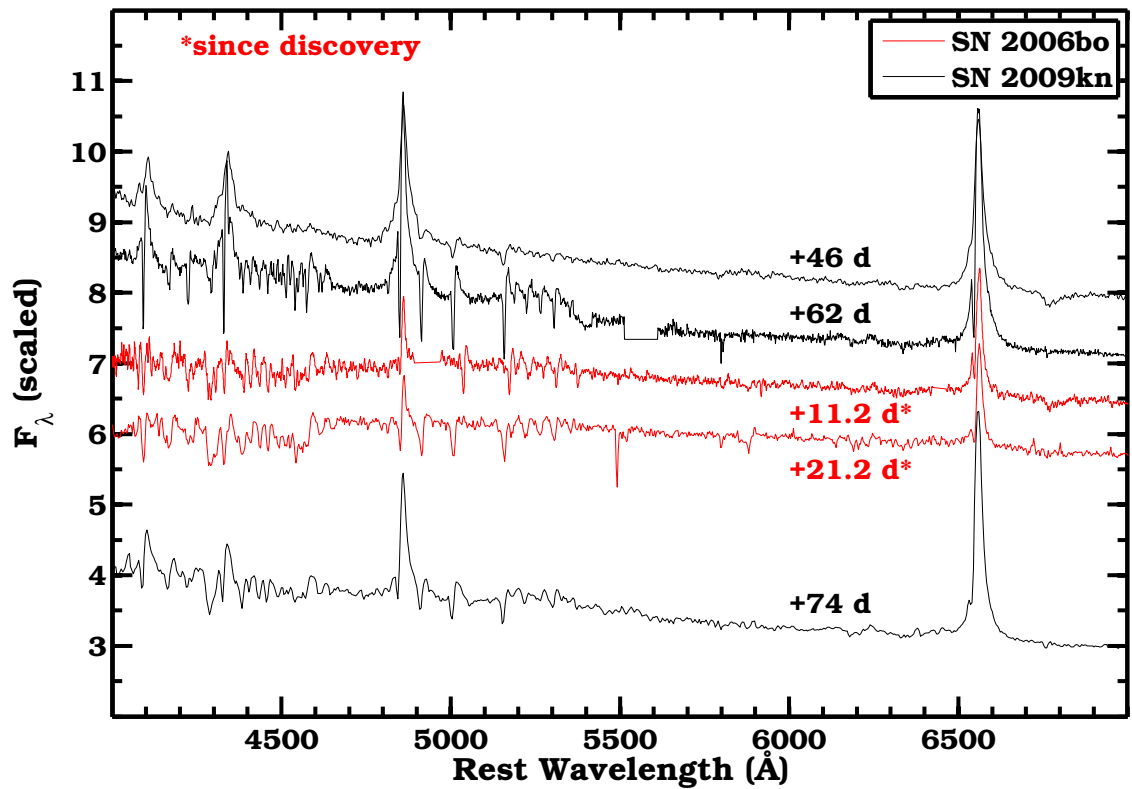}  
  \caption{\label{comp06bo}Spectral comparison between SN~2006bo and the 1994W-like SN~2009kn \citep{kankare12}.}
 \end{figure}

\clearpage
\begin{figure}
 \centering
\includegraphics[width=16cm]{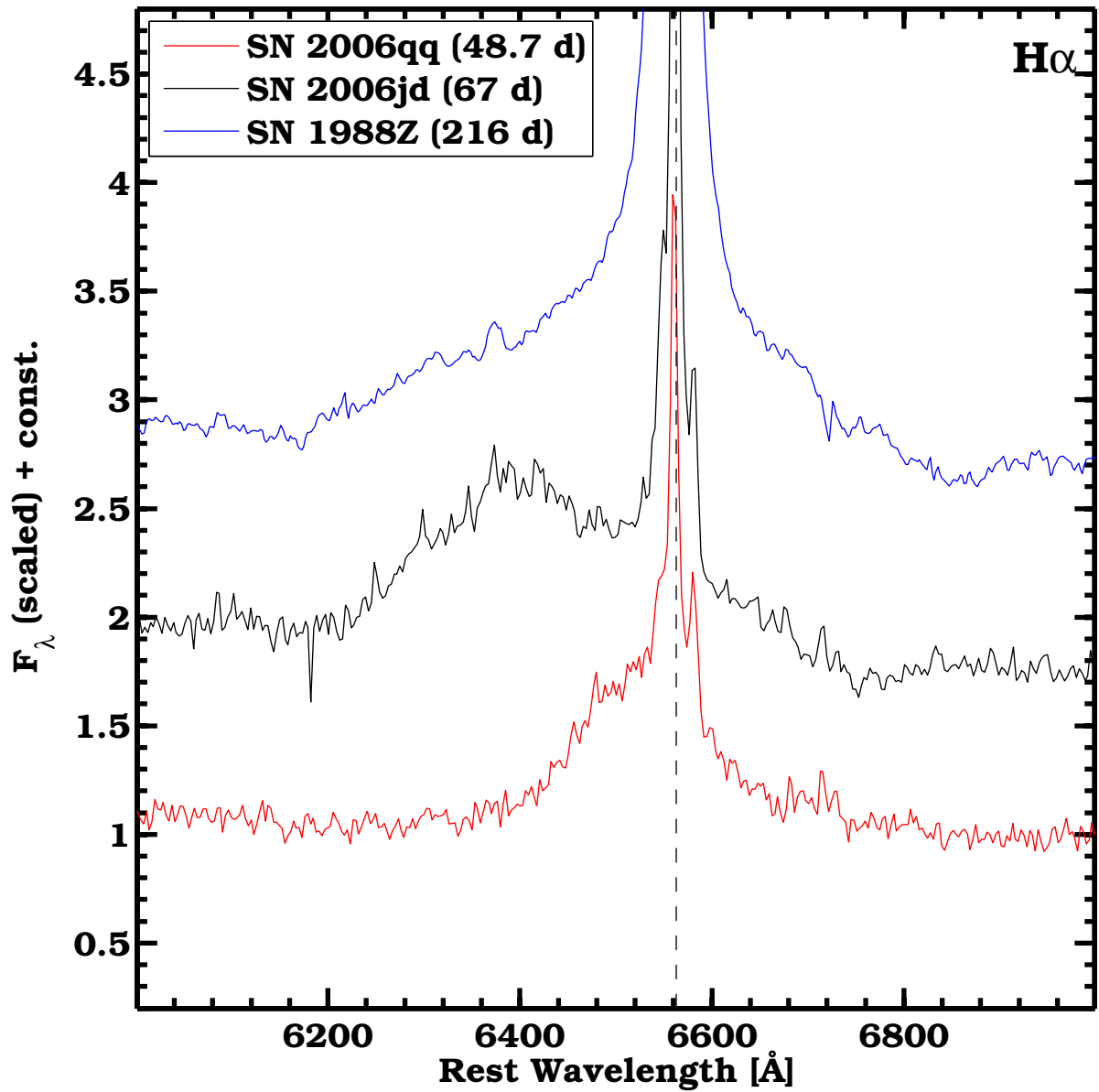}  
  \caption{\label{asym}H$\alpha$ comparison between SNe~1988Z, 2006jd, and 2006qq.  Each object exhibits preferential emission in the blue wing.}
 \end{figure}
\clearpage

\begin{figure}
 \centering
\includegraphics[width=16cm]{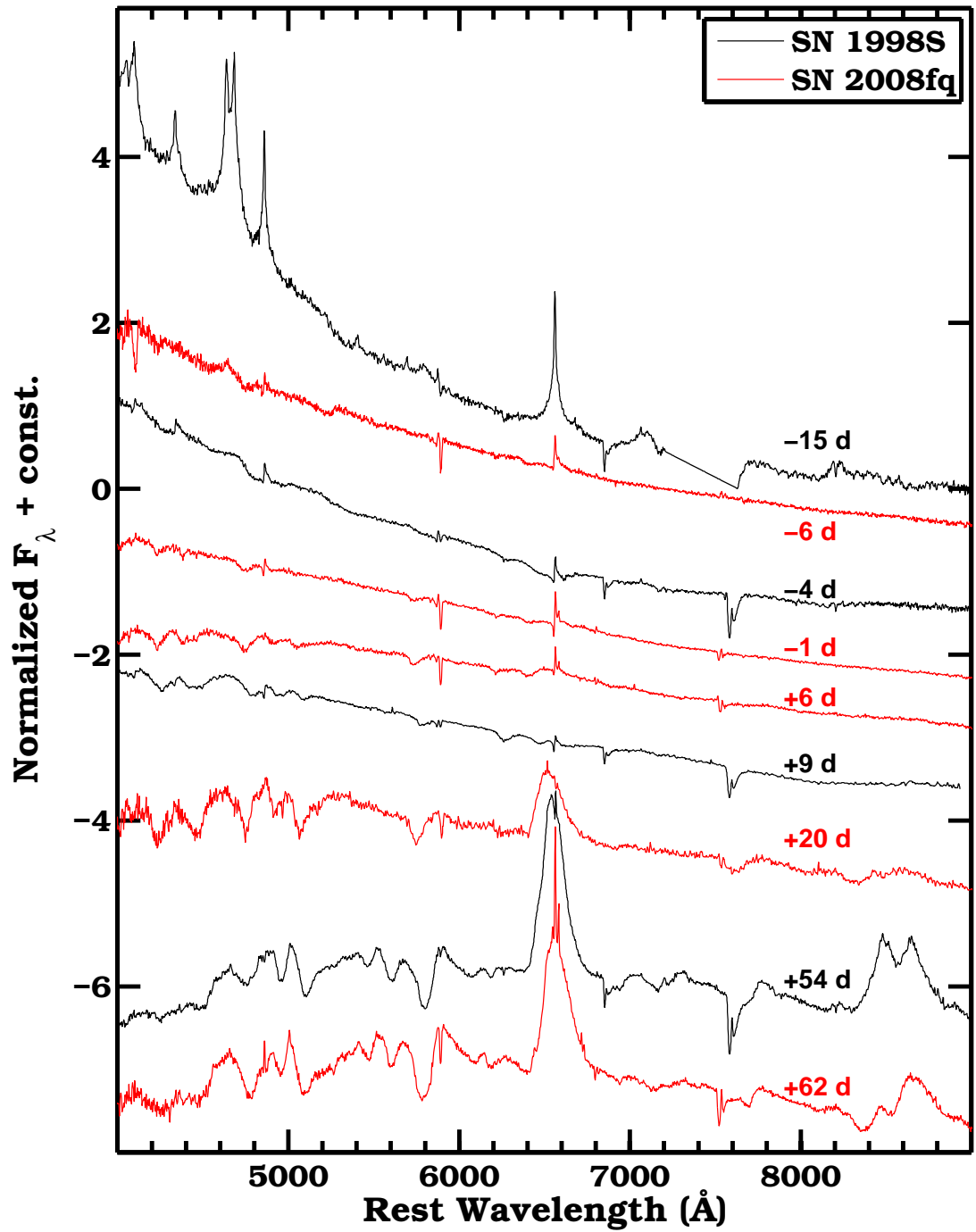}  
  \caption{\label{comp98S08fq}Spectral evolution of SNe~1998S and 2008fq. Temporal phase of each spectrum is given relative to  the time of  $V$-band maximum.}
 \end{figure}

\end{document}